\documentclass[10.5pt,a4paper,oneside]{extarticle}
\usepackage{amsmath}
\usepackage{amssymb}
\usepackage{mathrsfs}
\usepackage{braket}
\usepackage{bm}
\usepackage{bbm}
\usepackage{multirow}
\usepackage{mathtools}
\mathtoolsset{showonlyrefs=true}

\usepackage{latexsym}
\def\qed{\hfill $\Box$}
\usepackage{ntheorem}
\theoremstyle{break}
\newtheorem{thm}{Theorem}[section]
\newtheorem{lem}[thm]{Lemma}

\newtheorem{prop}[thm]{Proposition}
\newtheorem*{thm**}{Theorem}
\newtheorem*{lem**}{Lemma}
\newtheorem*{cor**}{Corollary}
\newtheorem*{prop**}{Proposition}

\theorembodyfont{\normalfont}

\newtheorem*{defi**}{Definition}
\newtheorem*{pf}{Proof}

\theoremheaderfont{\itshape}

\newtheorem*{eg**}{Example}
\newtheorem{rmk}[thm]{Remark}
\newtheorem*{rmk**}{Remark}

\usepackage{appendix}
\usepackage{setspace}

\newcommand{\Thm}{Theorem~}

\usepackage[margin=27mm]{geometry}
\usepackage{lineno}

\makeatletter
\renewcommand{\theequation}
{\arabic{section}.\arabic{equation}}
\@addtoreset{equation}{section}
\makeatother

\usepackage{hhline}

\usepackage[]{graphicx}

\usepackage[bf]{caption} 
\usepackage[subrefformat=parens]{subcaption}
\usepackage[]{hyperref}
\hypersetup{%
 bookmarksnumbered=true,%
 colorlinks=false,%
 setpagesize=false,%
 pdftitle={},%
 pdfauthor={Ryo Takakura},%
 pdfsubject={},%
 pdfkeywords={}
}

\newcommand{\R}{\mathbb{R}}
\newcommand{\C}{\mathbb{C}}

\newcommand{\Tr}{\mathrm{Tr}}

\newcommand{\ketbra}[2]{\ket{#1}\hspace{-0.25em}\bra{#2}}
\newcommand{\inter}[1]{\left\lceil{#1}\right\rfloor}

\newcommand{\ang}[1]{\left\langle{#1}\right\rangle}

\newcommand{\hi}{\mathcal{H}} 
 
\newcommand{\id}{\mathbbm{1}}

\newcommand{\A}{\mathsf{A}}
\newcommand{\B}{\mathsf{B}}
\newcommand{\E}{\mathsf{E}}

\newcommand{\x}{\mathrm{x}}

\newcommand{\Q}{\mathsf{Q}}

\usepackage{color}
\usepackage{xcolor}

\newcommand{\red}[1]{\textcolor{red}{#1}}

\usepackage{authblk}
\usepackage{comment}

\begin{document}  
\title{Optimal CHSH values for regular polygon theories in generalized probabilistic theories}

\author{Ryo Takakura\thanks{ryo.takakura@yukawa.kyoto-u.ac.jp}}
\affil{Yukawa Institute for Theoretical Physics\\
	Kyoto University\\
	Kitashirakawa Oiwakecho, Sakyo-ku, Kyoto, 606-8502, Japan}
\date{}
\maketitle

\begin{abstract}
In this study, we consider generalized probabilistic theories (GPTs) and focus on a class of theories called regular polygon theories, which can be regarded as natural generalizations of a two-level quantum system (a qubit system). In the usual CHSH setting for quantum theory, the CHSH value is known to be optimized by maximally entangled states. This research will reveal that the same observations are obtained also in regular polygon theories. Our result gives a physical meaning to the concept of ``maximal entanglement" in regular polygon theories.
\end{abstract}


\section{Introduction}
One of the most striking observations in quantum theory is the existence of entanglement.
Among its resulting phenomena, the violation of Bell inequality (or its specific form CHSH inequality) \cite{Bell_inequality,PhysRevLett.23.880} is particularly important because it dramatically changes our view of the world.
The importance lies not only in foundational aspects but also in applications in quantum physics such as quantum computations and quantum cryptography \cite{Brunner2014,Myrvold2021}.

Recently, much research has been given that aims to manifest what is essential in our world from perspectives beyond quantum theory.
In particular, studies on \textit{generalized probabilistic theories} (\textit{GPTs}) \cite{hardy2001quantum,PhysRevLett.99.240501,PhysRevA.75.032304,barnum2012teleportation,PhysRevA.81.062348,PhysRevA.84.012311,Masanes_Muller_2011} have been developing as one of those attempts: 
quantum foundations and applications such as uncertainty relations and teleportation protocols were generalized and their essence was examined in GPTs \cite{Plavala_2021_GPTs}.
In particular, entanglement or other non-local properties have been actively investigated in the field of GPTs \cite{Banik_steer,Barnum2013,pawlowski_information,PR-box,PhysRevA.89.022123,PhysRevA.96.052127} despite the indeterminacy in defining composite systems \cite{Aubrun2021}.
However, studies on the notion of ``maximally entangled states'' have not developed well although the mathematical definition of entangled states as non-separable states can be given in the same way as quantum theory.
There are studies on maximally entangled states in GPTs revealing their relations with local operations and classical communications (LOCCs) \cite{Chiribella2015} or steering \cite{Barnum2013}, but those results were obtained for certain classes of GPTs equipped with mathematical structures such as the possibility of ``purification".
Besides, there are studies where non-local properties of maximally entangled states in GPTs called \textit{regular polygon theories} were investigated \cite{1367-2630-13-6-063024,bipartite_polygon}.
Regular polygon theories can be naturally interpreted as intermediate theories between a classical trit system and a qubit system, and thus have been focused in the field of GPTs to find what is specific in classical and quantum theory \cite{PhysRevA.89.022123,PhysRevA.89.052124,PhysRevA.87.052131,Banik2019,Takakura_2019,PhysRevA.102.022203,Heinosaari_2022}.
Despite their geometrical simplicity, it has not been revealed yet whether maximally entangled states yield optimum CHSH values in regular polygon theories while those in quantum theory optimize it.

In the present study, we investigate maximal entanglement in regular polygon theories.
We consider a specific bipartite system (called the maximal tensor products) of a similar regular polygon theory, and focus on maximally entangled states in the composite system introduced by natural generalizations of those in quantum theory. 
Those maximally entangled states are the same ones in the previous study \cite{Barnum2013,1367-2630-13-6-063024} defined as order-isomorphisms between the cones of effects and states.
For those ``mathematically" introduced states, we prove that they are in fact ``physical" in the sense that they optimize the CHSH value similarly to the quantum case as conjectured in \cite{1367-2630-13-6-063024}. 
While only the simplest class of GPTs is treated, our result reveals relations between abstract and physical aspects of entanglement from a broader perspective of GPTs than quantum theory.

This paper is organized as follows.
In Sec.~\ref{sec:GPTs}, we give a brief review of GPTs.
There general formulation of GPTs and regular polygon theories are presented.
In particular, we introduce maximally entangled states in regular polygon theories in accord with \cite{Barnum2013,1367-2630-13-6-063024}.
In Sec.~\ref{sec:CHSH}, we review the CHSH scenario.
We apply the scenario to GPTs and rewrite it in terms of the so-called CHSH game \cite{Brunner2014,Lawson2010,Dey2013,Oppenheim1072}.
In Sec.~\ref{sec:main}, we present our main theorem and its proof.
It is revealed whether maximally entangled states optimize the CHSH value in regular polygon theories.

\section{Generalized probabilistic theories (GPTs)}
\label{sec:GPTs}
In this section, we present a brief explanation on the mathematical formulation of GPTs.
For its more detailed description, we recommend \cite{Plavala_2021_GPTs,Lami_PhD,Takakura_PhD}.

\subsection{Single systems}
\label{subsec:GPTs}
GPTs are physical theories where probabilistic mixtures of states and effects (observables) are possible.
Mathematically, a GPT is given by a pair of sets $(\Omega, \mathcal{E})$, where
\begin{itemize}
\item $\Omega$ is a compact convex set in a real and finite-dimensional Euclidean space $V$ with the standard inner product $\ang{\cdot,\cdot}$;
\item the origin $O$ of $V$ is not contained in $\Omega$ and the linear span $\mathit{span}(\Omega)$ of $\Omega$ is $V$;
\item $\mathcal{E}$ is the set of all elements $e$ in $V$ such that $\ang{e,\omega}\in[0,1]$ for all $\omega\in\Omega$;
\item in particular, there is an element $u\in \mathcal{E}$ such that $\ang{u,\omega}=1$ for all $\omega\in\Omega$.
\end{itemize}
The sets $\Omega$ and $\mathcal{E}$ are called the \textit{state space} and the \textit{effect space} of the theory, their elements \textit{states} and \textit{effects}, and their extreme points \textit{pure states} and \textit{pure effects} respectively.
The specific effect $u$ is called the \textit{unit effect}, and we call a family of effects $\{e_i\}_{i}$ an \textit{observable} if $\sum_i e_i=u$ (we only consider observables with finite outcomes in this paper).
States and effects (observables) are mathematical representations of preparations of systems and measurement procedures on them respectively, and their convexity represents the possibility of probabilistic mixtures.
We note that in the description above we made several assumptions such as 
the finite dimensionality of $V$ for mathematical simplicity.
In particular, we assume the \textit{no-restriction hypothesis} \cite{PhysRevA.81.062348} that any $e\in V$ such that $\ang{e,\forall\omega}\in[0,1]$ is an element of $\mathcal{E}$, i.e., it is physically realizable.
It is often convenient to introduce the set $V_+$ and $V_+^*$ of ``unnormalized" states and effects respectively defined as 
\begin{equation}
V_+=\{x\in V\mid x=\lambda\omega, \lambda\ge0, \omega\in\Omega\}
\end{equation}
and
\begin{equation}
V_+^*=\{f\in V\mid f=\tau e, \tau\ge0, e\in\mathcal{E}\}.
\end{equation}
The set $V_+$ is called the \textit{positive cone} and $V_+^*$ the \textit{dual cone} of the theory.
A GPT is called \textit{self-dual} if its positive cone $V_+$ and dual cone $V_+^*$ satisfy $V_+=V_+^*$, and called \textit{weakly self-dual} if there is a linear bijection $\phi\colon V\to V$ such that $\phi(V_+^*)=V_+$ \cite{barnum2012teleportation,Barnum2013,1367-2630-13-6-063024}.
The notion of (weak) self-duality plays an important role when discussing our main result.

We present two classes of GPTs as examples, classical and quantum theory.
From the perspective of GPTs, the convex hull $\Delta_n\subset\R^{n+1}$ of the $(n+1)$ vectors of an orthonormal basis $\{v_i\}_{i=1}^{n+1}$ in $\R^{n+1}$ (an $n$-simplex) expresses the state space of an $(n+1)$-level classical theory.
The corresponding effect space $\mathcal{E}(\Delta_n)$ is given by 
\[
\mathcal{E}(\Delta_n)=\{f\in\R^{n+1}\mid f=\sum_{i=1}^{n+1} \lambda_i v_i,\  0\le\forall\lambda_i\le1\}
\]
and the unit effect is $u=\sum_{i=1}^{n+1} v_i$ because the standard inner product of a pure state $v_i$ and an element $f=\sum_{i=1}^{n+1} \lambda_i v_i$ of $\mathcal{E}(\Delta_n)$ is calculated as $\ang{f, v_i}=\lambda_i$.
It is easy to see that the positive and dual cones generated respectively by $\Delta_n$ and $\mathcal{E}(\Delta_n)$ are identical, i.e., the classical theory $(\Delta_n, \mathcal{E}(\Delta_n))$ is self-dual.
On the other hand, from the perspective of GPTs, the finite-dimensional quantum theory associated with a $d$-dimensional Hilbert space $\mathcal{H}=\C^{d}$ ($d<\infty$) is expressed as $(\mathcal{S}(\mathcal{H}), \mathcal{E}(\mathcal{H}))$, where the state space $\mathcal{S}(\mathcal{H})$ is 
\[
\mathcal{S}(\mathcal{H})=\{\rho\in M_d(\C)\mid \rho\ge0,\ \Tr[\rho]=1\}
\]
and the effect space $\mathcal{E}(\mathcal{H})$ is
\[
\mathcal{E}(\mathcal{H})=\{E\in M_d(\C)\mid 0\le E\le I\}
\]
including the identity operator $I$ on $\hi$ as the unit effect.
The positive and dual cones are the set of all positive operators on $\hi$, and thus the quantum theory $(\mathcal{S}(\mathcal{H}), \mathcal{E}(\mathcal{H}))$ is also self-dual.
It is known that $(\mathcal{S}(\mathcal{H}), \mathcal{E}(\mathcal{H}))$ can be embedded into $\R^{d^2}$ equipped with the Hilbert-Schmidt inner product as the standard inner product, and is consistent with the formulation presented at the beginning of this subsection.

\subsection{Bipartite systems and entanglement}
\label{subsec_bipartite}
In this part, we explain how to describe bipartite systems and introduce the notion of entanglement in GPTs.
Let $(\Omega_A, \mathcal{E}_A)$ and $(\Omega_B, \mathcal{E}_B)$ be two GPTs and consider their composite system.
A fundamental assumption is that the composite is also a GPT, which we write by $(\Omega_{AB}, \mathcal{E}_{AB})$.
In addition, requiring physically natural axioms such as the no-signaling principle, we have 
\begin{itemize}
\item the embedding vector space $V_{AB}=\mathit{span}(\Omega_{AB})$ of the state space $\Omega_{AB}$ is given by the tensor product of Euclidean spaces $V_A=\mathit{span}(\Omega_{A})$ and $V_B=\mathit{span}(\Omega_{B})$, i.e., $V_{AB}=V_A\otimes V_B$ (we write by $\ang{\cdot,\cdot}_{AB}$ the standard inner product of $V_{AB}$);
\item the effect space $\mathcal{E}_{AB}$ is also embedded into $V_{AB}$ by $\mathcal{E}_{AB}=\{\xi\in V_{AB}\mid\ang{\xi, \mu}_{AB}\in[0,1]\ \forall\mu\in\Omega_{AB}$\};
	\item the independent preparation of states $\omega\in \Omega_A$ and $\xi\in \Omega_B$ in each system is given by $\omega\otimes\xi$, and the independent measurement of effects $e\in \mathcal{E}_A$ and $f\in \mathcal{E}_B$ is $e\otimes f$;
	\item the unit effect $u_{AB}\in\mathcal{E}_{AB}$ for $\Omega_{AB}$ is given by $u_{AB}=u_A\otimes u_B$, where $u_A$ and $u_B$ are the respective unit effect for $\Omega_A$ and $\Omega_B$; 
	\item $\Omega_A\otimes_{min}\Omega_B\subseteq\Omega_{AB}\subseteq\Omega_A\otimes_{max}\Omega_B$, where 
	\begin{equation*}
	\label{eq:min tensor}
\begin{aligned}
\Omega_{A}\otimes_{min}\Omega_{B}
=\{\mu\in V_A\otimes V_B\mid\ &\mu=\sum_{i=1}^{n}p_i \omega_i\otimes\xi_i,\ \omega_i\in\Omega_{A},
\\&\xi_i\in\Omega_B,\ p_i\ge0,\ \sum_{i=1}^{n}p_i=1,\ n:\mbox{finite}\}
\end{aligned}
	\end{equation*}
	and
\begin{equation*}
	\label{eq:max tensor}
	\begin{aligned}
			\Omega_{A}\otimes_{max}\Omega_{B}
=\{\mu\in V_A\otimes V_B\mid&\ang{u_A\otimes u_B,\mu}_{AB}=1,
\\&\ \ \ang{e\otimes f,\mu}_{AB}\ge0\ \mbox{for all $e\in\mathcal{E}_{A}, f\in\mathcal{E}_{B}$}\}
\end{aligned}
	\end{equation*}
(similarly $\mathcal{E}_A\otimes_{min}\mathcal{E}_B\subseteq\mathcal{E}_{AB}\subseteq\mathcal{E}_A\otimes_{max}\mathcal{E}_B$).
\end{itemize}
In the description above, the set $\Omega_{A}\otimes_{min}\Omega_{B}$ is called the \textit{minimal tensor product} of $\Omega_A$ and $\Omega_B$, and its elements are called \textit{separable states}.
On the other hand, the set $\Omega_{A}\otimes_{max}\Omega_{B}$ is called the \textit{maximal tensor product} and elements in $\Omega_{A}\otimes_{max}\Omega_{B}\backslash\Omega_{A}\otimes_{min}\Omega_{B}$ are called \textit{entangled states} (separable and entangled effects are defined in the same way).
It was shown in \cite{Aubrun2021} that $\Omega_{A}\otimes_{max}\Omega_{B}\backslash\Omega_{A}\otimes_{min}\Omega_{B}\neq\emptyset$ if and only if neither $\Omega_A$ nor $\Omega_B$ is a simplex. 
We note that $\Omega_{A}\otimes_{min}\Omega_{B}$ and $\Omega_{A}\otimes_{max}\Omega_{B}$ are compact convex sets \cite{Aubrun_2019preprint}.

There is a one-to-one correspondence between elements in $\Omega_{A}\otimes_{max}\Omega_{B}$ and normalized and cone-preserving maps between the cones $V_{A+}^*$ and $V_{B+}$ generated by $\mathcal{E}_A$ and $\Omega_B$ respectively.
In fact, an element $\eta\in\Omega_{A}\otimes_{max}\Omega_{B}$ defines a linear map $\hat{\eta}\colon V_{A}\to V_{B}$ that maps $e\in V_A$ to $\hat{\eta}(e)\in V_B$ by
\begin{equation}
\label{state-map}
\ang{f, \hat{\eta}(e)}_B=\ang{e\otimes f, \eta}_{AB}\quad(\forall f\in V_B),
\end{equation} 
where $\ang{\cdot,\cdot}_B$ denotes the inner product in $V_B$.
By virtue of the relation \eqref{state-map}, the linear map $\hat{\eta}$ is understood as giving a conditional state of Bob after a local measurement by Alice on the bipartite state $\eta$.
We can find that the map $\hat{\eta}$ is normalized and cone-preserving (or \textit{positive}), i.e., $\ang{u_B, \hat{\eta}(u_A)}_B=1$ and $\hat{\eta}(V_{A+}^*)\subset V_{B+}$.
Conversely, if a linear map $\hat{\eta}\colon V_{A}\to V_B$ satisfies $\ang{u_B, \hat{\eta}(u_A)}_B=1$ and $\hat{\eta}(V_{A+}^*)\subset V_{B+}$, then we can construct $\eta\in V_A\otimes V_B$ through $\hat{\eta}$ as $\ang{e\otimes f, \eta}_{AB}=\ang{f, \hat{\eta}(e)}_B$ ($\forall e\in V_A,\ f\in V_B$), and it is easy to show that $\eta\in\Omega_{A}\otimes_{max}\Omega_{B}$ holds.
It is interesting to investigate entangled states in quantum theory from this viewpoint.
For simplicity, let us consider a bipartite quantum system $(\mathcal{S}(\mathcal{H}\otimes\mathcal{H}), \mathcal{E}(\mathcal{H}\otimes\mathcal{H}))$ with $\hi=\C^d$, whose subsystems are both described by $(\mathcal{S}(\mathcal{H}), \mathcal{E}(\mathcal{H}))$, and focus on a maximally entangled state $\rho_\star=\frac{1}{d}\sum_{i,j=1}^d\ketbra{i}{j}\otimes\ketbra{i}{j}$, where $\{\ket{i}\}_{i=1}^d$ is an orthonormal basis of $\mathcal{H}$.
Through the formula \eqref{state-map}, the bipartite state $\rho_\star$ can be expressed also as a linear map $\hat{\rho}_\star\colon V\to V$ between the corresponding embedding space $V=\mathit{span}(\mathcal{S}(\mathcal{H}))=\R^{d^2}$ for $(\mathcal{S}(\mathcal{H}), \mathcal{E}(\mathcal{H}))$ that maps $E\in\mathcal{E}(\mathcal{H})$ to its transpose $\frac{1}{d}E^{T}$ with respect to the basis $\{\ket{i}\}_{i=1}^d$.
This implies that the (unnormalized) local state of Bob after Alice's measuring the effect $E$ locally on the bipartite state $\rho_\star$ is $\frac{1}{d}E^{T}$.
Such observation is a concrete example of results in quantum measurement theory \cite{doi:10.1063/1.526000,Busch_quantummeasurement}.
We can find an important property of the maximally entangled state $\hat{\rho}_\star$ that it is an \textit{order-isomorphism} \cite{Barnum2013} between the dual cone $V_+^*$ and positive cone $V_+$  generated respectively by $\mathcal{E}(\mathcal{H})$ and $\mathcal{S}(\mathcal{H})$.
That is, the map $\hat{\rho}_\star$ is a bijective linear map such that $\hat{\rho}_\star(V_+^*)=V_+$, and it thus maps effects in rays of $V_+^*$ to states in rays of $V_+$.
It is also important that $\hat{\rho}_\star$ is norm-preserving and maps the identity $I$ (unit effect) to the maximally mixed state $\frac{I}{d}$.
These observations will be used for generalizing the notion of maximally entangled states in Subsec.~\ref{subsec:RPT}.

\subsection{Regular polygon theories}
\label{subsec:RPT}
\begin{figure}[h]
	\centering
	\includegraphics[scale=0.39]{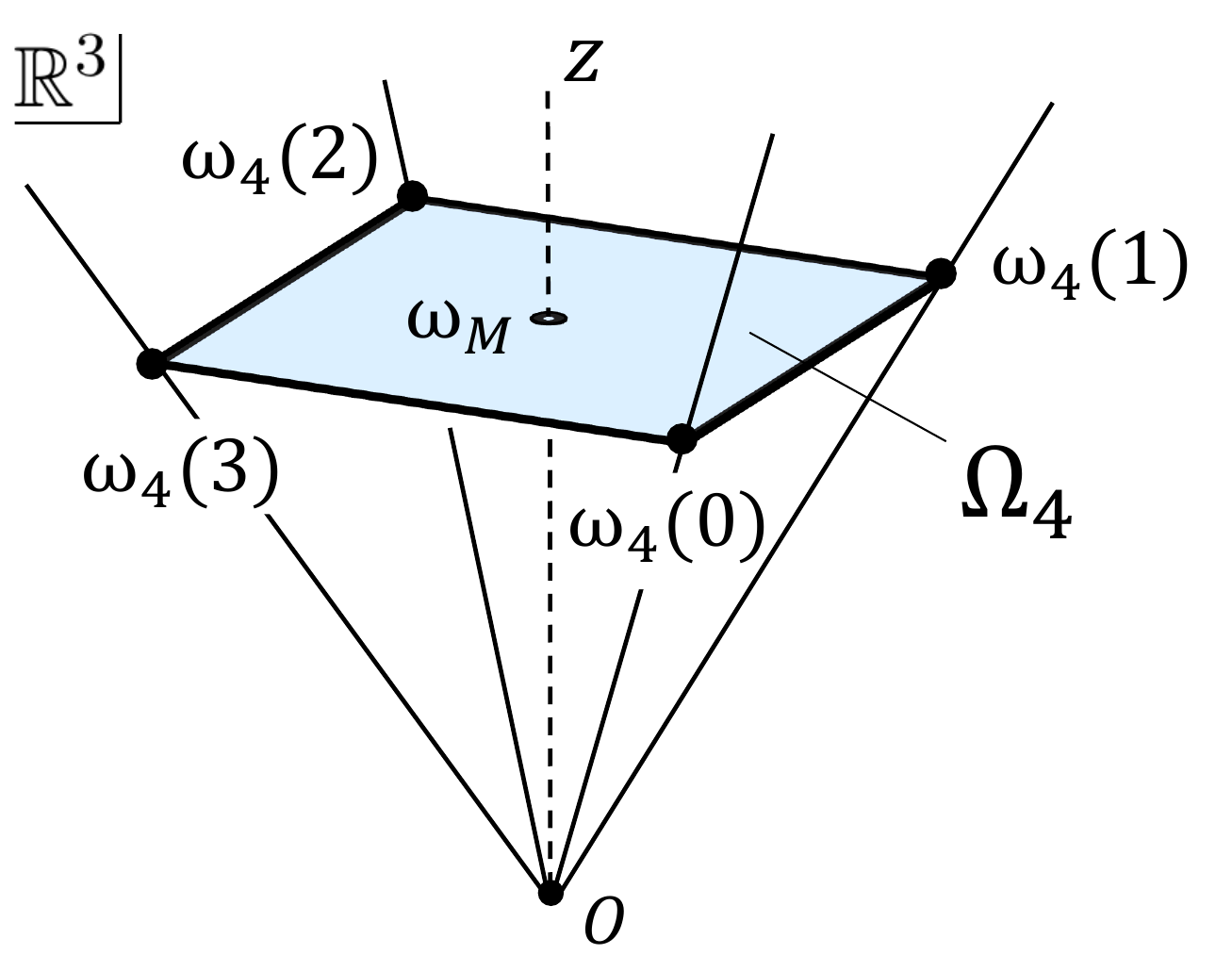}
	\caption{
		Illustration of the state space $\Omega_4$ (colored in blue) on the hyperplane $z=1$ in $\R^3$.
		There are four pure states $\{\omega_4(i)\}_{i=0}^3$ and the maximally mixed state $\omega_M=(0,0,1)^{T}$ in the square state space.}
	\label{fig:Omega4}
\end{figure}

We introduce a specific class of GPTs called regular polygon theories \cite{1367-2630-13-6-063024}.
As we will see, these theories can be considered as intermediate theories between certain classical and quantum theories.
Our main result is about the optimal CHSH value in a composite of regular polygon theories.
A GPT is called a regular polygon theory if its state space is a two-dimensional regular polygon on the hyperplane $z=1$ in $\R^3$.
Concretely, with an integer $n\ge3$, the state space of a regular polygon theory is given by the convex hull $\Omega_n$ of $n$ ``vertices" (pure states) $\{\omega_n(i)\}_{i=0}^{n-1}$, 
where
\begin{align}
	\label{def:polygon pure state}
	\omega_{n}(i)=
	\left( 
	\begin{array}{c}
		r_{n}\cos({\frac{2\pi i}{n}})\\
		r_{n}\sin({\frac{2\pi i}{n}})\\
		1
	\end{array}
	\right)\quad\mbox{with}\quad r_{n}=\sqrt{\frac{1}{\cos({\frac{\pi}{n}})}}.
\end{align}
There is an important state $\omega_M$ in the state space $\Omega_n$ called the maximally mixed state defined as
\begin{equation}
	\label{eq:omega_M}
	\omega_M=\frac{1}{n}\sum_{i=1}^n\omega_{n}(i)=	\left(
	\begin{array}{c}
		0\\
		0\\
		1
	\end{array}
	\right).
\end{equation}
We illustrate the state space $\Omega_4$ in Fig.~\ref{fig:Omega4}.
For the state space $\Omega_n$, the corresponding effect space $\mathcal{E}_n$ is given by the convex hull of its pure effects $\mathcal{E}_n^{\mathrm{ext}}$, where
\begin{equation}
	\label{def:polygon pure effect0}
	\mathcal{E}^{\mathrm{ext}}_n=
	\left\{
	\begin{aligned}
		&\ \{e_{n}(i)\}_{i=0}^{n-1}\quad(n:\mbox{even}),\\
		&\ \{e_{n}(i)\}_{i=0}^{n-1} \cup \{\overline{e_{n}(i)}\}_{i=0}^{n-1}\quad(n:\mbox{odd})
	\end{aligned}
\right.
\end{equation}
with
\begin{equation}
	\label{def:polygon pure effect}
	e_{n}(i)=\left\{
	\begin{aligned}
		&\ \frac{1}{2}
		\left(
		\begin{array}{c}
			r_{n}\cos({\frac{(2i+1)\pi}{n}})\\
			r_{n}\sin({\frac{(2i+1)\pi}{n}})\\
			1
		\end{array}
		\right)\quad(n:\mbox{even})\\
		&\frac{1}{1+r_{n}^{2}}
		\left(
		\begin{array}{c}
			r_{n}\cos({\frac{2i\pi}{n}})\\
			r_{n}\sin({\frac{2i\pi}{n}})\\
			1
		\end{array}
		\right)\quad(n:\mbox{odd}),
	\end{aligned}
\right.
\quad
\overline{e_{n}(i)}=u-e_{n}(i),\quad
	u=	\left(
\begin{array}{c}
	0\\
	0\\
	1
\end{array}
\right).
\end{equation}
We note that $\overline{e_{n}(i)}=u-e_{n}(i)=e_n(i+\frac{n}{2})$ holds for even $n$.
Comparing the states \eqref{def:polygon pure state} and effects \eqref{def:polygon pure effect0} or \eqref{def:polygon pure effect} for a regular polygon theory $(\Omega_n, \mathcal{E}_n)$, we can find that $(\Omega_n, \mathcal{E}_n)$ is either weakly self-dual or self-dual contingent on the parity of $n$. 
In fact, the positive cone $V_{n+}$ and dual cone $V_{n+}^*$ generated respectively by $\Omega_n$ and  $\mathcal{E}_n$ satisfy 
\begin{equation}\label{def_max ent00}
	T_n(e_n(i))=\left\{
		\begin{aligned}
			&\ \frac{1}{2}~\omega_n(i)&&(n:\mbox{even})\\
			&\ \frac{1}{1+r_{n}^{2}}~\omega_n(i)&&(n:\mbox{odd})
		\end{aligned}
	\right.
\quad\mbox{and}\quad	T_n(V_{n+}^*)=V_{n+},
\end{equation}
where $T_n\colon\R^3\to\R^3$ is an order-isomorphism between the cones $V_{n+}^*$ and $V_{n+}$ defined as
\begin{equation}
	\label{def_max ent0}
	T_n=\left\{\ 
	\begin{aligned}
		&\begin{pmatrix}
			\cos\frac{\pi}{n} & \sin\frac{\pi}{n} & 0\\
			-\sin\frac{\pi}{n} & \cos\frac{\pi}{n}  & 0\\
			0 & 0 & 1\\
		\end{pmatrix}
		\quad(\mbox{$n$: even})
		\\
		&\begin{pmatrix}
			1 & 0 & 0\\
			0 & 1 & 0\\
			0 & 0 & 1\\
		\end{pmatrix}
		\quad(\mbox{$n$: odd}),
	\end{aligned}
	\right.
\end{equation}
and thus $(\Omega_n, \mathcal{E}_n)$ is weakly self-dual for even $n$ and self-dual for odd $n$. 
We can also consider the limiting theory $(\Omega_\infty, \mathcal{E}_\infty)$, whose set of pure states $\{\omega_\infty(\theta)\}_{\theta\in[0,2\pi)}$ and effects $\{e_\infty(\theta)\}_{\theta\in[0,2\pi)}$ are given respectively by
\begin{equation}
	\label{def:polygon pure effect_infty}
		\omega_{\infty}(\theta)=
		\left(
		\begin{array}{c}
			\cos\theta\\
			\sin\theta\\
			1
		\end{array}
		\right),\quad
		e_{\infty}(\theta)=\frac{1}{2}
		\left(
		\begin{array}{c}
			\cos\theta\\
			\sin\theta\\
			1
		\end{array}
		\right)
\end{equation}
(the maximally mixed state is the same as \eqref{eq:omega_M}).
It is easy to see from \eqref{def:polygon pure effect_infty} that the theory $(\Omega_\infty, \mathcal{E}_\infty)$ is self-dual.

Regular polygon theories can be regarded as including certain classical and quantum theories.
In fact, the theory $(\Omega_3, \mathcal{E}_3)$ represents a classical trit system: any point in the triangle state space $\Omega_3$ has a unique convex decomposition into the three distinguishable pure states.
On the other hand, the theory $(\Omega_\infty, \mathcal{E}_\infty)$ is also important because it represents a qubit system with real coefficients.
The disc state space $\Omega_\infty$ corresponds to the equatorial plane of the Bloch ball, which is the set of all qubit states without components for the Pauli operator $\sigma_y$.
Regular polygon theories thus can be regarded as physical theories between primitive classical and quantum systems.

In the composite system $\Omega_n\otimes_{max}\Omega_n$ $(n>3)$, we can naturally introduce generalizations of maximally entangled states in quantum theory through the identification of $\Omega_n\otimes_{max}\Omega_n$ with the set of all normalized and cone-preserving maps.
With the set $GL(\Omega_n)$ of all linear bijection $T\colon\R^3\to\R^3$ such that $T(\Omega_n)=\Omega_n$ and the isomorphism $T_n$ introduced in \eqref{def_max ent00} and \eqref{def_max ent0}, 
we call an element $\eta\in\Omega_n\otimes_{max}\Omega_n$ a \textit{maximally entangled state} if its inducing normalized and cone-preserving map $\hat{\eta}\colon V_{n+}^*\to V_{n+}$ between the positive cone $V_{n+}$ and dual cone $V_{n+}^*$ belongs to the set $T_{n}\cdot GL(\Omega_n)$.
We apply this definition also to the case $n=\infty$ by setting 
\[
T_\infty=\begin{pmatrix}
	1 & 0 & 0\\
	0 & 1 & 0\\
	0 & 0 & 1\\
\end{pmatrix},
\]
that is, elements belonging to $T_{\infty}\cdot GL(\Omega_\infty)(=GL(\Omega_\infty))$ are called maximally entangled states in the composite $\Omega_\infty\otimes_{max}\Omega_\infty$.
We note that $GL(\Omega_n)$ is composed of orthogonal transformations in $\R^3$ that keep the maximally mixed state $\omega_M$ in \eqref{eq:omega_M} invariant
\cite{Dummit_abstractalgebra}.
Our definition of maximally entangled states is the same one introduced in \cite{1367-2630-13-6-063024}.
As discussed there, elements in $T_{n}\cdot GL(\Omega_n)$ satisfy similar properties to quantum maximally entangled states reviewed in Subsec.~\ref{subsec:GPTs} (for example, it maps the unit effect $u$ to the maximally mixed state $\omega_M$), and, in addition, are pure in $\Omega_n\otimes_{max}\Omega_n$ whenever $n>3$ \cite{Barnum2013}.
Therefore, they can be regarded as reasonable generalizations of maximally entangled states in quantum theory to GPTs: in particular, the usual quantum maximally entangled state (Bell state) $\frac{1}{2}\sum_{i,j=0,1}\ketbra{i}{j}\otimes\ketbra{i}{j}\in\Omega_\infty\otimes_{max}\Omega_\infty$ belongs to $GL(\Omega_\infty)$ since its inducing transposition map with respect to the z-basis $\{\ket{0}, \ket{1}\}$ is just the identity map $T_\infty$.
We remark that such maximally entangled states cannot always be introduced in GPTs.
In fact, if we consider a regular cube in $\R^4$ as the state space of a GPT, then the corresponding dual cone is given by the conic hull of a regular octahedron \cite{Grunbaum_polytope}, which is not isomorphic to the cube, and there is no order-isomorphism between the dual and positive cones.

\begin{rmk}\label{rmk1}
For the classical case $n=3$, we have $\Omega_3\otimes_{max}\Omega_3=\Omega_3\otimes_{min}\Omega_3$ and thus all composite states are separable.
In this case, each $\hat{\sigma}\in GL(\Omega_3)$ is a permutation on the set $\{0,1,2\}$ or considered as $\sigma=\frac{1}{3}\sum_{i=0}^2\omega_3(\hat{\sigma}(i))\otimes\omega_3(i)\in \Omega_3\otimes_{min}\Omega_3$.
\end{rmk}

\section{CHSH values}
\label{sec:CHSH}
In this section, we explain the CHSH scenario and introduce the CHSH value.
\subsection{Preliminaries}
\label{subsec:CHSH}
In the CHSH scenario, two spacelike separated parties, Alice and Bob, share an input-output apparatus, and they input $s$ and $t$ ($s, t\in\{0,1\}$) independently and randomly to each subapparatus to obtain outputs $a$ and $b$ ($a,b\in\{1,-1\}$) respectively.
It results in a set of probabilities $\mathbf{p}:=\{p(a,b|s,t)\}_{a,b\in\{1,-1\},s,t\in\{0,1\}}$, where $p(a,b|s,t)$ represents the probability of observing outcomes $a$ and $b$ when  $s$ and $t$ are input by Alice and Bob respectively.
We note that we set $p(s,t)=p(s)p(t)$ and $p(s)=p(t)=\frac{1}{2}$ for any $(s,t)$ to reflect the independent and random choice of the input.
We define the CHSH value $C[\mathbf{p}]$ for $\mathbf{p}=\{p(a,b|s,t)\}_{a,b,s,t}$ as 
\begin{equation}
\label{def:CHSH}
C[\mathbf{p}]=E(00)+E(01)+E(10)-E(11),
\end{equation}
where
\begin{equation*}
E(st)=\sum_{a,b\in\{1,-1\}} ab~p(a,b|s,t)\quad((s,t)\in\{0,1\}^2).
\end{equation*}
It is a well-known result that if $\mathbf{p}$ is a hidden variable model satisfying the Bell-locality condition, then $|C[\mathbf{p}]|\le2$ holds (the Bell-CHSH inequality) \cite{Bell_inequality,PhysRevLett.23.880,Brunner2014,Myrvold2021}.

We apply the above argument to the state-observable formulation in GPTs.
Let Alice and Bob be with systems described respectively by GPTs $(\Omega_A, \mathcal{E}_A)$ and $(\Omega_B, \mathcal{E}_B)$, and $(\Omega_{AB}, \mathcal{E}_{AB})$ be their composite.
They have two binary observables $(\A_0, \A_1)$ and $(\B_0, \B_1)$ respectively, and share a bipartite state $\eta\in\Omega_{AB}$.
In this setting, Alice chooses randomly one of the observables $\A_s$ ($s\in\{0,1\}$) and then performs its measurement locally on her subsystem of the state $\eta$ to obtain an outcome $a$ ($a\in\{1,-1\}$), and similarly, Bob chooses and measures $\B_t$ ($t\in\{0,1\}$) on his subsystem to obtain an outcome $b$ ($b\in\{1,-1\}$).
Now an input-output scheme is constructed and the resulting probabilities $\mathbf{p}=\{p(a, b|s,t)\}_{a,b,s,t}$ is obtained through
\begin{equation}
\label{prob}
p(a, b|s,t)=\ang{\A_s^a\otimes \B_t^b,\eta}_{AB},
\end{equation}
where $\A_s^a$ is the effect of the observable $\A_s$ corresponding to the outcome $a$ (similarly for $\B_t^b$) and $\ang{\cdot,\cdot}_{AB}$ is the inner product in $V_{AB}=\mathit{span}(\Omega_{AB})$.
We note that the order of measurements performed by Alice and Bob does not affect the observations due to the no-signaling condition.
An important example of such set of probabilities $\mathbf{p}$ is obtained in quantum theory: suitable choices of an entangled state and observables yield $|C[\mathbf{p}]|=2\sqrt{2}$ \cite{Brunner2014,Tsirelson1980}.

\subsection{CHSH values via CHSH games}
\label{subsec:CHSH game}
We can study the CHSH scenario in another way known as the CHSH (or nonlocal) game \cite{Brunner2014,Lawson2010,Dey2013,Oppenheim1072}.
It will be found that this setting is more useful than the original one for deriving our main result.
In the description of a CHSH game, both Alice and Bob again have two choices of input values $\{0,1\}$.
They choose independently and randomly their input $s$ and $t$ ($s, t\in\{0,1\}$), and then obtain output values $a$ and $b$ ($a,b\in\{0,1\}$), which are slightly different from the previous scenario, respectively.
The scheme is characterized by a set of probabilities $\mathbf{p}=\{p(a, b|s,t)\}_{a,b,s,t}$, where $p(a, b|s,t)$ is the probability of obtaining outputs $a$ and $b$ when $s$ and $t$ are input by Alice and Bob respectively.
In this setting, we say that Alice and Bob win the CHSH game if the values satisfy $a\oplus b=s\cdot t$.
The winning probability $P_{\mathrm{win}}[\mathbf{p}]$ is calculated as (remember that $p(s,t)=p(s)p(t)$ and $p(s)=p(t)=\frac{1}{2}$ hold)
\begin{equation}
	\label{win}
	P_{\mathrm{win}}[\mathbf{p}]=\frac{1}{4}\sum_{s,t\in\{0,1\}}\sum_{a,b\in\{0,1\}}V(a, b|s,t)p(a, b|s,t)
\end{equation}
with
\begin{equation}
	\label{eq:V}
V(a, b|s,t)=\left\{
\begin{aligned}
	&1\quad(a\oplus b=s\cdot t)\\
	&0\quad(\mbox{otherwise}).
\end{aligned}
\right.
\end{equation}
It is known that the winning probability \eqref{win} is equivalent to the CHSH value \eqref{def:CHSH} in the sense that 
\begin{equation}
	\label{win-CHSH}
	C[\mathbf{p}]=4~(2P_{\mathrm{win}}[\mathbf{p}]-1),
\end{equation}
where we identify the two $\mathbf{p}$ in \eqref{def:CHSH} and \eqref{win} by the replacement of the values for $(a,b)$: $1\to0$ and $-1\to1$.

As in subsec.~\ref{subsec:CHSH}, let us analyze the winning probability \eqref{win} from the perspective of GPTs.
We consider the same situation as \eqref{prob}.
In this case, a bipartite state $\eta\in\Omega_{AB}$ and binary observables $(\A_0, \A_1)$ and $(\B_0, \B_1)$ respectively of Alice and Bob determine the set of probabilities $\mathbf{p}=\{p(a,b|s,t)\}_{a,b,s,t}$ through \eqref{prob}, and thus we write $\mathbf{p}=(\eta; \A_0, \A_1; \B_0, \B_1)$ instead of the original expression.
Then we can rewrite \eqref{win} as
\begin{align}
		\label{win_GPT0}
		P_{\mathrm{win}}[\eta; \A_0, \A_1; \B_0, \B_1]
		&=\frac{1}{4}\sum_{s,t\in\{0,1\}}\sum_{a,b\in\{0,1\}}V(a, b|s,t)\ang{\A_s^a\otimes \B_t^b,\eta}_{AB}\notag\\
		&=\frac{1}{4}\sum_{s,t\in\{0,1\}}\sum_{a,b\in\{0,1\}}V(a, b|s,t)\ang{\B_t^b,\hat{\eta}(\A_s^a)}_{B},
\end{align}
where $\hat{\eta}$ is the normalized and cone-preserving map from $V_A=\mathit{span}(\Omega_A)(=\mathit{span}(\mathcal{E}_A))$ to $V_B=\mathit{span}(\Omega_B)$ induced by the state $\eta$ through \eqref{state-map} and $\ang{\cdot,\cdot}_B$ is the standard inner product in $V_B$.
To make \eqref{win_GPT0} simpler, we introduce 
\begin{equation}
	\label{eq:Q}
	\mathsf{Q}_s^a=\frac{1}{2}\sum_{t\in\{0,1\}}\sum_{b\in\{0,1\}}V(a, b|s,t)\B_t^b
\end{equation}
for each $(s,a)\in\{0,1\}^2$.
They are explicitly given as
\begin{equation}\label{eq:Q(s=0)}
	\mathsf{Q}_{s=0}^{a=0}=\frac{1}{2}\left(\B_{t=0}^{b=0}+\B_{t=1}^{b=0}\right),
	\ \ 
	\mathsf{Q}_{s=0}^{a=1}=\frac{1}{2}\left(\B_{t=0}^{b=1}+\B_{t=1}^{b=1}\right)
\end{equation}
and
\begin{equation}\label{eq:Q(s=1)}
		\mathsf{Q}_{s=1}^{a=0}=\frac{1}{2}\left(\B_{t=0}^{b=0}+\B_{t=1}^{b=1}\right),\ \ 
		\mathsf{Q}_{s=1}^{a=1}=\frac{1}{2}\left(\B_{t=0}^{b=1}+\B_{t=1}^{b=0}\right)
\end{equation}
by means of \eqref{eq:V}.
We note that  $\Q_s:=\{\Q_s^a\}_{a\in\{0,1\}}$ is an observable on $\Omega_B$ for each $s\in\{0,1\}$ because $\sum_{a\in\{0,1\}}\Q_s^a=u_B$ holds, where $u_B$ is the unit effect for $\Omega_B$.
The equation \eqref{win_GPT0} now becomes
\begin{equation}\label{win_GPT00}
		P_{\mathrm{win}}[\eta; \A_0, \A_1; \B_0, \B_1]=\frac{1}{2}\sum_{s\in\{0,1\}}\sum_{a\in\{0,1\}}\ang{\mathsf{Q}_s^a,\hat{\eta}(\A_s^a)}_{B}.
\end{equation}
Following \cite{Oppenheim1072}, we also introduce
\begin{equation}\label{eq:ensemble}
	p(a|s)=\ang{u_B,\hat{\eta}(\A_s^a)}_B,\quad \omega_s^a=\frac{\hat{\eta}(\A_s^a)}{\ang{u_B,\hat{\eta}(\A_s^a)}_B}
\end{equation}
for each $(s,a)\in\{0,1\}^2$.
Since $\hat{\eta}$ is normalized and cone-preserving, $p(a|s)\ge0$ and $\sum_{a\in\{0,1\}}p(a|s)=1$, and $\omega_s^a\in\Omega_B$ hold for each $(s,a)$.
As explained in Subsec.~\ref{subsec_bipartite}, the family
$(p(a|s); \omega_s^a)_{a}$ represents the ``assemblage'' \cite{Pusey2013} of Bob's system after Alice's measurements of $\A_s$ on her local system.
The fact that $\sum_{a\in\{0,1\}}p(a|s)\omega_s^a(=\hat{\eta}(u_B))$ does not depend on Alice's choice $s$ ensures the no-signaling condition.
Overall, we obtain a simpler form of \eqref{win_GPT0} as
\begin{equation}
	\label{win_GPT}
	P_{\mathrm{win}}[\eta; \A_0, \A_1; \B_0, \B_1]=\frac{1}{2}\sum_{s\in\{0,1\}}\sum_{a\in\{0,1\}}p(a|s)\ang{\mathsf{Q}_s^a,\omega_s^a}_{B}.
\end{equation}
This equation implies that the winning probability $P_{\mathrm{win}}$ is determined by observables $(\Q_{0}, \Q_{1})$ and assemblages $(p(a|0); \omega_{0}^a)_{a}$, $(p(a|1); \omega_{1}^a)_{a}$ satisfying 
$\sum_{a\in\{0,1\}}p(a|0)\omega_0^a=\sum_{a\in\{0,1\}}p(a|1)\omega_1^a$ on Bob's system.
We note that we can develop similar argument to express $P_{\mathrm{win}}[\eta; \A_0, \A_1; \B_0, \B_1]$ in terms of notions on the other subsystem $(\Omega_A, \mathcal{E}_A)$.
For example, we can express \eqref{win_GPT00} also as 
\begin{equation}\label{win_GPT00A}
	P_{\mathrm{win}}[\eta; \A_0, \A_1; \B_0, \B_1]=\frac{1}{2}\sum_{t\in\{0,1\}}\sum_{b\in\{0,1\}}\ang{\mathsf{R}_t^b,\check{\eta}(\B_t^b)}_{A}
\end{equation}
in terms of the inner product $\ang{\cdot,\cdot}_A$ in $V_A$.
In this expression, we introduced observables 
\begin{equation}
	\label{eq:R}
	\mathsf{R}_t^b=\frac{1}{2}\sum_{s\in\{0,1\}}\sum_{a\in\{0,1\}}V(a,b|s,t)\A_s^a
\end{equation}
on $\Omega_A$ and considered the bipartite state $\eta$ as a normalized and cone-preserving map $\check{\eta}\colon V_B\to V_A$ instead of $\hat{\eta}\colon V_A\to V_B$.
We can prove that $\check{\eta}$ is the transposition of the former $\hat{\eta}$:
\begin{equation}\label{eq:transpose}
	\check{\eta}=\hat{\eta}^{T}.
\end{equation}
This follows from the elemental formula 
\[
\ang{v_B,\hat{\eta}(v_A)}_{B}=\ang{v_A,\hat{\eta}^{T}(v_B)}_{A}
\]
for any $v_A\in V_A$ and $v_B\in V_B$, and is explicitly confirmed as
\begin{align*}
	\frac{1}{2}\sum_{s}\sum_{a}\ang{\mathsf{Q}_s^a,\hat{\eta}(\A_s^a)}_{B}
	&=\frac{1}{4}\sum_{s,t}\sum_{a,b}V(a, b|s,t)\ang{\B_t^b,\hat{\eta}(\A_s^a)}_{B}\\
	&=\frac{1}{4}\sum_{s,t}\sum_{a,b}V(a, b|s,t)\ang{\A_s^a,\hat{\eta}^{T}(\B_t^b)}_{A}\\
	&=\frac{1}{2}\sum_{t}\sum_{b}\ang{\frac{1}{2}\sum_s\sum_aV(a, b|s,t)\A_s^a,\ \hat{\eta}^{T}(\B_t^b)}_{A}\\
	&=\frac{1}{2}\sum_{t}\sum_{b}\ang{\mathsf{R}_t^b,\hat{\eta}^{T}(\B_t^b)}_{A}.
\end{align*}
The replacement \eqref{win_GPT00A} and \eqref{eq:transpose} will be used when discussing our main result.

\section{Main result: optimal CHSH values for regular polygon theories}
\label{sec:main}
In this section, we present our main result on what bipartite states exhibit optimal CHSH values in a composite of regular polygon theories.
We consider the same situations as \cite{1367-2630-13-6-063024} and prove the conjecture given there to be true.

\subsection{CHSH values for regular polygon theories}
We explained CHSH games with the state-observable formulation in GPTs in the last section. 
In this section, we apply the settings to regular polygon theories.
Let us consider the same situation as Subsec.~\ref{subsec:CHSH game}.
Following the previous study \cite{1367-2630-13-6-063024}, we assume that the two parties Alice and Bob are both with the same local systems described by a regular polygon theory $(\Omega_n, \mathcal{E}_n)$ ($n\ge3$ including $n=\infty$) reviewed in Subsec.~\ref{subsec:RPT}, and the state space of their composite system is given by the maximal tensor $\Omega_n\otimes_{max}\Omega_n$.
The purpose of this study is to find a bipartite state $\eta\in\Omega_n\otimes_{max}\Omega_n$ and observables $(\A_0,\A_1)$ and $(\B_0,\B_1)$ that maximize the CHSH value $|C[\eta;\A_0,\A_1;\B_0,\B_1]|$.
This problem was initially studied in the previous study \cite{1367-2630-13-6-063024}.
There the authors proposed a natural conjecture that the maximum of $|C[\eta;\A_0,\A_1;\B_0,\B_1]|$ is attained by maximally entangled states in $T_{n}\cdot GL(\Omega_n)$ and certain sets of observables following the intuition in the quantum CHSH setting.
In the remaining of this section, we investigate whether the conjecture is true or not.

In this study, in terms of \eqref{win-CHSH}, we evaluate the winning probability $P_{\mathrm{win}}[\eta;\A_0,\A_1;\B_0,\B_1]$ rather the CHSH value $C[\eta;\A_0,\A_1;\B_0,\B_1]$ itself.
To deal with this problem, because the winning probability is a convex quantity with respect to local observables of both parties, we assume that all observables are composed of pure effects.
We remark that our setting is a generalization of the usual CHSH setting for two-qubit system, where the subsystems of Alice and Bob are identical and rank-1 PVMs (composed of pure effects) are measured.
In addition, we can also assume that the observables are of the form $\E_n(i)=\{e_n(i), \overline{e_n(i)}\}$ (see \eqref{def:polygon pure effect}): for example, Alice's observable $\A_0=\{\A_0^0, \A_0^{1}\}$ is given by $\A_0^0=e_n(i)$ and $\A_0^1=\overline{e_n(i)}$ with some integer $i\in\{0,1,\ldots, n-1\}$.
Although observables of the form $\{\overline{e_n(i)}, e_n(i)\}$ also seem to be appropriate for our argument, this assumption is clearly justified when $n$ is even because we have $\overline{e_n(i)}=e_n(i+\frac{n}{2})$ for even $n$ (similarly for $n=\infty$).
On the other hand, when $n$ is odd, we return to \eqref{win_GPT00} to verify the assumption.
The expansion of its r.h.s. is of the form
\begin{equation}\label{eq:expand}
	\begin{aligned}
		\frac{1}{2}\left[
		\ang{\frac{e_n(i)+e_n(j)}{2}, ~\eta(f_n(k))}
		+\ang{\frac{\overline{e_n(i)}+\overline{e_n(j)}}{2},~ \eta(u-f_n(k))}\right.\qquad\qquad\\
		\left.
		+\ang{\frac{e_n(i)+\overline{e_n(j)}}{2},~ \eta(f_n(l))}
		+\ang{\frac{\overline{e_n(i)}+e_n(j)}{2}, ~\eta(u-f_n(l))}
		\right],
	\end{aligned}
\end{equation}
where $i,j,k,l\in\{0,\ldots, n-1\}$ and $f_n(k)$ is either $e_n(k)$ or $\overline{e_n(k)}$ and $f_n(l)$ either $e_n(l)$ or $\overline{e_n(l)}$.
Here the symbol $\ang{\cdot,\cdot}$ denotes the standard inner product in $\R^3(=\mathit{span}(\Omega_n))$.
If $f_n(k)=e_n(k)$, then \eqref{eq:expand} equals $P_{\mathrm{win}}[\eta;\A_0,\A_1;\B_0,\B_1]$ with 
\[
\A_0=\E_n(k),\quad \A_1=\E_n(l),\quad\B_0=\E_n(i)\quad\B_1=\E_n(j)
\]
or 
\[
\A_0=\E_n(k),\quad \A_1=\E_n(l),\quad\B_0=\E_n(j)\quad\B_1=\E_n(i)
\]
corresponding respectively to the case $f_n(l)=e_n(l)$ or $f_n(l)=\overline{e_n(l)}$.
If $f_n(k)=\overline{e_n(k)}$, then using $u-\overline{e_n(l)}=e_n(l)$ and $\ang{u,u}=1$, we rewrite \eqref{eq:expand} as
\begin{align*}
	&\begin{aligned}
		\frac{1}{2}\left[
		1-\left(\ang{\frac{e_n(i)+e_n(j)}{2}, \eta(e_n(k))}
		+\ang{\frac{\overline{e_n(i)}+\overline{e_n(j)}}{2}, \eta(\overline{e_n(k)})}\right)\right.\qquad\qquad\\
		\left.
		+1-\left(\ang{\frac{e_n(i)+\overline{e_n(j)}}{2}, \eta(u-f_n(l))}
		+\ang{\frac{\overline{e_n(i)}+e_n(j)}{2}, \eta(f_n(l))}\right)
		\right]
	\end{aligned}\\
&=1-P_{\mathrm{win}}[\eta;\A_0,\A_1;\B_0,\B_1],
\end{align*}
where 
\[
\A_0=\E_n(k),\quad \A_1=\E_n(l),\quad\B_0=\E_n(j)\quad\B_1=\E_n(i)
\]
or 
\[
\A_0=\E_n(k),\quad \A_1=\E_n(l),\quad\B_0=\E_n(i)\quad\B_1=\E_n(j)
\]
corresponding respectively to the case $f_n(l)=e_n(l)$ or $f_n(l)=\overline{e_n(l)}$.
Because the replacement $P_{\mathrm{win}}[\eta;\A_0,\A_1;\B_0,\B_1]\to1-P_{\mathrm{win}}[\eta;\A_0,\A_1;\B_0,\B_1]$ causes $C[\eta;\A_0,\A_1;\B_0,\B_1]\to-C[\eta;\A_0,\A_1;\B_0,\B_1]$ (see \eqref{win-CHSH}) and thus does not change the value $|C[\eta;\A_0,\A_1;\B_0,\B_1]|$, our assumption that all observables are of the form $\E_n(i)=\{e_n(i), \overline{e_n(i)}\}$ (not $\{\overline{e_n(i)}, e_n(i)\}$) is now verified.

Let us consider optimizing the winning probability $P_{\mathrm{win}}[\eta;\A_0,\A_1;\B_0,\B_1]$.
As explained above, observables $(\A_0,\A_1;\B_0,\B_1)$ that yield the optimum of $P_{\mathrm{win}}$ are of the form $(\E_n(i),\E_n(j);\E_n(k),\E_n(l))$ with integers $i,j,k,l\in\{0,1,\ldots, n-1\}$.
Also, since $\Omega_n\otimes_{max}\Omega_n$ is a compact set, there does exist a bipartite state optimizing the quantity.
We note that the word ``optimize" here means either ``minimize" or ``maximize" according to the remainder of $n$ divided by 8: it will be proved that $(\eta;\A_0,\A_1;\B_0,\B_1)$ minimizing $P_{\mathrm{win}}[\eta;\A_0,\A_1;\B_0,\B_1]$ give a greater CHSH value $|C[\eta;\A_0,\A_1;\B_0,\B_1]|$ than that maximizing $P_{\mathrm{win}}$ for $n\equiv 3,5$ while the converse situation holds for $n\equiv 1,7$ (mod 8).
It will be also found that the analytical method of optimizing $P_{\mathrm{win}}$ varies dramatically with the parity of $n$.
Although we have to develop various ways of analysis depending on the value of $n$, we eventually reach the following theorem.
\begin{thm}
	\label{thm:main}
For any element $\hat{\eta}_\star\in T_{n}\cdot GL(\Omega_n)$ and its inducing maximally entangled state $\eta_\star$, 
there exist integers $i_\star,j_\star,k_\star,l_\star\in\{0,1,\ldots, n-1\}$ such that 
\begin{equation}
	|C[\eta;\A_0,\A_1;\B_0,\B_1]|\le |C[\eta_\star;\E_n(i_\star),\E_n(j_\star);\E_n(k_\star),\E_n(l_\star)]|
\end{equation}
holds for any bipartite state $\eta\in\Omega_n\otimes_{max}\Omega_n$ and pairs of binary observables $(\A_{0},\A_{1})$ and $(\B_{0},\B_{1})$ on $\Omega_n$.
\end{thm}
Theorem \ref{thm:main} is the main result of this study revealing the exact bipartite states in $\Omega_n\otimes_{max}\Omega_n$ that optimize the CHSH value.
It proves the conjecture in the previous study \cite{1367-2630-13-6-063024} to be true: maximally entangled states in regular polygon theories optimize the CHSH value as in quantum theory.
We present the proof of the theorem for even $n$ and $n=\infty$ in Subsec.~\ref{subsec:proof for even} and that for odd $n$ in Subsec.~\ref{subsec:proof for odd}.

\begin{rmk}
	\label{rmk:3}
The claim of Theorem~\ref{thm:main} holds in particular for the case $n=3$ although there is no entangled state in $\Omega_3\otimes_{max}\Omega_3=\Omega_3\otimes_{min}\Omega_3$.
To verify this, among elements in $GL(\Omega_3)$, let us focus on the state $\sigma_\star=\frac{1}{3}\sum_{i=0}^2\omega_3(i)\otimes\omega_3(i)$ induced by the permutation $\hat{\sigma}_{\star}\colon\{0,1,2\}\to\{0,1,2\}$ such that $\hat{\sigma}_{\star}(i)=i$ (see Remark~\ref{rmk1}).
For this $\sigma_\star$, we can indeed confirm that $(\E_3(i_\star),\E_3(j_\star);\E_3(k_\star),\E_3(l_\star))=(\E_3(1),\E_3(0);\E_3(2),\E_3(0))$
yield $P_{\mathrm{win}}=\frac{1}{4}$ or $C=-2$ (the classical CHSH bound).
\end{rmk}

\subsection{Optimal CHSH values for even-sided regular polygon theories}
\label{subsec:proof for even}
In this part, we prove Theorem \ref{thm:main} for even $n$.
The proof proceeds by finding a bipartite state $\eta\in\Omega_n\otimes_{max}\Omega_n$ and integers $i,j,k,l\in\{0,\ldots, n-1\}$ that optimize $P_{\mathrm{win}}[\eta;\E_n(i),\E_n(j);\E_n(k),\E_n(l)]$.
We use \eqref{win_GPT} for its analysis. 
By virtue of \eqref{def:polygon pure effect}, \eqref{eq:Q(s=0)}, and \eqref{eq:Q(s=1)}, the observables $\Q_{0}$ and $\Q_{1}$ in \eqref{win_GPT} are written respectively as
\begin{equation}\label{eq:Q0_even1}
		\mathsf{Q}_{s=0}^{a=0}=\left(
	\begin{array}{c}
		\vec{q}_0\\
		\frac{1}{2}
	\end{array}
	\right),\ \ 
	\mathsf{Q}_{s=0}^{a=1}=\left(
	\begin{array}{c}
		-\vec{q}_0\\
		\frac{1}{2}
	\end{array}
	\right)
\end{equation}
with 
\begin{equation}\label{eq:Q0_even2}
	\vec{q}_0=	\frac{r_{n}\cos((k-l)\theta_n)}{2}\cdot
	\left(
	\begin{array}{c}
		\cos((k+l+1)\theta_n)\\
		\sin((k+l+1)\theta_n)
	\end{array}
	\right)
\end{equation}
and 
\begin{equation}\label{eq:Q0_odd1}
	\mathsf{Q}_{s=1}^{a=0}=\left(
	\begin{array}{c}
		\vec{q}_1\\
		\frac{1}{2}
	\end{array}
	\right),\ \ 
	\mathsf{Q}_{s=1}^{a=1}=\left(
	\begin{array}{c}
		-\vec{q}_1\\
		\frac{1}{2}
	\end{array}
	\right)
\end{equation}
with 
\begin{equation}\label{eq:Q0_odd2}
	\vec{q}_1=\frac{r_{n}\sin((k-l)\theta_n)}{2}\cdot
	\left(
	\begin{array}{c}
		\cos((k+l+1)\theta_n+\frac{\pi}{2})\\
		\sin((k+l+1)\theta_n+\frac{\pi}{2})
	\end{array}
	\right),
\end{equation}
where we introduced $\theta_n=\frac{\pi}{n}$.
We derive tight bounds for \eqref{win_GPT} in terms of the equations above.
The expressions of those bounds vary depending on whether $n\equiv0,4$ or $n\equiv2,6$ (mod 8), but their derivations are given in the same way.
Thus, letting the modulo be 8 in the following, we only treat the case $n\equiv2,6$, which is a little more complicated than the case $n\equiv0,4$.

\begin{figure}[h]
	\centering
	\includegraphics[scale=0.365]{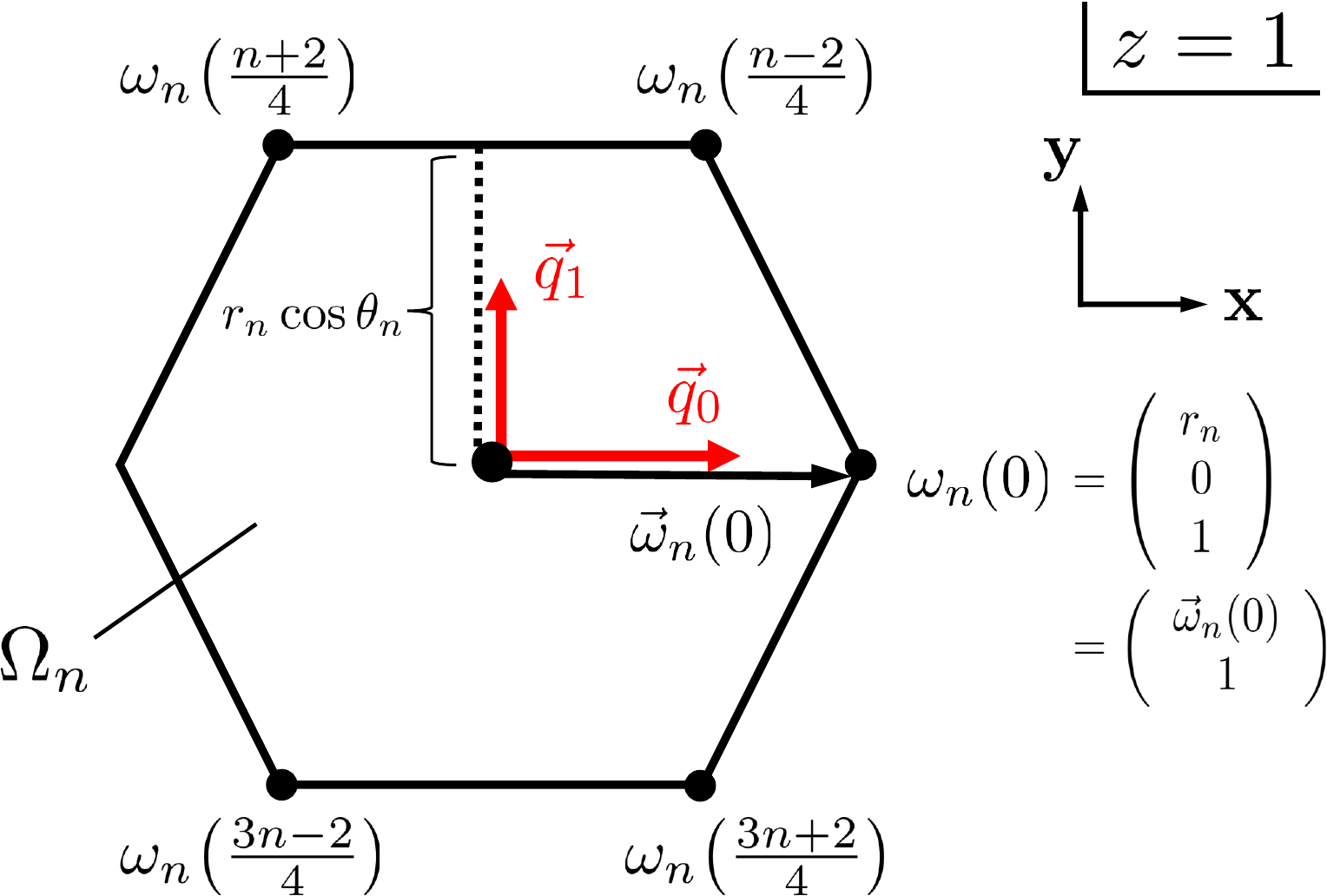}
	\caption{Illustrations of vectors $\vec{q}_0$ and $\vec{q}_1$ for $n=6$.
	The same description can be presented for general $n\equiv2,6$.}
	\label{fig_even_vec}
\end{figure}
Now assume that $n\equiv2,6$.
We rewrite the conditional states in \eqref{win_GPT} as
\begin{equation}
	\label{eq:vec_omega_s^a}
	\omega_s^a=\left(
	\begin{array}{c}
		\vec{w}_s^a\\
		1
	\end{array}
	\right)
\end{equation}
to obtain
\begin{multline}
	\label{eq:Pwin_even}
	2P_{\mathrm{win}}[\eta;\E_n(i),\E_n(j);\E_n(k),\E_n(l)]=\left[p(0|0)\ang{\vec{q}_0, \vec{w}_0^0}+p(1|0)\ang{-\vec{q}_0, \vec{w}_0^1}\right]\\
	+\left[p(0|1)\ang{\vec{q}_1, \vec{w}_1^0}+p(1|1)\ang{-\vec{q}_1, \vec{w}_1^1}\right]+1.
\end{multline}
In this equation, $\ang{\cdot, \cdot}$ denotes the standard inner product in $\R^2$.
Although it is the same notation as the inner products in $\R^3$, hereafter we do not explicitly distinguish them.
Let us consider replacing $l$ by $l'=l+\frac{n}{2}$.
This replacement $l\to l'$ causes $\vec{q}_0\to\vec{q}_1,\ \vec{q}_1\to\vec{q}_0$, or 
\[
\Q_{0}^0\to\Q_{1}^0,\ \ \Q_{0}^1\to\Q_{1}^1,\ \ \Q_{1}^0\to\Q_{0}^0,\ \ \Q_{1}^1\to\Q_{0}^1,
\]
where $\{\Q_s^a\}_{s,a}$ are defined in \eqref{eq:Q0_even2} and \eqref{eq:Q0_odd2}.
It means
\begin{equation*}
	P_{\mathrm{win}}[\eta;\E_n(i),\E_n(j);\E_n(k),\E_n(l)]=P_{\mathrm{win}}\left[\eta;\E_n(j),\E_n(i);\E_n(k),\E_n\!\left(l+\frac{n}{2}\right)\right],
\end{equation*}
and thus, because $\frac{n}{2}$ is an odd integer, we can assume without loss of generality that $k+l$ is odd (equivalently $k-l$ is odd).
With a suitable rotation and reflection, we can in addition assume $l\in\{0,\ldots,\frac{n-2}{4}\}$, $k\in\{\frac{3n-2}{4},\ldots,n-1\}$, and $k+l=n-1$ so that
\begin{equation}
	\label{eq:Q0,1_even00}
	\begin{aligned}
		\vec{q}_0
		&=-\frac{r_{n}\cos((k-l)\theta_n)}{2}\cdot
		\left(
		\begin{array}{c}
			1\\
			0
		\end{array}
		\right),
	\end{aligned}
\quad
\begin{aligned}
	\vec{q}_1
	&=\frac{r_{n}\sin((k-l)\theta_n)}{2}\cdot
	\left(
	\begin{array}{c}
		0\\
		1
	\end{array}
	\right).
\end{aligned}
\end{equation}
We note that 
\[
-\cos((k-l)\theta_n)=\cos\left ((2l+1)\frac{\pi}{n}\right )\ge0,\quad\sin((k-l)\theta_n)=\sin\left((2l+1)\frac{\pi}{n}\right)\ge0
\]
hold because $l\in\{0,\ldots,\frac{n-2}{4}\}$ implies $(2l+1)\frac{\pi}{n}\in[\frac{\pi}{n}, \frac{\pi}{2}]$.
In Fig.~\ref{fig_even_vec}, the vectors $\vec{q}_0, \vec{q}_1\in\R^2$ are illustrated on the hyperplane $z=1$ together with the state space $\Omega_n$.
There we can see that $\vec{q}_0$ is in the direction of the state $\omega_n(0)$.
That is, rewriting \eqref{def:polygon pure state} in a similar way to \eqref{eq:Q0_even1} and \eqref{eq:Q0_odd1} as
\begin{equation}\label{eq:def_vec_pure}
	\omega_n(i)=\left(
	\begin{array}{c}
		\vec{\omega}_n(i)\\
		1
	\end{array}
	\right),
\end{equation}
we have $\vec{q}_0\propto\vec{\omega}_n(0)(=(r_n,0)^{T})$.
Let us evaluate \eqref{win_GPT} in this simplified situation.
For the first term of \eqref{eq:Pwin_even}, since a geometrical consideration in Fig.~\ref{fig_even_vec} implies 
\[
	\ang{\vec{q}_0, \vec{w}_0^0}\le
	\ang{\vec{q}_0, \vec{\omega}_n(0)}
	=-\frac{r_n^2}{2}\cos((k-l)\theta_n),
\]
we have
\begin{align}
	p(0|0)\ang{\vec{q}_0, \vec{w}_0^0}+p(1|0)\ang{-\vec{q}_0, \vec{w}_0^1}
	&\le
	(p(0|0)+ p(1|0))\cdot\left(-\frac{r_n^2}{2}\cos((k-l)\theta_n)\right)\notag\\
	&=-\frac{r_n^2}{2}\cos((k-l)\theta_n).\label{eq:bound_even_s=0}
\end{align}
The equality holds for arbitrary $p(0|0)$ and $p(1|0)$, and $\vec{w}_0^0=\vec{\omega}_n(0)$ and $\vec{w}_0^1=-\vec{\omega}_n(0)$, i.e., $\omega_0^0=\omega_n(0)$ and $\omega_0^1=\omega_n(\frac{n}{2})$.
To evaluate the second term of \eqref{eq:Pwin_even}, we confirm from Fig.~\ref{fig_even_vec} that 
\begin{align*}
	\ang{\vec{q}_1, \vec{w}_1^0}
	&=\frac{r_{n}\sin((k-l)\theta_n)}{2}\ang{
		\left(
		\begin{array}{c}
			0\\
			1
		\end{array}
		\right), \vec{w}_1^0}\\
	&\leq
	\frac{r_{n}\sin((k-l)\theta_n)}{2}\cdot r_n\cos\theta_n=\frac{r_n^2}{2}\cos\theta_n\sin((k-l)\theta_n)
\end{align*}
and the equality holds for $\vec{w}_1^0=p\vec{\omega}_n\!\left(\frac{n-2}{4}\right)+(1-p)\vec{\omega}_n\!\left(\frac{n+2}{4}\right)$ with some $p\in[0,1]$.
It follows that the second term of \eqref{eq:Pwin_even} can be evaluated as
\begin{equation}\label{eq:bound_even_s=1}
	p(0|1)\ang{\vec{q}_1, \vec{w}_1^0}+p(1|1)\ang{-\vec{q}_1, \vec{w}_1^1}
	\leq\frac{r_n^2}{2}\cos\theta_n\sin((k-l)\theta_n),
\end{equation}
where the equality holds for arbitrary $p(0|1)$ and $p(1|1)$, and   $\omega_1^0=p{\omega}_n\!\left(\frac{n-2}{4}\right)+(1-p){\omega}_n\!\left(\frac{n+2}{4}\right)$ and $\omega_1^1=q{\omega}_n\!\left(\frac{3n-2}{4}\right)+(1-q){\omega}_n\!\left(\frac{3n+2}{4}\right)$ with some $p, q\in[0,1]$.
Overall, the probability \eqref{eq:Pwin_even} is bounded in terms of \eqref{eq:bound_even_s=0} and \eqref{eq:bound_even_s=1} as
\begin{equation}\label{eq:Pwin_bound}
	\begin{aligned}
			2P_{\mathrm{win}}[\eta;\E_n(i),\E_n(j);\E&_n(k),\E_n(l)]-1\\
		&\le\frac{r_n^2}{2}\left[-\cos((k-l)\theta_n)+\cos\theta_n\sin((k-l)\theta_n)\right]\\
		&=\frac{r_n^2}{2}\left[\cos((2l+1)\theta_n)+\cos\theta_n\sin((2l+1)\theta_n)\right].
	\end{aligned}
\end{equation}
We study when the upper bound \eqref{eq:Pwin_bound} is realizable.
Let a bipartite state $\eta_\star\in\Omega_n\otimes_{max}\Omega_n$ and integers $i_\star,j_\star\in\{0,\ldots,n-1\}$ (or observables $\E_n(i_\star)=\{e_n(i_\star), \overline{e_n(i_\star)}\}$,\ $\E_n(j_\star)=\{e_n(j_\star), \overline{e_n(j_\star)}\}$) realize the upper bound, i.e., they satisfy 
\begin{equation}\label{eq:Pwin_bound0}
	\begin{aligned}
		2P_{\mathrm{win}}[\eta_\star;\E_n(i_\star),\E_n(j_\star);\E&_n(k),\E_n(l)]-1\\
		&=\frac{r_n^2}{2}\left[\cos((2l+1)\theta_n)+\cos\theta_n\sin((2l+1)\theta_n)\right].
	\end{aligned}
\end{equation}
As we have seen, $(\eta_\star; \E_n(i_\star),\E_n(j_\star))$ satisfy
\begin{equation}\label{eq:even_s=0_bound condition0}
	\eta_\star(e_n(i_\star))=p_\star(0|0)\omega_n(0),\quad 	\eta_\star(\overline{e_n(i_\star)})=p_\star(1|0)\omega_n\!\left(\frac{n}{2}\right)
\end{equation}
with probabilities $\{p_\star(0|0),p_\star(1|0)\}$ and
\begin{equation}\label{eq:even_s=0_bound condition1}
	\eta_\star(e_n(j_\star))=p_\star(0|1)\omega_\star(0|1),\quad 	\eta_\star(\overline{e_n(j_\star)})=p_\star(1|1)\omega_\star(1|1)
\end{equation}
with probabilities $\{p_\star(0|1),p_\star(1|1)\}$ and states $\omega_\star(0|1)\in \Omega_n[\frac{n\pm2}{4}]$ and $\omega_\star(1|1)\in \Omega_n[\frac{3n\pm2}{4}]$, where 
\begin{equation}\label{eq:n_pm}
	\begin{aligned}	&\Omega_n\!\left[\frac{n\pm2}{4}\right]:=\mathit{conv}\left\{\omega_n\!\left(\frac{n-2}{4}\right),\omega_n\!\left(\frac{n+2}{4}\right)\right\},\\ &\Omega_n\!\left[\frac{3n\pm2}{4}\right]:=\mathit{conv}\left\{\omega_n\!\left(\frac{3n-2}{4}\right),\omega_n\!\left(\frac{3n+2}{4}\right)\right\}.
	\end{aligned}
\end{equation}
It follows from \eqref{eq:even_s=0_bound condition0} that 
\begin{equation}\label{key}
	\eta_\star(u)=\eta_\star(e_n(i_\star))+\eta_\star(\overline{e_n(i_\star)})\in\mathit{conv}\left\{\omega_n\!\left(0\right), \omega_n\!\left(\frac{n}{2}\right)\right\}.
\end{equation}
In particular, the y-coordinate $\eta_\star(u)|_\mathrm{y}$ of $\eta_\star(u)$ satisfies
\[
\left.\eta_\star(u)\right| _\mathrm{y}=0,
\]
and the relation
\[
\eta_\star(u)=\eta_\star(e_n(j_\star))+\eta_\star(\overline{e_n(j_\star)})
\]
implies
\begin{equation}\label{key}
	\left.\eta_\star(e_n(j_\star))\right| _\mathrm{y}=- \left.\eta_\star(\overline{e_n(j_\star)})\right| _\mathrm{y}.
\end{equation}
Because we have
\[
\omega_\star(0|1)| _\mathrm{y}=-\omega_\star(1|1)| _\mathrm{y}\left(=\cos\theta_n\right),
\]
we obtain from \eqref{eq:even_s=0_bound condition1}
\begin{equation}\label{eq:s=1:p=1/2}
	p_\star(0|1)=p_\star(1|1)=\frac{1}{2}.
\end{equation}

\begin{figure}[h]
	\centering
	\includegraphics[scale=0.38]{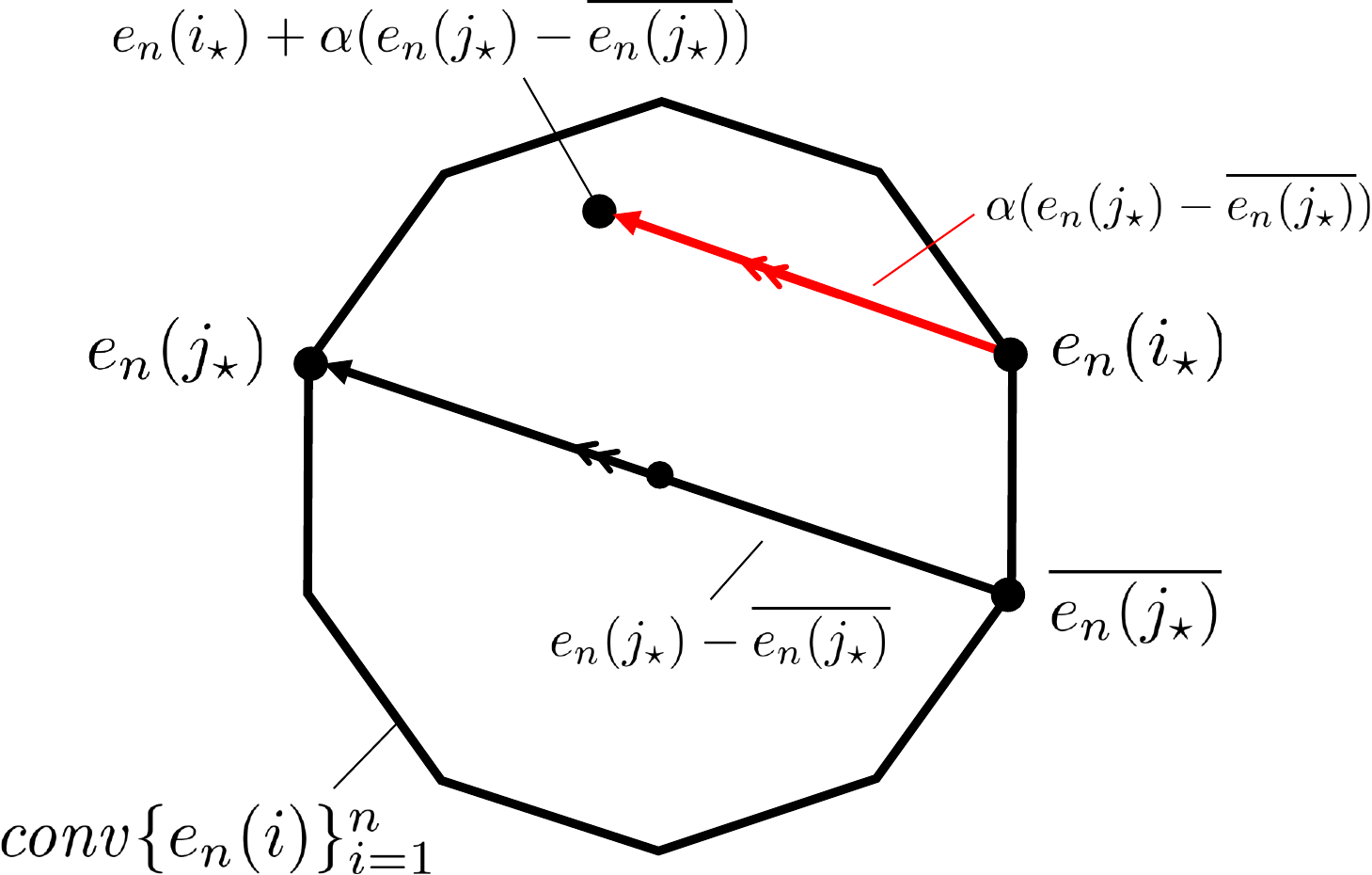}
	\caption{The existence of $\alpha\in\R$ such that $e_n(i_\star)+\alpha(e_n(j_\star)-\overline{e_n(j_\star)})\in\mathit{conv}\{e_n(i)\}_{i=1}^n$ for $n=10$.
	The same argument can be given for general $n\equiv2,6$ and $(i_\star, j_\star)$ that are not located as illustrated in the figure.}
	\label{fig_exist_alpha}
\end{figure}
\begin{figure}[h]
	\centering
	\includegraphics[scale=0.382]{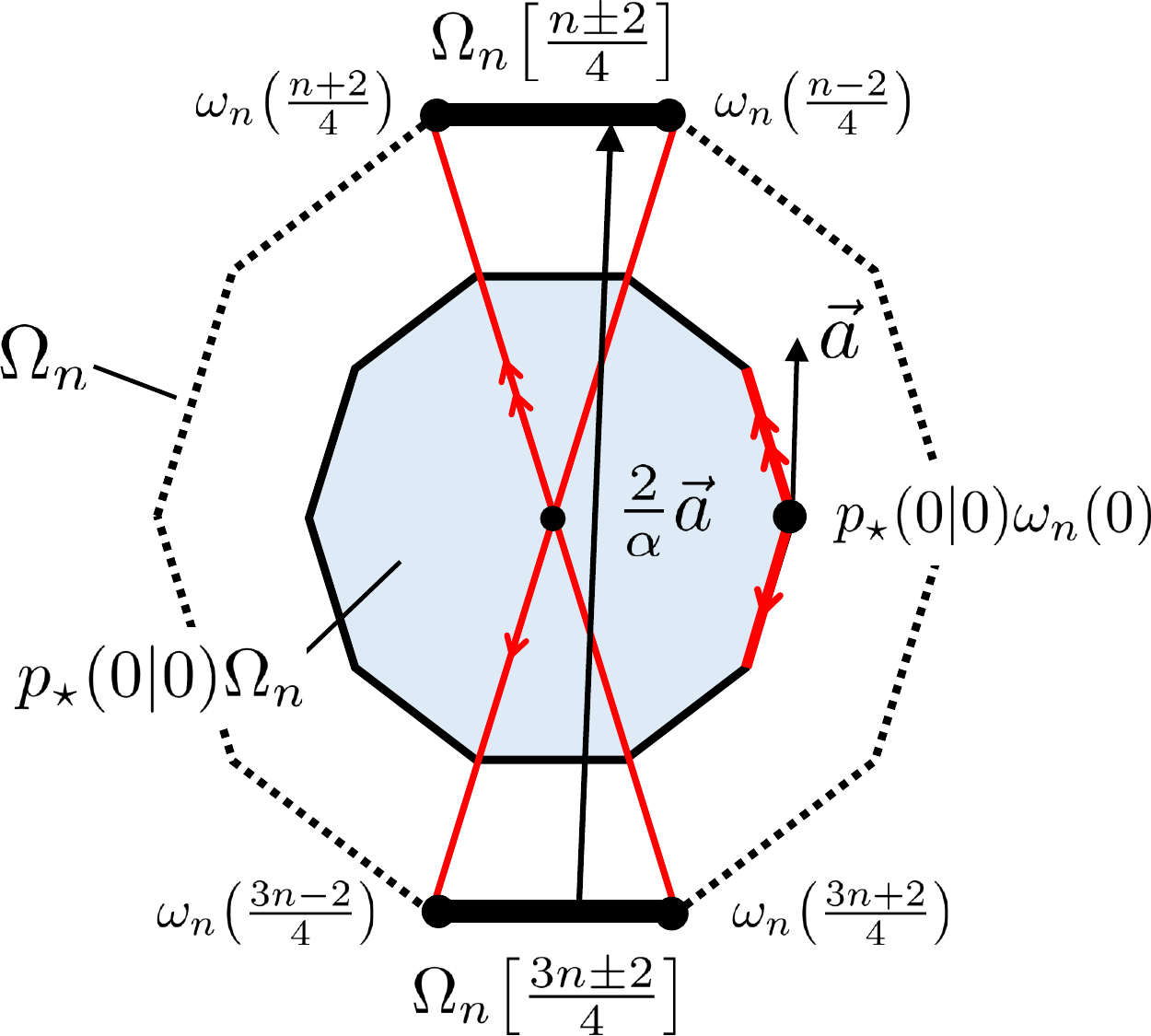}
	\caption{Geometrical derivations of \eqref{eq::omega(0|0)} and \eqref{eq:s=0:p=1/2} for $n=10$.
		The colored domain represents $p_\star(0|0)\Omega_n$ while the dotted line encloses  $\Omega_n$.
	As shown here, $p_\star(0|0)\omega_n(0)+\vec{a}\notin p_\star(0|0)\Omega_n$ if $\vec{a}$ is neither parallel to $\omega_n(\frac{n-2}{4})-\omega_n(\frac{3n-2}{4})$ nor $\omega_n(\frac{3n+2}{4})-\omega_n(\frac{n+2}{4})$.}
	\label{fig_vec_a}
\end{figure}

We can present a further analysis for the assemblage \eqref{eq:even_s=0_bound condition1}.
It is important that there exists a nonzero $\alpha\in\R$ such that $e_n(i_\star)+\alpha(e_n(j_\star)-\overline{e_n(j_\star)})$ belongs to the subset $\mathit{conv}\{e_n(i)\}_{i=1}^n$ of the effect space $\mathcal{E}_n$ (see Fig.~\ref{fig_exist_alpha}).
This indicates  $\eta_\star(e_n(i_\star))+\alpha(\eta_\star(e_n(j_\star))-\eta_\star(\overline{e_n(j_\star)}))\in V_{n+}$, or explicitly
\begin{equation}\label{eq:vec_in}
	p_\star(0|0)\omega_n(0)+\frac{\alpha}{2}\left(\omega_\star(0|1)-\omega_\star(1|1)\right)\in V_{n+}.
\end{equation}
The vector $\frac{\alpha}{2}\left(\omega_\star(0|1)-\omega_\star(1|1)\right)$ is of the form
\begin{equation}\label{key}
	\frac{\alpha}{2}\left(\omega_\star(0|1)-\omega_\star(1|1)\right)=
	\left(\begin{array}{c}
		\vec{a}\\
		0
	\end{array}
\right)
\end{equation}
with $\vec{a}\in\R^2$, and thus $p_\star(0|0)\omega_n(0)+\frac{\alpha}{2}\left(\omega_\star(0|1)-\omega_\star(1|1)\right)$ belongs to the intersection of $V_{n+}$ and the hyperplane $z=p_\star(0|0)$, which forms a contracted regular polygon $p_\star(0|0)\Omega_n$.
However, as shown in Fig.~\ref{fig_vec_a}, this is possible if and only if $\vec{a}$ is parallel to the vector $\vec{\omega}_n(\frac{n-2}{4})-\vec{\omega}_n(\frac{3n-2}{4})$ or $\vec{\omega}_n(\frac{3n+2}{4})-\vec{\omega}_n(\frac{n+2}{4})$ (remember the notation in \eqref{eq:def_vec_pure}).
Hence we obtain 
\begin{equation}\label{eq::omega(0|0)}
	(\omega_\star(0|1),\omega_\star(1|1))=\left(\omega_n\!\left(\frac{n\pm2}{4}\right), \omega_n\!\left(\frac{3n\pm2}{4}\right)\right),\ \left(\omega_n\!\left(\frac{3n\pm2}{4}\right), \omega_n\!\left(\frac{n\pm2}{4}\right)\right).
\end{equation}
It follows that $\eta_\star(u)=(0,0,1)^{T}$ and thus 
\begin{equation}\label{eq:s=0:p=1/2}
	p_\star(0|0)=p_\star(0|1)=\frac{1}{2}
\end{equation}
hold because $\eta_\star(u)=p_\star(0|0)\omega_n(0)+p_\star(0|1)\omega_n(\frac{n}{2})$.
We can now easily confirm that the state $\eta_\star$ that realizes \eqref{eq:even_s=0_bound condition0}, \eqref{eq:even_s=0_bound condition1}, \eqref{eq::omega(0|0)}, and \eqref{eq:s=0:p=1/2} is an element of $T_n\cdot GL(\Omega_n)$.
The observables $(\E_n(i_\star), \E_n(j_\star))$ can be chosen according to $\eta_\star$ as $e_n(i_\star)=\eta_\star^{-1}T_n^{-1}e_n(0)$ and $e_n(j_\star)=\eta_\star^{-1}T_n^{-1}e_n(\frac{n-2}{4})$, for example.
We note that once we find such $(\eta_\star; \E_n(i_\star), \E_n(j_\star))$, then
for any $\eta_\star'\in T_n\cdot GL(\Omega_n)$ there exist observables $\E_n(i_\star'), \E_n(j_\star')$ such that $(\eta_\star'; \E_n(i_\star'), \E_n(j_\star'))$ optimize the winning probability:
\[
P_{\mathrm{win}}[\eta_\star';\E_n(i_\star'),\E_n(j_\star');\E_n(k),\E_n(l)]=P_{\mathrm{win}}[\eta_\star;\E_n(i_\star),\E_n(j_\star);\E_n(k),\E_n(l)].
\]
In fact, with an orthogonal transformation $T:=(\eta_\star')^{-1}\eta_\star\in GL(\Omega_n)$, we can construct observables $(\E_n(i_\star'),\E_n(j_\star'))$ by $e_n(i_\star')=Te_n(i_\star)$ and $e_n(j_\star')=Te_n(j_\star)$, which satisfy $\eta_\star'(e_n(i_\star'))=\eta_\star(e_n(i_\star))$ and $\eta_\star'(e_n(j_\star'))=\eta_\star(e_n(j_\star))$ respectively.
The same argument can be clearly applied to the case $n=\infty$ because of the geometric symmetry of its disc state space, and Theorem \ref{thm:main} has been proved for even $n$ and $n=\infty$.

\begin{rmk}
	\label{rmk:2}
In the proof above, we only investigated the maximum of the quantity \eqref{eq:Pwin_bound}.
For its minimum, we can find that its absolute value equals to that of the maximum.
This can be easily verified by the same geometrical argument as above: for example, the first term of \eqref{eq:Pwin_bound} can be evaluated as
\begin{align*}
	p(0|0)\ang{\vec{q}_0, \vec{w}_0^0}+p(1|0)\ang{-\vec{q}_0, \vec{w}_0^1}
	&\ge
	(p(0|0)+ p(1|0))\cdot(r_n^2\cos((k-l)\theta_n))\notag\\
	&=\frac{r_n^2}{2}\cos((k-l)\theta_n),
\end{align*}
and the equality holds if and only if $\omega_0^0=\omega_n(\frac{n}{2})$ and $\omega_0^1=\omega_n(0)$ instead of \eqref{eq:bound_even_s=0}.
Hence there is no essential difference between investigating the minimum and maximum for the  optimization of the winning probability (the CHSH value) when $n$ is even or $n=\infty$.
\end{rmk}

\begin{rmk}
We can further consider maximizing the r.h.s. of \eqref{eq:Pwin_bound} with respect to $k$ and $l$ such that $l\in\{0,\ldots,\frac{n-2}{4}\}$, $k\in\{\frac{3n-2}{4},\ldots,n-1\}$, and $k+l=n-1$.
A lengthy calculation shows
\begin{align}
		\frac{r_n^2}{2}\left[\cos((2l+1)\theta_n)+\cos\theta_n\sin((2l+1)\theta_n)\right]
		\qquad\qquad\qquad\qquad\qquad\qquad\qquad\\
		\le
		\left\{
		\begin{aligned}
			&\frac{r_n^2}{2}\left[3\cos\left(\frac{n+2}{4n}\pi\right)+\sin\left(\frac{n+6}{4n}\pi\right)\right]\quad(n\equiv 2)\\
			&\frac{r_n^2}{2}\left[3\sin\left(\frac{n+2}{4n}\pi\right)+\cos\left(\frac{n+6}{4n}\pi\right)\right]\quad(n\equiv 6),
		\end{aligned}
		\right.
\end{align}
where the equality holds if and only if 
\begin{equation}
	l=
	\left\{
	\begin{aligned}
		&\frac{n-2}{8}\quad(n\equiv 2)\\
		&\frac{n-6}{8}\quad(n\equiv 6)
	\end{aligned}
	\right.
	\qquad
	\left(
	k=
	\left\{
	\begin{aligned}
		&\frac{7n-6}{8}\quad(n\equiv 2)\\
		&\frac{7n-2}{8}\quad(n\equiv 6)
	\end{aligned}
	\right.
	\right).
\end{equation}
These optimal values and observables are consistent with those in \cite{1367-2630-13-6-063024}.
\end{rmk}

\subsection{Optimal CHSH values for odd-sided regular polygon theories}
\label{subsec:proof for odd}
We study odd $n$ cases in this part.
Since the case $n=3$ was already examined (see Remark~\ref{rmk:3}), we focus on $n>3$.
As in Subsec.~\ref{subsec:proof for even}, we seek $\eta\in\Omega_n\otimes_{max}\Omega_n$ and $i,j,k,l\in\{0,\ldots, n-1\}$ optimizing $P_{\mathrm{win}}[\eta;\E_n(i),\E_n(j);\E_n(k),\E_n(l)]$, and if we find one such $\eta\in GL(\Omega_n)(=T_n\cdot GL(\Omega_n))$, then it proves Theorem~\ref{thm:main} (see the argument above Remark~\ref{rmk:2}). 
For this purpose, the expression \eqref{win_GPT} is again used.
We write the winning probability $P_{\mathrm{win}}$ as 
\begin{equation}\label{eq:Pwin_ens}
P_{\mathrm{win}}[\{(p(a|s); \omega^a_s)_a\}_{s};\E_n(k),\E_n(0)]
\end{equation}
by means of the observables $\{\E_n(k),\E_n(0)\}$ of Bob and the state assemblages $\{(p(a|s); \omega^a_s)_{a}\}_s=\{(p(a|0); \omega^a_0)_{a},(p(a|1); \omega^a_1)_{a}\}$ of Bob induced by the bipartite state $\eta$ and observables $\A_{s=0}=\E_n(i)$ and $\A_{s=1}=\E_n(j)$ of Alice (see \eqref{eq:ensemble}).
The assemblages satisfy
\begin{equation}\label{eq:consistent}
	\sum_{a=0,1}p(a|0)\omega^a_0=\sum_{a=0,1}p(a|1)\omega^a_1(=\hat{\eta}(u)).
\end{equation}
In the evaluation of \eqref{eq:Pwin_ens}, we can set
\begin{equation}\label{eq:l and k}
	k\in\left\{0,\ldots, \frac{n-1}{2}\right\}, \quad l=0
\end{equation}
without loss of generality in terms of a suitable rotation about the z-axis and the reflection about the x-axis.
We similarly write the CHSH values resulting from $P_{\mathrm{win}}$ through \eqref{win-CHSH} as
\begin{equation}\label{eq:CHSH_ens}
	C[\{(p(a|s); \omega^a_s)_a\}_{s};\E_n(k),\E_n(0)].
\end{equation}
We remark that not all assemblages $\{(p(a|s); \omega^a_s)_{a}\}_s$ are realizable by $(\eta; \E_n(i),\E_n(j))$.
We introduce two classes of pairs of assemblages
	\begin{align}
				&\mathsf{Ens}^{\mathrm{All}}
				=\Bigl\{\{(p(a|s); \omega^a_s)_{a}\}_s\ \Big|\ p(a|s)\ge0,~\sum_{a=0,1}p(a|s)=1,
		\\
		&\qquad\qquad\qquad\qquad\qquad\qquad\qquad\qquad
		\omega^a_s\in\Omega_n,~\sum_{a=0,1}p(a|0)\omega^a_0=\sum_{a=0,1}p(a|1)\omega^a_1\Bigr\};\label{eq:Ens^all}
		\\
&\mathsf{Ens}^{\mathrm{ME}}
=\Bigl\{
\{(p(a|s); \omega^a_s)_{a}\}_s\ \Big|\ 
\exists\hat{\eta}\in GL(\Omega_n),~ \exists i,j\in\{0,\ldots,n-1\}\ ~\mbox{s.t.}
\\
&\qquad\qquad\qquad\qquad\qquad\qquad\qquad\qquad
\hat{\eta}(e_n(i))=p(0|0)\omega^0_0,~\hat{\eta}(\overline{e_n(i)})=p(1|0)\omega^1_0,\\
&\qquad\qquad\qquad\qquad\qquad\qquad\qquad\qquad
\hat{\eta}(e_n(j))=p(0|1)\omega^0_1,~\hat{\eta}(\overline{e_n(j)})=p(1|1)\omega^1_1
\Bigr\}.\label{eq:Ens^ME}
\end{align}
The first set $\mathsf{Ens}^{\mathrm{All}}$ is the set of all pairs of assemblages satisfying \eqref{eq:consistent}.
On the other hand, the second set $\mathsf{Ens}^{\mathrm{ME}}$ is the set of all assemblages realized by a maximally entangled state in $\Omega_n\otimes_{max}\Omega_n$ and observables of the form $\{\E_n(i), \E_n(j)\}$.
It clearly holds that $\mathsf{Ens}^{\mathrm{ME}}\subseteq\mathsf{Ens}^{\mathrm{All}}$.
We can consider optimum values of $P_{\mathrm{win}}$ in these classes.
We define 
\begin{equation}\label{eq:G}
			G_n(k)=\underset{\{(p(a|s); \omega^a_s)_a\}_{s}\in\mathsf{Ens}^{\mathrm{All}}}{\max}~|C[\{(p(a|s); \omega^a_s)_a\}_{s};\E_n(k),\E_n(0)]|
\end{equation}
and
\begin{equation}\label{eq:H}
			H_n(k)=\underset{\{(p(a|s); \omega^a_s)_a\}_{s}\in\mathsf{Ens}^{\mathrm{ME}}}{\max}~|C[\{(p(a|s); \omega^a_s)_a\}_{s};\E_n(k),\E_n(0)]|
\end{equation}
The following lemma is crucial for our analysis (the proof is given in \ref{appA}).
\begin{lem}
	\label{lem:max_k0}
For each $n>3$, there exists a unique integer $n_\star\in\{0,\ldots,\frac{n-1}{2}\}$ such that 
\begin{equation}\label{eq:lem0}
		\underset{k\in\{0,\ldots,\frac{n-1}{2}\}}{\operatorname{argmax}} G_n(k)=
	\underset{k\in\{0,\ldots,\frac{n-1}{2}\}}{\operatorname{argmax}} H_n(k)=\{n_\star\},
\end{equation}
i.e.,
\begin{align}
	&\max\left\{G_n(k)\ \middle|\ k\in\left\{0,\ldots,\frac{n-1}{2}\right\},~k\neq n_\star\right\}< G_n(n_\star),\label{eq:lem10}
	\\
	&\max\left\{H_n(k)\ \middle|\ k\in\left\{0,\ldots,\frac{n-1}{2}\right\},~k\neq n_\star\right\}< H_n(n_\star).\label{eq:lem1}
\end{align}
holds.
The unique integer $n_\star$ is given by
\begin{equation}\label{eq:n_star}
	n_\star=\left\{
	\begin{aligned}
		&\frac{n-1}{4}\quad(n\equiv 1,5)\\
		&\frac{n+1}{4}\quad(n\equiv 3,7).
	\end{aligned}
	\right.
\end{equation}
In addition, it holds that
\begin{equation}\label{eq:lem2}
	\max\left\{G_n(k)\ \middle|\ k\in\left\{0,\ldots,\frac{n-1}{2}\right\},~k\neq n_\star\right\}< H_n(n_\star)<G_n(n_\star).
\end{equation}
\end{lem}
This lemma indicates that for a bipartite state to yield a greater CHSH value than maximally entangled states, Bob's observables $(\B_{0},\B_{1})$ need to be $(\E_n(n_\star),\E_n(0))$.
In other words, a quintuple of the form $(\eta;\E_n(i),\E_n(j);\E_n(n_\star),\E_n(0))$ optimizes the CHSH value.
We consider exchanging the roles of Alice and Bob.
As shown in \eqref{eq:R}, the bipartite state $\eta$ also can be seen as a map $\check{\eta}$ from the dual cone of Bob to the positive cone of Alice.
For this map $\check{\eta}$, we can develop the same argument as above and conclude that Alice's observables need to be of the form $(\E_n(j+n_\star),\E_n(j))$ to optimize the CHSH value.
We now obtain the following proposition.
\begin{prop}\label{prop_self-adj}
	There exists $\eta_\star\in\Omega_n\otimes_{max}\Omega_n$ such that the induced map $\hat{\eta}_\star$ is self-adjoint ($\hat{\eta}_\star=\hat{\eta}_\star^{T}$) in the Euclidean space $\R^3$ and the quintuple $(\eta_\star;\E_n(n_\star),\E_n(0);\E_n(n_\star),\E_n(0))$ with $n_\star$ in \eqref{eq:n_star} optimizes the CHSH value.
	That is, 
	\begin{equation}\label{eq:claim_prop}
			|C[\eta;\A_0,\A_1;\B_0,\B_1]|\le |C[\eta_\star;\E_n(n_\star),\E_n(0);\E_n(n_\star),\E_n(0)]|
	\end{equation}
holds for any $(\eta;\A_0,\A_1;\B_0,\B_1)$.
\end{prop}
\begin{pf}
We can assume that the optimum is realized by $(\eta;\E_n(j+n_\star),\E_n(j);\E_n(n_\star),\E_n(0))$ as mentioned above.
We define a rotation operator $T_{2j\theta_n}$ about the z-axis in $\R^3$ by
\[
T_{2j\theta_n}=
\begin{pmatrix}
	\cos(2j\theta_n) & -\sin(2j\theta_n)& 0\\
	\sin(2j\theta_n)& \cos(2j\theta_n) & 0\\
	0 & 0& 1
\end{pmatrix},
\]
where $\theta_n=\frac{\pi}{n}$.
We note that $T_{2j\theta_n}\in GL(\Omega_n)$ clearly holds and it preserves the effect space $\mathcal{E}_n$  as well as the state space $\Omega_n$.
The composite $\hat{\eta}_j:=\hat{\eta}\circ T_{2j\theta_n}$ is again a normalized and cone-preserving map between $V_{n+}^*$ and $V_{n+}$ and thus defines a bipartite state $\eta_j\in\Omega_n\otimes_{max}\Omega_n$.
We can rewrite \eqref{win_GPT00} in terms of this state $\eta_j$.
Since
\[
T_{2j\theta_n}(e_n(0))=e_n(j),\quad T_{2j\theta_n}(\overline{e_{n}(0)})=\overline{e_{n}(j)}
\]
and 
\[
T_{2j\theta_n}(e_n(n_\star))=e_n(j+n_\star),\quad T_{2j\theta_n}(\overline{e_{n}(n_\star)})=\overline{e_{n}(j+n_\star)}
\]
holds, we have
\begin{equation}\label{eq:re_1}
	\ang{\mathsf{Q}_s^a,\hat{\eta}(\A_s^a)}=\ang{\mathsf{Q}_s^a,\hat{\eta}_j({\A'}_s^{a})}
\end{equation}
with observables $(\A_0, \A_1)=(\E_n(j+n_\star), \E_n(j))$, $(\A'_0, \A'_1)=(\B_0, \B_1)=(\E_n(n_\star), \E_n(0))$, and $\mathsf{Q}_s^{a}=\frac{1}{2}\sum_{t,b}V(a,b|s,t)\B_t^{b}$.
It implies 
\begin{equation}\label{eq:re_2}
	C[\eta;\E_n(j+n_\star),\E_n(j);\E_n(n_\star),\E_n(0)]=C[\eta_j;\E_n(n_\star),\E_n(0);\E_n(n_\star),\E_n(0)].
\end{equation}
We make a further rewriting of \eqref{win_GPT00} in addition to \eqref{eq:re_1}.
We use \eqref{win_GPT00A} and \eqref{eq:transpose} to obtain
\[
\sum_{s,a}\ang{\mathsf{Q}_s^a,\hat{\eta}_j({\A'}_s^{a})}=\sum_{t,b}\ang{{\mathsf{R}'}_t^{b}, \hat{\eta}_j^{T}(\B_t^{b})},
\]
where ${\mathsf{R}'}_t^{b}=\frac{1}{2}\sum_{s,a}V(a,b|s,t){\A'}_s^{a}$.
Because $(\A'_0, \A'_1)=(\B_0, \B_1)$ (and thus ${\mathsf{R}'}_t^{b}=\mathsf{Q}_t^{b}$) holds in this case, the r.h.s. can be rewritten as
\[
\sum_{t,b}\ang{{\mathsf{R}'}_t^{b}, \hat{\eta}_j^{T}(\B_t^{b})}=\sum_{t,b}\ang{\mathsf{Q}_t^b,\hat{\eta}_j^{T}({\A'}_t^{b})}=\sum_{s,a}\ang{\mathsf{Q}_s^a,\hat{\eta}_j^{T}({\A'}_s^{a})}.
\]
Therefore, letting $\eta_j^{T}$ be the induced state by $\hat{\eta}_j^{T}$, we obtain
\begin{align}
		C[\eta;\E_n(j+n_\star),\E_n(j);\E_n(n_\star),\E_n(0)]
		&=C[\eta_j;\E_n(n_\star),\E_n(0);\E_n(n_\star),\E_n(0)]\\
		&=C[\eta_j^{T};\E_n(n_\star),\E_n(0);\E_n(n_\star),\E_n(0)]\\
		&=C\left[\frac{\eta_j+\eta_j^{T}}{2};\E_n(n_\star),\E_n(0);\E_n(n_\star),\E_n(0)\right],\label{eq:sa}
\end{align}
which proves the claim.
\qed
\end{pf}

Similarly to the claim of Proposition~\ref{prop_self-adj}, the optimal CHSH value $H_n(n_\star)$ for maximally entangled states is realized by a quintuple $(\eta^{\mathrm{ME}}_\star;\E_n(n_\star),\E_n(0);\E_n(n_\star),\E_n(0))$ with a maximally entangled state $\eta^{\mathrm{ME}}_\star\in GL(\Omega_n)$ whose inducing linear map $\hat{\eta}^{\mathrm{ME}}_\star$ is self-adjoint.
In fact, letting $\eta^{\mathrm{ME}}_\star\in GL(\Omega_n)$ be
\begin{equation}\label{eq:opt ME}
		\hat{\eta}^{\mathrm{ME}}_\star=T_{2K\theta_n}\circ T_{\mathrm{x}}=
		\begin{pmatrix}
			\cos(2K_n\theta_n) & \sin(2K_n\theta_n) & 0\\
			\sin(2K_n\theta_n)& -\cos(2K_n\theta_n) & 0\\
			0 & 0 & 1
		\end{pmatrix},
\end{equation}
where
\begin{equation}\label{key}
	T_{\mathrm{x}}=\begin{pmatrix}
		1 & 0 & 0\\
		0& -1 & 0\\
		0 & 0 & 1
	\end{pmatrix}
\in GL(\Omega_n)
\end{equation}
is the reflection about the x-axis and 
\begin{equation}\label{key}
T_{2K_n\theta_n}=\begin{pmatrix}
		\cos(2K_n\theta_n) & -\sin(2K_n\theta_n) & 0\\
		\sin(2K_n\theta_n)& \cos(2K_n\theta_n) & 0\\
		0 & 0 & 1
	\end{pmatrix}
\in GL(\Omega_n)
\end{equation}
is the $(2K_n\theta_n)$-rotation about the z-axis with an integer $K_n$ (remember $\theta_n=\frac{\pi}{n}$), we can prove that the optimum $H_n(n_\star)$ is attained by $(\eta^{\mathrm{ME}}_\star;\E_n(n_\star),\E_n(0);\E_n(n_\star),\E_n(0))$.
The concrete value of $H_n(n_\star)$ for each $n$ is summarized in Table~\ref{table:H_opt}.
	\begin{table}[h]
		\begin{center}
			\begin{tabular}{|c|c|c|c|c|}
				\hline
				$n$ & $n_\star$ & $K_n$ & $k_n$& $H_n(n_\star)$ \rule[-3.5mm]{0mm}{10.5mm}                       
				\\ \hhline{|=|=|=|=|=|}
				$n\equiv1$ & $\frac{n-1}{4}$ &  $\frac{3n-3}{8}$ & $-\frac{n-1}{8}$&
				$2R_n^2\left[
				1+r_n^2\left\{2\cos\left(\frac{\pi}{4}+\frac{3\theta_n}{4}\right)+6\sin\left(\frac{\pi}{4}+\frac{\theta_n}{4}\right)+r_n^2-2\right\}\right]$  \rule[-3.5mm]{0mm}{10.5mm}          
				\\ \hline
				$n\equiv3$ & $\frac{n+1}{4}$ & $-\frac{n-3}{8}$ & $\frac{3n-1}{8}$ &  $2R_n^2\left[
				-1+r_n^2\left\{2\sin\left(\frac{\pi}{4}+\frac{3\theta_n}{4}\right)+6\cos\left(\frac{\pi}{4}+\frac{\theta_n}{4}\right)+2-r_n^2\right\}\right]$  \rule[-3.5mm]{0mm}{10.5mm}      
				\\			\hline
				$n\equiv5$ & $\frac{n-1}{4}$ & $-\frac{n+3}{8}$ & $\frac{3n+1}{8}$ & $2R_n^2\left[
				-1+r_n^2\left\{2\cos\left(\frac{\pi}{4}+\frac{3\theta_n}{4}\right)+6\sin\left(\frac{\pi}{4}+\frac{\theta_n}{4}\right)+2-r_n^2\right\}\right]$    \rule[-3.5mm]{0mm}{10.5mm}          
				\\ \hline
				$n\equiv7$ & $\frac{n+1}{4}$ & $\frac{3n+3}{8}$ & $-\frac{n+1}{8}$ & $2R_n^2\left[
				1+r_n^2\left\{2\sin\left(\frac{\pi}{4}+\frac{3\theta_n}{4}\right)+6\cos\left(\frac{\pi}{4}+\frac{\theta_n}{4}\right)+r_n^2-2\right\}\right]$  \rule[-3.5mm]{0mm}{10.5mm}      
				\\			\hline
			\end{tabular}
		\end{center}
		\caption{Explicit expressions of $H_n(n_\star)$.
			In the table, we set $\theta_n=\frac{\pi}{n}$, $r_n=\sqrt{\sec\theta_n}$, and $R_n=\frac{1}{1+r_n^2}$.}
		\label{table:H_opt}
	\end{table}
The derivation of these values is presented in \ref{appA} (see \eqref{eq:I_hat_e+}, \eqref{eq:J_hat_o+}, and Table~\ref{table:H(k)}).

Let $(\eta_\star;\E_n(n_\star),\E_n(0);\E_n(n_\star),\E_n(0))$ be the quintuple in Proposition~\ref{prop_self-adj} optimizing the CHSH value.
In \ref{appA}, we introduced another coordinate system whose orthonormal basis $\{\mathbf{f}_\mathrm{x}, \mathbf{f}_\mathrm{y},\mathbf{f}_\mathrm{z}\}$ is given by acting an orthogonal transformation (rotation)
\begin{equation}\label{key}
	W=\begin{pmatrix}
		\cos(n_\star\theta_n) & -\sin(n_\star\theta_n) &0\\
		\sin(n_\star\theta_n) & \cos(n_\star\theta_n) & 0\\
		0 & 0 & 1
	\end{pmatrix}
\end{equation}
on the original basis $\{\mathbf{e}_\mathrm{x}, \mathbf{e}_\mathrm{y},\mathbf{e}_\mathrm{z}\}$ of $\R^3$:
\begin{equation}\label{eq:coor2}
	\mathbf{f}_\mathrm{x}=W \mathbf{e}_\mathrm{x},\quad 
	\mathbf{f}_\mathrm{y}=W \mathbf{e}_\mathrm{y},\quad
	\mathbf{f}_\mathrm{z}=W\mathbf{e}_\mathrm{z}(=\mathbf{e}_\mathrm{z}).
\end{equation}
The system \eqref{eq:coor2} is chosen so that the observables $\Q_{0}$ and $\Q_{1}$ are respectively of the form
\begin{equation}\label{Q0}
	\Q_{0}^0
	=\left(
	\begin{array}{c}
		R_nr_n\cos(n_\star\theta_n)\\
		0\\
		R_n
	\end{array}
	\right)
	,\quad\Q_0^1
	=\left(
	\begin{array}{c}
		-R_nr_n\cos(n_\star\theta_n)\\
		0\\
		1-R_n
	\end{array}
	\right)
\end{equation}
and
\begin{equation}\label{Q1}
	\Q_{1}^0
	=\left(
	\begin{array}{c}
		0\\
		R_nr_n\sin(n_\star\theta_n)\\
		\frac{1}{2}
	\end{array}
	\right)
	,\quad\Q_1^1
	=\left(
	\begin{array}{c}
		0\\
		-R_nr_n\sin(n_\star\theta_n)\\
		\frac{1}{2}
	\end{array}
	\right)
\end{equation}
(see \eqref{eq:coor}).
To simplify the problem, we use this coordinate system in the following.
The maximally entangled state $\hat{\eta}^{\mathrm{ME}}_\star$ in \eqref{eq:opt ME} becomes
\begin{align}
		\hat{\eta}^{\mathrm{ME}}_\star&=
	W^{-1}\begin{pmatrix}
		\cos(2K_n\theta_n) & \sin(2K_n\theta_n) & 0\\
		\sin(2K_n\theta_n)& -\cos(2K_n\theta_n) & 0\\
		0 & 0 & 1
	\end{pmatrix}W\notag\\
&=\begin{pmatrix}
	\cos(2(K_n-n_\star)\theta_n) & \sin(2(K_n-n_\star)\theta_n) & 0\\
	\sin(2(K_n-n_\star)\theta_n)& -\cos(2(K_n-n_\star)\theta_n) & 0\\
	0 & 0 & 1
\end{pmatrix}.\label{opt ME2}
\end{align}
We explicitly parameterize the self-adjoint linear map $\hat{\eta}_\star\colon\R^3\to\R^3$ in the coordinates \eqref{eq:coor2} as
\begin{equation}\label{eq:eta_star0}
	\hat{\eta}_\star=\begin{pmatrix}
		a & b & c\\
		b & d & e\\
		c & e & 1
	\end{pmatrix}
\end{equation}
with $a,b,c,d,e\in\R$.
We note that the normalization condition $\ang{u,\hat{\eta}_\star(u)}=1$ is reflected in this expression.
Let us explicitly write down the CHSH value $C[\eta_\star;\E_n(n_\star),\E_n(0);\E_n(n_\star),\E_n(0)]$ in terms of $(a,b,c,d,e)$.
We use the expression \eqref{win-CHSH} and \eqref{win_GPT00}.
The assemblages $\{(p_\star(a|s); \omega^a_{\star s})_a\}_{s}$ on Bob induced by the state and Alice's observables $(\eta_\star;\E_n(n_\star),\E_n(0))$ are given as
\begin{equation}\label{eq:ens0}
	\begin{aligned}
		&
		\begin{aligned}
			p_\star(0|0)\omega^0_{\star 0}=\hat{\eta}_\star(e_n(n_\star))
			&=
			\begin{pmatrix}
				a & b & c\\
				b & d & e\\
				c & e & 1
			\end{pmatrix}
			W^{-1}
			\begin{pmatrix}
				R_nr_n\cos(2n_\star\theta_n)\\
				R_nr_n\sin(2n_\star\theta_n)\\
				R_n
			\end{pmatrix}\\
			&=R_n\begin{pmatrix}
				r_n  \cos (n_\star\theta_n)a+r_n  \sin (n_\star\theta_n)b+c\\
				r_n  \cos (n_\star\theta_n)b+r_n  \sin (n_\star\theta_n)d+e\\
				r_n  \cos (n_\star\theta_n)c+r_n \sin (n_\star\theta_n)e+1
			\end{pmatrix},
		\end{aligned}\\
		&p_\star(1|0)\omega^1_{\star 0}=\hat{\eta}_\star(\overline{e_n(n_\star)})
			=	R_n\begin{pmatrix}
				-r_n  \cos (n_\star\theta_n)a-r_n  \sin (n_\star\theta_n)b+r_n^2c\\
				-r_n  \cos (n_\star\theta_n)b-r_n  \sin (n_\star\theta_n)d+r_n^2e\\
				-r_n  \cos (n_\star\theta_n)c-r_n \sin (n_\star\theta_n)e+r_n^2
			\end{pmatrix},
	\end{aligned}
\end{equation}
and
\begin{equation}\label{eq:ens1}
	\begin{aligned}
		&p_\star(0|1)\omega^0_{\star 1}=\hat{\eta}_\star(e_n(0))
		=R_n\begin{pmatrix}
				r_n  \cos (n_\star\theta_n)a-r_n  \sin (n_\star\theta_n)b+c\\
				r_n  \cos (n_\star\theta_n)b-r_n  \sin (n_\star\theta_n)d+e\\
				r_n  \cos (n_\star\theta_n)c-r_n \sin (n_\star\theta_n)e+1
			\end{pmatrix},
	\\
		&p_\star(1|1)\omega^1_{\star 1}=\hat{\eta}_\star(\overline{e_n(0)})
			=	R_n\begin{pmatrix}
				-r_n  \cos (n_\star\theta_n)a+r_n  \sin (n_\star\theta_n)b+r_n^2c\\
				-r_n  \cos (n_\star\theta_n)b+r_n  \sin (n_\star\theta_n)d+r_n^2e\\
				-r_n  \cos (n_\star\theta_n)c+r_n \sin (n_\star\theta_n)e+r_n^2
			\end{pmatrix}.
		\end{aligned}
\end{equation}
The CHSH value $C[\eta_\star;\E_n(n_\star),\E_n(0);\E_n(n_\star),\E_n(0)]\equiv C[\{(p_\star(a|s); \omega^a_{\star s})_a\}_{s}; \E_n(n_\star), \E_n(0)]$ is calculated through Bob's observables $(\E_n(n_\star),\E_n(0))$ (i.e., \eqref{Q0} and \eqref{Q1}) as
\begin{align}
	C[\{(p_\star(a|s); \omega^a_{\star s})_a\}_{s}; \E_n(n_\star), \E_n(0)]
	&=4\left(\sum_{s,a}p_\star(a|s)\ang{\mathsf{Q}_s^a,\omega_{\star s}^a}-1\right)\\
	&=4R_n^2 r_n(\vec{C}\cdot(a,b,c,d,e)^{T})+2(1-2R_n)^2\label{eq:CHSH explicit}
\end{align}
with
\begin{equation}\label{eq:vec C}
	\vec{C}=
	\begin{pmatrix}
		r_n [1+\cos (2n_\star \theta_n)]\\
		2r_n \sin (2n_\star \theta_n)\\
		2(1-r_n^2)\cos (n_\star \theta_n)\\
		r_n [-1+\cos (2n_\star \theta_n)]\\
		2(1-r_n^2)\sin (n_\star \theta_n)
	\end{pmatrix}.
\end{equation}
Note that if we set 
\begin{equation}\label{ME_elements}
	\begin{aligned}
		(a,b,c,d,e)&=(\cos(2(K_n-n_\star)\theta_n), \sin(2(K_n-n_\star)\theta_n), 0, -\cos(2(K_n-n_\star)\theta_n),0)\\
		(&=:(a_\star,b_\star,c_\star,d_\star,e_\star))
	\end{aligned}
\end{equation}
in \eqref{eq:CHSH explicit} according to \eqref{opt ME2}, then we can successfully recover the CHSH value $H_n(n_\star)$ shown in Table~\ref{table:H_opt}.
That is, we have
\begin{equation}\label{opt_}
	H_n(n_\star)=
	\left\{
	\begin{aligned}
		&4R_n^2 r_n(\vec{C}\cdot(a_\star,b_\star,c_\star,d_\star,e_\star)^{T})+2(1-2R_n)^2&&(n\equiv1,7)\\
		&-[4R_n^2 r_n(\vec{C}\cdot(a_\star,b_\star,c_\star,d_\star,e_\star)^{T})+2(1-2R_n)^2]\ \ &&(n\equiv3,5)
	\end{aligned}
	\right.
\end{equation}
by substituting $n_\star$ and $K_n$ shown in Table~\ref{table:H_opt} for each case.

The problem is to find real numbers $(a,b,c,d,e)$ that optimize \eqref{eq:CHSH explicit}.
The following proposition is important (the proof is presented in \ref{app_min}).
\begin{prop}\label{prop:min_max}
	The CHSH value $H_n(n_\star)$ for the maximally entangled state $\hat{\eta}^{\mathrm{ME}}_\star$ satisfies\\
	$(n\equiv1,7)$
	\begin{equation}
		\begin{aligned}
			&H_n(n_\star)\\
			&\qquad>-\min\left\{C[\eta;\E_n(i),\E_n(j);\E_n(k),\E_n(l)]\ \middle|\ \eta\in \Omega_n\otimes_{max}\Omega_n,\ i,j,k,l\in\{0,\ldots,n-1\}\right\};
		\end{aligned}
		\label{min_max(1)}
	\end{equation}
$(n\equiv3,5)$
\begin{equation}
	\begin{aligned}
		&H_n(n_\star)\\
		&\qquad>\max\left\{C[\eta;\E_n(i),\E_n(j);\E_n(k),\E_n(l)]\ \middle|\ \eta\in \Omega_n\otimes_{max}\Omega_n,\ i,j,k,l\in\{0,\ldots,n-1\}\right\}.
	\end{aligned}
	\label{min_max(2)}
\end{equation}
\end{prop}
For simplicity, we assume $n\equiv1$.
(the argument below can be applied similarly to the other cases).
According to Proposition~\ref{prop:min_max}, we can concentrate on finding
\begin{equation}\label{eq:eta_star}
	\hat{\eta}_\star=\begin{pmatrix}
		a & b & c\\
		b & d & e\\
		c & e & 1
	\end{pmatrix}\in\Omega_n\otimes_{max}\Omega_n
\end{equation}
that maximizes the CHSH value.
In the expression \eqref{eq:eta_star}, the coordinates $\{\mathbf{f}_\mathrm{x}, \mathbf{f}_\mathrm{y},\mathbf{f}_\mathrm{z}\}$ (see \eqref{eq:coor2}) is applied and we continue following them in this part.
To prove \Thm\ref{thm:main}, suppose that the maximum CHSH value given by $\eta_\star$ satisfies
\begin{equation}
	 C[\eta_\star;\E_n(n_\star),\E_n(0);\E_n(n_\star),\E_n(0)]>	H_n(n_\star).
	\label{suppose}
\end{equation}
For $\epsilon\in(0,1)$, we introduce another state $\eta_\star^\epsilon$ by
\begin{equation}\label{key}
	\eta_\star^\epsilon=(1-\epsilon)\eta^{\mathrm{ME}}_\star+\epsilon \eta_\star,
\end{equation}
where $\eta^{\mathrm{ME}}_\star$ is the maximally entangled state \eqref{opt ME2} explicitly given as
\begin{equation}
	\hat{\eta}^{\mathrm{ME}}_\star
	=\begin{pmatrix}
		\cos\left(\frac{n-1}{4}\theta_n\right) & \sin\left(\frac{n-1}{4}\theta_n\right) & 0\\
		\sin\left(\frac{n-1}{4}\theta_n\right)& -\cos\left(\frac{n-1}{4}\theta_n\right) & 0\\
		0 & 0 & 1
	\end{pmatrix}.\label{opt ME2_n=1}
\end{equation}
Considering $\frac{n-1}{4}\theta_n=\frac{\pi}{4}-\frac{\pi}{4n}\in[0,\frac{\pi}{4}]$, we take sufficiently small $\epsilon\in(0,1)$ so that the matrix expression of $\eta_\star^\epsilon$ is given as
\begin{equation}\label{eq:eta_epsilon}
	\hat{\eta}^\epsilon_\star=\begin{pmatrix}
		a' & b' & c'\\
		b' & d' & e'\\
		c' & e' & 1
	\end{pmatrix}
\end{equation}
with
\begin{equation}\label{cond0}
	a'\ge0,\quad  b'\ge0,\quad  d'\le0.
\end{equation}
Note that the state $\eta_\star^\epsilon$ also satisfies 
\begin{equation}
	C[\eta_\star^\epsilon;\E_n(n_\star),\E_n(0);\E_n(n_\star),\E_n(0)]>	H_n(n_\star)
	\label{suppose2}
\end{equation}
due to the convexity.

Let us introduce other conditions that the state $\eta_\star^\epsilon$ should satisfy.
One restriction is introduced based on the observation in \ref{appA}. 
There we derived that for assemblages $\{(p(a|s); \omega^a_s)_a\}_{s}\in\mathsf{Ens}^{\mathrm{All}}$ such that the $\mathbf{f}_\mathrm{x}$-component $x$ of the average state $\sum_{a}p(a|0)\omega_0^a=\sum_{a}p(a|1)\omega_1^a$ satisfies 
\[
x\in[r_n\cos((2M+2)\theta_n), r_n\cos(2M\theta_n)]\quad\left(M=0,\ldots, \frac{n-3}{2}\right),
\] 
its CHSH value $C[\{(p(a|s); \omega^a_s)_a\}_{s};\E_n(n_\star), \E_n(0)]$ is bounded as
\begin{equation}\label{eq:S_bound3main}
	|C[\{(p(a|s); \omega^a_s)_a\}_{s};\E_n(n_\star), \E_n(0)]|\le 4(\beta_n^e(n_\star;M)x+\hat{\beta}_n^e(n_\star;M))-2
\end{equation}
with
\begin{equation}\label{eq:beta_emain}
	\begin{aligned}
		&\beta_n^e(n_\star;M)=R_nr_n\!\left[R_n(r_n^2-1)(\cos(n_\star\theta_n)-1)-\frac{\sin(n_\star\theta_n)}{\tan((2M+1)\theta_n)}\right],
		\\
		&\hat{\beta}_n^e(n_\star;M)=
		R_n\!\left[2R_nr_n^2\cos(n_\star\theta_n)+\frac{\sin(n_\star\theta_n)}{\sin((2M+1)\theta_n)}\right]\!+R_n^2(1+r_n^4).
	\end{aligned}
\end{equation}
The term $\beta_n^e(n_\star;M)x+\hat{\beta}_n^e(n_\star;M)$ as a function of $x$ is illustrated in Fig.~\ref{fig:graph} in \ref{appA}.
We note that for assemblages $\{(p(a|s); \omega^a_s)_a\}_{s}\in\mathsf{Ens}^{\mathrm{ME}}$ generated by a maximally entangled state, we have $x=0\in[r_n\cos((2M_0+2)\theta_n), r_n\cos(2M_0\theta_n)]$ with $M_0=\frac{n-1}{4}$ and thus 
\begin{equation}\label{eq:optimal bound CHSH}
	|C[\{(p(a|s); \omega^a_s)_a\}_{s};\E_n(n_\star), \E_n(0)]|\le 4\hat{\beta}_n^e(n_\star;M_0)-2.
\end{equation}
The equality in \eqref{eq:optimal bound CHSH} is realized by the maximally entangled state $\hat{\eta}^{\mathrm{ME}}_\star$ in \eqref{eq:opt ME}.
According to Fig.~\ref{fig:graph}, for the assemblages $\{(p(a|s); \omega^a_s)_a\}_{s}\in\mathsf{Ens}^{\mathrm{All}}$ to give a greater (or equal) CHSH value than the optimal bound in \eqref{eq:optimal bound CHSH} for maximally entangled states, $x\ge0$ needs to hold.
Based on this argument, we can observe that for \eqref{suppose2} to hold, the state $\eta_\star^\epsilon$ and its inducing assemblages $\{(p_\star^\epsilon(a|s); \omega^{\epsilon a}_{\star s})_a\}_{s}$ need to satisfy 
\begin{equation}\label{eq:cond0}
	c'=\hat{\eta}_\star^\epsilon(u)|_{\mathrm{x}}\left(=\sum_{a}p_\star^\epsilon(a|0)\omega^{\epsilon a}_{\star 0}|_{\mathrm{x}}=\sum_{a}p_\star^\epsilon(a|1)\omega^{\epsilon a}_{\star 1}|_{\mathrm{x}}\right)\ge0,
\end{equation}
where $\cdot|_{\mathrm{x}}$ denotes the $\mathbf{f}_\mathrm{x}$-component of the vector concerned.
Another series of conditions derives from a natural requirement that $\hat{\eta}_\star^\epsilon$ maps effects to (unnormalized) states. 
Due to its definition, the state $\eta_\star^\epsilon$ is ``close" to the maximally entangled state $\eta_\star^{\mathrm{ME}}$.
In particular, the map $\hat{\eta}_\star^\epsilon\circ\hat{\eta}_\star^{\mathrm{ME}}$ (note that $\hat{\eta}_\star^{\mathrm{ME}}=(\hat{\eta}_\star^{\mathrm{ME}})^{-1}$) is expected to map the effect $e_n(\frac{n_\star}{2})(=R_n(r_n\cos(n\star\theta_n),0,1)^T)$ to a neighborhood of the hyperplanes $L_1(\frac{n_\star}{2})$ and $L_2(\frac{n_\star}{2})$ in $\R^3$ spanned respectively by $\{\omega_n(\frac{n_\star}{2}), \omega_n(\frac{n_\star}{2}+1), O\}$ and $\{\omega_n(\frac{n_\star}{2}), \omega_n(\frac{n_\star}{2}-1), O\}$, where $O$ is the origin.
Introducing ``outward" normal vectors $l_1(\frac{n_\star}{2})$ and $l_2(\frac{n_\star}{2})$ of the hyperplanes $L_1(\frac{n_\star}{2})$ and $L_2(\frac{n_\star}{2})$ as
\begin{equation}\label{normal 0}
	l_1\!\left(\frac{n_\star}{2}\right)\!=\begin{pmatrix}
		1\\
		\frac{1}{\sin2\theta_n}-\frac{1}{\tan2\theta_n}\\
		-r_n
	\end{pmatrix},\quad
l_2\!\left(\frac{n_\star}{2}\right)\!=\begin{pmatrix}
	1\\
	-\frac{1}{\sin2\theta_n}+\frac{1}{\tan2\theta_n}\\
	-r_n
\end{pmatrix}
\end{equation}
respectively, we regard 
\begin{equation}\label{cond1}
	\ang{l_1\!\left(\frac{n_\star}{2}\right)\!,~ \hat{\eta}_\star^\epsilon\circ\hat{\eta}_\star^{\mathrm{ME}}\left(e_n\!\left(\frac{n_\star}{2}\right)\!\right)}\le0,\quad
	\ang{l_2\!\left(\frac{n_\star}{2}\right)\!,~ \hat{\eta}_\star^\epsilon\circ\hat{\eta}_\star^{\mathrm{ME}}\left(e_n\!\left(\frac{n_\star}{2}\right)\!\right)}\le0,
\end{equation}
equivalently,
\begin{equation}\label{d1,2_0}
	\vec{\gamma}_{1}^{(1)T}\cdot(a',b',c',d',e')^T\le r_n,\quad \vec{\gamma}_{2}^{(1)T}\cdot(a',b',c',d',e')^T\le r_n
\end{equation}
with
\begin{equation}\label{d1,2}
	\begin{aligned}
		\vec{\gamma}_{1}^{(1)}
		=
		\begin{pmatrix}
			r_n\cos(2m\theta_n)\\
			r_n^3\sin((2m+1)\theta_n)\\
			1-r_n^2\cos(2m\theta_n)\\
			r_n^3\sin(2m\theta_n)\sin\theta_n\\
			-r_n^2[\sin(2m\theta_n)-\sin\theta_n]
		\end{pmatrix}, \quad 
	\vec{\gamma}^{(1)}_{2}
		=\begin{pmatrix}
			r_n\cos(2m\theta_n)\\
			r_n^3\sin((2m-1)\theta_n)\\
			1-r_n^2\cos(2m\theta_n)\\
			-r_n^3\sin(2m\theta_n)\sin\theta_n\\
			-r_n^2[\sin(2m\theta_n)+\sin\theta_n]
		\end{pmatrix}
	\end{aligned}
\end{equation}
as natural constraints for $(a',b',c',d',e')$, where we introduced $m=\frac{n-1}{8}$.
We apply similar arguments for effects $e_n(\frac{n_\star}{2}+\alpha_i)$ ($i=1,2$) with
\begin{equation}\label{alpha_i}
	\alpha_1=2m,\quad \alpha_2=n-2m.
\end{equation}
That is, following \eqref{cond1}, we additionally impose 
\begin{equation}\label{cond2,3}
	\begin{aligned}
			&\ang{l_2\!\left(\frac{n_\star}{2}+\alpha_1\right)\!,~ \hat{\eta}_\star^\epsilon\circ\hat{\eta}_\star^{\mathrm{ME}}\left(e_n\!\left(\frac{n_\star}{2}+\alpha_1\right)\!\right)}\le0,\\
			&\ang{l_1\!\left(\frac{n_\star}{2}+\alpha_2\right)\!,~ \hat{\eta}_\star^\epsilon\circ\hat{\eta}_\star^{\mathrm{ME}}\left(e_n\!\left(\frac{n_\star}{2}+\alpha_2\right)\!\right)}\le0,
	\end{aligned}
\end{equation}
where each $l_j(\frac{n_\star}{2}+\alpha_i)$ ($i=1,2, j=1,2$) is a rotated normal vector given as $l_j(\frac{n_\star}{2}+\alpha_i)=T_{2\alpha_i\theta_n}l_j(\frac{n_\star}{2})$ with
\[
T_{2\alpha_i\theta_n}=\begin{pmatrix}
	\cos(2\alpha_i\theta_n) & -\sin(2\alpha_i\theta_n) & 0\\
	\sin(2\alpha_i\theta_n)& \cos(2\alpha_i\theta_n) & 0\\
	0 & 0 & 1
\end{pmatrix}.
\]
The conditions \eqref{cond2,3} are expanded respectively as
\begin{equation}\label{d3,4_0}
			\vec{\gamma}_{3}^{(1)T}\cdot(a',b',c',d',e')^T\le r_n,\quad \vec{\gamma}_{4}^{(1)T}\cdot(a',b',c',d',e')^T\le r_n
\end{equation}
with
\begin{equation}\label{d3,4}
		\vec{\gamma}^{(1)}_{3}
=\begin{pmatrix}
	r_n^3\cos((4m-1)\theta_n)\cos(2m\theta_n)\\
	r_n^3\sin((2m-1)\theta_n)\\
	r_n^2[\cos((4m-1)\theta_n)-\cos(2m\theta_n)]\\
	-r_n^3\sin((4m-1)\theta_n)\sin(2m\theta_n)\\
	r_n^2[\sin((4m-1)\theta_n)+\sin(2m\theta_n)]
\end{pmatrix},\quad
		\vec{\gamma}^{(1)}_{4}
		=\begin{pmatrix}
			r_n^3\cos(6m\theta_n)\cos((4m-1)\theta_n)\\
			r_n^3\sin((2m+1)\theta_n)\\
			r_n^2[\cos((4m-1)\theta_n)-\cos(6m\theta_n)]\\
			-r_n^3\sin((4m-1)\theta_n)\sin(6m\theta_n)\\
			-r_n^2[\sin((4m-1)\theta_n)+\sin(6m\theta_n)]
		\end{pmatrix}.
\end{equation}
Now we consider the following problem:
	\begin{equation}\label{key}
		\left[\quad		
		\begin{aligned}
			&\mbox{maximize}&&\vec{C}^{T}\cdot(a',b',c',d',e')^{T}\\
			&\mbox{subject to}
			&&a'\ge0,\ \ b'\ge0,\ \ c'\ge0,\ \ d'\le0,\\
			&&
			&\begin{pmatrix}
				& \vec{\gamma}_1^{(1)T} &\\
				& \vec{\gamma}_2^{(1)T} &\\
				& \vec{\gamma}_3^{(1)T} &\\
				& \vec{\gamma}_4^{(1)T} &
			\end{pmatrix}\cdot
			(a',b',c',d',e')^{T}
			\le
			\begin{pmatrix}
				r_n\\
				r_n\\
				r_n\\
				r_n
			\end{pmatrix}.
		\end{aligned}
	\quad\right]
	\end{equation}
For later use, we set $d'=-d'_1$ and $e'=e'_1-e'_2$ ($e'_1,e'_2\ge0$) and rewrite the problem as an equivalent form
\begin{equation}\label{primal_main}
	\left[\quad
		\begin{aligned}
		&\mbox{maximize}&&\vec{C}'^{T}\cdot(a',b',c',d',e'_1,e'_2)^{T}\\
		&\mbox{subject to}
		&&a'\ge0,\ \ b'\ge0,\ \ c'\ge0,\ \ d'\ge0,\ \ e'_1\ge0,\ \ e'_2\ge0,\\
		&&
		&\Gamma\cdot
		(a',b',c',d',e'_1,e'_2)^{T}
		\le
		\vec{r},
	\end{aligned}
\quad\right],
\end{equation}
where we introduced vectors
\begin{equation}\label{C'}
	\vec{C}'=\begin{pmatrix}
	r_n [1+\cos (4m \theta_n)]\\
	2r_n \sin (4m \theta_n)\\
	2(1-r_n^2)\cos (2m \theta_n)\\
	r_n [1-\cos (4m \theta_n)]\\
	2(1-r_n^2)\sin (4m \theta_n)\\
	-2(1-r_n^2)\sin (4m \theta_n)
	\end{pmatrix},\quad\vec{r}=\begin{pmatrix}
	r_n\\
	r_n\\
	r_n\\
	r_n
\end{pmatrix}
\end{equation}
and a matrix
\begin{equation}\label{key}
	\Gamma=\begin{pmatrix}
		& {\vec{\gamma}_1^{(1)'T}} &\\
		& {\vec{\gamma}_2}^{(1)'T}&\\
		& {\vec{\gamma}_3}^{(1)'T} &\\
		& {\vec{\gamma}_4}^{(1)'T} &
	\end{pmatrix}
\end{equation}
with
\begin{equation}\label{d1,2'}
	\begin{aligned}
		\vec{\gamma}_{1}^{(1)'}
		=
		\begin{pmatrix}
			r_n\cos(2m\theta_n)\\
			r_n[\sin(2m\theta_n)+\cos(2m\theta_n)\tan\theta_n]\\
			1-r_n^2\cos(2m\theta_n)\\
			-r_n\sin(2m\theta_n)\tan\theta_n\\
			-r_n^2\sin(2m\theta_n)+\tan\theta_n\\
			r_n^2\sin(2m\theta_n)-\tan\theta_n
		\end{pmatrix}, \quad 
		\vec{\gamma}_{2}^{(1)'}
		=\begin{pmatrix}
			r_n\cos(2m\theta_n)\\
			r_n[\sin(2m\theta_n)-\cos(2m\theta_n)\tan\theta_n]\\
			1-r_n^2\cos(2m\theta_n)\\
			r_n\sin(2m\theta_n)\tan\theta_n\\
			-r_n^2\sin(2m\theta_n)-\tan\theta_n\\
			r_n^2\sin(2m\theta_n)+\tan\theta_n
		\end{pmatrix}
	\end{aligned}
\end{equation}
and
\begin{equation}\label{d3,4'}
	\vec{\gamma}_{3}^{(1)'}
	=\begin{pmatrix}
		r_n^3\cos((4m-1)\theta_n)\cos(2m\theta_n)\\
		r_n^3\sin((2m-1)\theta_n)\\
		r_n^2[\cos((4m-1)\theta_n)-\cos(2m\theta_n)]\\
		r_n^3\sin((4m-1)\theta_n)\sin(2m\theta_n)\\
		r_n^2[\sin((4m-1)\theta_n)+\sin(2m\theta_n)]\\
		-r_n^2[\sin((4m-1)\theta_n)+\sin(2m\theta_n)]
	\end{pmatrix},\quad
	\vec{\gamma}_{4}^{(1)'}
	=\begin{pmatrix}
		r_n^3\cos(6m\theta_n)\cos((4m-1)\theta_n)\\
		r_n^3\sin((2m+1)\theta_n)\\
		r_n^2[\cos((4m-1)\theta_n)-\cos(6m\theta_n)]\\
		r_n^3\sin((4m-1)\theta_n)\sin(6m\theta_n)\\
		-r_n^2[\sin((4m-1)\theta_n)+\sin(6m\theta_n)]\\
		r_n^2[\sin((4m-1)\theta_n)+\sin(6m\theta_n)]
	\end{pmatrix}.
\end{equation}
In \ref{app_LP}, we prove that 
\begin{equation}\label{ME_elements3_main}
	\begin{aligned}
		(a',b',c',d'_1,e'_1,e'_2)=(a_\star,b_\star,c_\star,-d_\star,0,0),
	\end{aligned}
\end{equation}
i.e.,
\begin{equation}\label{ME_elements2_main}
	\begin{aligned}
		(a',b',c',d',e')=(a_\star,b_\star,c_\star,d_\star,e_\star)
		=(\cos(2m\theta_n), \sin(2m\theta_n), 0, -\cos(2m\theta_n),0)
	\end{aligned}
\end{equation}
is an optimal solutions for this linear programming problem, but it contradicts \eqref{suppose2}.
Therefore, we conclude that the maximally entangled state $	\hat{\eta}^{\mathrm{ME}}_\star$ shows the maximum CHSH value for odd polygon theories with $n\equiv1$:
\begin{equation}\label{conclude max}
	\begin{aligned}
		H_n(n_\star)
		&=4R_n^2 r_n(\vec{C}\cdot(a_\star,b_\star,c_\star,d_\star,e_\star)^{T})+2(1-2R_n)^2\\
		&=\max\left\{C[\eta;\E_n(i),\E_n(j);\E_n(k),\E_n(l)]\ \middle|\ \eta\in \Omega_n\otimes_{max}\Omega_n,\ i,j,k,l\in\{0,\ldots,n-1\}\right\}.
	\end{aligned}
\end{equation}
We can develop similar arguments for the other cases $n\equiv3,5,7$ by modifying parameters in the linear programming problem (the explicit formulations are presented in \ref{app_LP_others}).
In this way, \Thm\ref{thm:main} has been proved for arbitrary odd polygon theories.

\section{Conclusion}
\label{sec:conc}
In the present research, we studied the CHSH scenario in a class of GPTs called regular polygon theories, and investigated whether maximally entangled states in regular polygon theories as natural generalizations of quantum ones realize the optimal CHSH values.
As a consequence, similarly to the quantum result, where maximally entangled states optimize the CHSH value, we found that the generalized maximally entangled states give the optimal CHSH values also in regular polygon theories.
In our study, the extension of maximally entangled states to regular polygon theories was given in terms of  abstract order-isomorphisms between effects and states.
Our result thus manifests that such an abstract notion of maximal entanglement indeed has a physical meaning: it optimizes the CHSH correlation.
We expect our result to contribute to revealing what is essential for entanglement to realize phenomena impossible in classical theory. 
While we successfully proved that maximal entanglement is also necessary for optimizing the CHSH value in even-sided polygon theories, our method in this paper does not imply whether similar observation is obtained in odd-sided theories.
Future study will be needed to give the complete characterization of the optimal CHSH values for odd-sided polygon theories or more general class of GPTs.

\section*{Acknowledgment}
The author thanks Takayuki Miyadera for giving insightful comments.
The author is supported by MEXT QLEAP.

\appendix
\def\thesection{Appendix\ \Alph{section}}
\section{Proof of Lemma~\ref{lem:max_k0}}
\label{appA}
\renewcommand{\theequation}{A.\arabic{equation}}
\renewcommand{\thesection}{\Alph{section}}
\setcounter{equation}{0}
\setcounter{subsection}{0}
\renewtheorem{lem}[subsection]{Lemma}
\subsection{Part 1: Concrete values of $G_n(k)$ and $H_n(k)$}\label{appA1}
\begin{figure}[h]
	\centering
	\includegraphics[scale=0.4]{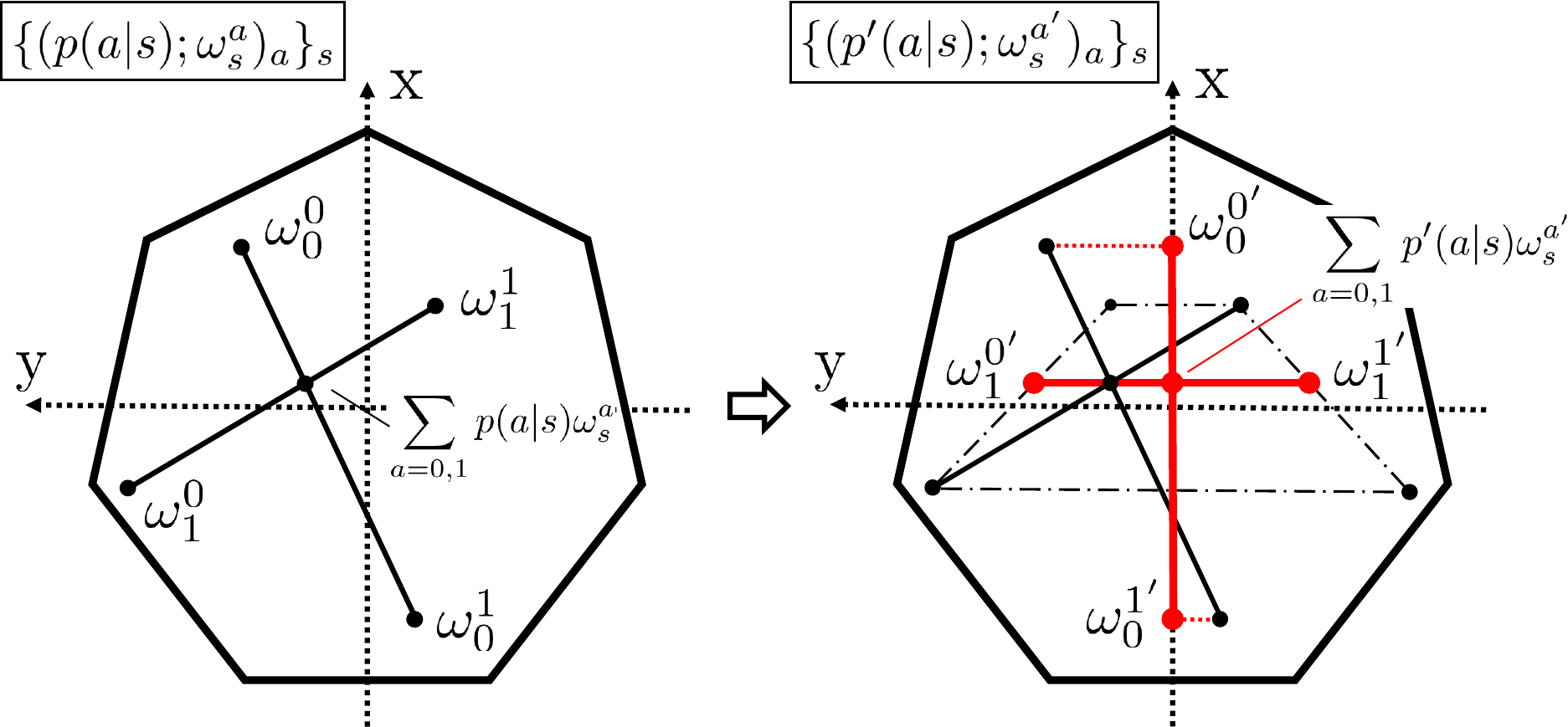}
	\caption{Simplification of assemblages $\{(p(a|s); \omega^{a}_s)_a\}_{s}$.
	The original assemblages $\{(p(a|s); \omega^{a}_s)_a\}_{s}$ are converted to ``orthogonalized" ones $\{(p'(a|s); {\omega'}^{a}_s)_a\}_{s}$ with the same CHSH value.}
	\label{fig_ortho}
\end{figure}

In this proof, we often write a set of sequential integers $\{m',m'+1,\ldots,m-1, m\}$ by $\inter{m',m}$.
We fix $k\in\inter{0,\frac{n-1}{2}}$ and consider optimizing the quantity 
\begin{equation}\label{eq:2Pwin}
	2P_{\mathrm{win}}[\{(p(a|s); \omega^a_s)_a\}_{s}; \E_n(k), \E_n(0)]=\sum_{s=0,1}\sum_{a=0,1}p(a|s)\ang{\mathsf{Q}_s^a,\omega_s^a}
\end{equation}
for $\{(p(a|s); \omega^a_s)_a\}_{s}\in\mathsf{Ens}^{\mathrm{All}}$ and $\{(p(a|s); \omega^a_s)_a\}_{s}\in\mathsf{Ens}^{\mathrm{ME}}$.
The observables $\Q_{0}$ and $\Q_{1}$ are respectively of the form (see \eqref{eq:Q(s=0)} and \eqref{eq:Q(s=1)}) 
\begin{equation}
	\Q_{0}^0
	=\left(
	\begin{array}{c}
		\vec{q}_0\\
		R_n
	\end{array}
\right)
	,\quad\Q_0^1
		=\left(
\begin{array}{c}
	-\vec{q}_0\\
	1-R_n
\end{array}
		\right)
\end{equation}
with 
\begin{equation}\label{key}
	\vec{q}_0=R_nr_n\cos(k\theta_n)\left(
	\begin{array}{c}
		\cos(k\theta_n)\\
		\sin(k\theta_n)
	\end{array}
	\right)
\end{equation}
and
\begin{equation}
	\Q_{1}^0
	=\left(
	\begin{array}{c}
		\vec{q}_1\\
		\frac{1}{2}
	\end{array}
	\right)
	,\quad\Q_0^1
	=\left(
	\begin{array}{c}
		-\vec{q}_1\\
		\frac{1}{2}
	\end{array}
	\right)
\end{equation}
with 
\begin{equation}\label{key}
	\vec{q}_1=R_nr_n\sin(k\theta_n)\left(
	\begin{array}{c}
		\cos(k\theta_n+\frac{\pi}{2})\\
		\sin(k\theta_n+\frac{\pi}{2})
	\end{array}
	\right),
\end{equation}
where $R_n=\frac{1}{1+r_n^2}(\le\frac{1}{2})$ and $\theta_n=\frac{\pi}{n}$.
To simplify the argument, we reset the x-axis and y-axis in the direction of the vectors $(\cos(k\theta_n),
\sin(k\theta_n))^{T}$ and $(\cos(k\theta_n+\frac{\pi}{2}),
\sin(k\theta_n+\frac{\pi}{2}))^{T}$ respectively so that
\begin{equation}\label{eq:coor}
		\vec{q}_0=R_nr_n\cos(k\theta_n)\left(
	\begin{array}{c}
		1\\
		0
	\end{array}
	\right),\quad
		\vec{q}_1=R_nr_n\sin(k\theta_n)\left(
	\begin{array}{c}
	0\\
	1
	\end{array}
	\right)
\end{equation}
(see also \eqref{eq:coor2}).
Under this setting, suppose that a pair of assemblages $\{(p(a|s); \omega^a_s)_a\}_{s}\in\mathsf{Ens}^{\mathrm{All}}$ optimizes \eqref{eq:2Pwin}.
We introduce another $\{(p'(a|s); \omega^{a'}_s)_a\}_{s}$ defined as
\begin{equation}\label{eq:ortho_Ens1}
	p'(a|s)=\left\{
	\begin{aligned}
		&p(a|s)&&(s=0)\\
		&\ \ \frac{1}{2}&&(s=1)
	\end{aligned}
	\right.		
\end{equation}
and
\begin{equation}\label{eq:ortho_Ens2}
	{\omega'}^{a}_s=
	\left\{
	\begin{aligned}
		&\qquad\qquad
		\begin{pmatrix}
			\omega^a_s |_\mathrm{x}\\
			0\\
			1
		\end{pmatrix}
	&&(s=0)\\
		&\begin{pmatrix}
				p(0|1)\omega^0_s |_\mathrm{x}+p(1|1)\omega^{1}_s |_\mathrm{x}\\
				p(a|s)\omega^a_s |_\mathrm{y}-p(\bar{a}|s)\omega^{\bar{a}}_s |_\mathrm{y}\\
				1
		\end{pmatrix}
	&&(s=1),
	\end{aligned}
	\right.
\end{equation}
where $\bar{a}=a\oplus1$ and $\omega^a_s |_\mathrm{x}$ denotes the x-component of the vector $\omega^a_s\in\R^3$ (similarly for $\omega^a_s |_\mathrm{y}$).
Because ${\omega'}^{a}_s\in\Omega_n$ due to the geometrical symmetry of $\Omega_n$ and
\[
\sum_{a=0,1}p'(a|0){\omega'}_0^{a}=\sum_{a=0,1}p'(a|1){\omega'}_1^{a}\left(=
\begin{pmatrix}
	\sum_{a=0,1}p(a|s)\omega^a_s|_\mathrm{x}\\
	0\\
	1
\end{pmatrix}\right)
\]
holds, we can see $\{(p'(a|s); {\omega'}^{a}_s)_a\}_{s}\in\mathsf{Ens}^{\mathrm{All}}$.
Moreover, \eqref{eq:ortho_Ens1} and \eqref{eq:ortho_Ens2} imply
\begin{equation}\label{eq:Inv}
	\sum_{a=0,1}p'(a|s)\ang{\mathsf{Q}_s^a,{\omega'}_s^{a}}=\sum_{a=0,1}p(a|s)\ang{\mathsf{Q}_s^a,\omega_s^{a}}
\end{equation}
for both $s=0,1$.
Hence, to investigate the quantity $P_{\mathrm{win}}$ for $\mathsf{Ens}^{\mathrm{All}}$, it is enough to focus on such ``orthogonalized" assemblages as \eqref{eq:ortho_Ens1} and \eqref{eq:ortho_Ens2} (see Fig.~\ref{fig_ortho}).

We first assume that $k$ is even.
We consider specifying the assemblage $(p(a|0);\omega_0^a)_a$ optimizing the sum
\begin{equation}\label{eq:S_0}
	S_0:=\sum_{a=0,1}p(a|0)\ang{\mathsf{Q}_0^a,\omega_0^a}
\end{equation}
for $s=0$ in \eqref{eq:2Pwin}.
Fix $\sum_{a=0,1}p(a|0)\omega_0^a|_\x=x$ and let $p(0|0)\omega_0^0|_\x=x_0$.
The vector $p(0|0)\omega_0^0$ is in the intersection of the positive cone $V_{n+}$ and the hyperplane $z=p(0|0)$, so 
\begin{equation}\label{eq:bound1}
	-\frac{p}{r_n}\le x_0\le pr_n
\end{equation}
holds, where we simply write $p(0|0)$ by $p$.
We apply the same argument for $p(1|0)\omega_0^1|_\x=x-x_0$ to obtain $-\frac{1-p}{r_n}\le x-x_0\le (1-p)r_n$ or 
\begin{equation}\label{eq:bound2}
	x-(1-p)r_n\le x_0\le x+\frac{1-p}{r_n}.
\end{equation}
We calculate \eqref{eq:S_0} as
\begin{align}
	S_0
	&=R_nr_n\cos(k\theta_n)x_0+pR_n-R_nr_n\cos(k\theta_n)(x-x_0)+(1-p)(1-R_n)\\
	&=2R_nr_n\cos(k\theta_n)x_0-R_nr_n\cos(k\theta_n)x+(2p-1)R_n+1-p\label{eq:S01}.
\end{align}
Note that $\cos(k\theta_n)\ge0$ ($k\in\inter{0,\frac{n-1}{2}}$) holds.
We have
\begin{equation}\label{eq:eva}
	2R_nr_n\cos(k\theta_n)x_0\le\left\{
	\begin{aligned}
		&2R_nr_n\cos(k\theta_n)(pr_n)&&\left(pr_n\le x+\frac{1-p}{r_n}\right)\\
		&2R_nr_n\cos(k\theta_n)\left(x+\frac{1-p}{r_n}\right)&&\left(pr_n\ge x+\frac{1-p}{r_n}\right)
	\end{aligned}
	\right.
\end{equation}
by means of \eqref{eq:bound1} and \eqref{eq:bound2}.
Since $pr_n\le x+\frac{1-p}{r_n}$ iff $p\le\frac{r_nx+1}{r_n^2+1}$, it follows from \eqref{eq:S01} that 
\begin{equation}\label{eq:eva1}
	S_0\le\left\{
	\begin{aligned}
		&(2R_nr_n^2\cos(k\theta_n)+2R_n-1)p-R_nr_n\cos(k\theta_n)x+1-R_n&&\left(p\le\frac{r_nx+1}{r_n^2+1}\right)\\
		&(-2R_n\cos(k\theta_n)+2R_n-1)p+R_n(r_nx+2)\cos(k\theta_n)+1-R_n&&\left(p\ge\frac{r_nx+1}{r_n^2+1}\right).
	\end{aligned}
	\right.
\end{equation}
The former coefficient of $p$ satisfies $(2R_nr_n^2\cos(k\theta_n)+2R_n-1)\ge0$ because $\cos(k\theta_n)\in\left[\sin(\frac{\pi}{2n}),1\right]$, while the latter one clearly satisfies $(2R_n-1-2R_n\cos(k\theta_n))\le0$.
Thus we obtain the tight upper bound for $S_0$ as
\begin{equation}\label{eq:S_0_upper(cos>0)}
	S_0\leq R_n^2r_n(r_n^2-1)(\cos(k\theta_n)-1)x+2R_n^2r_n^2\cos(k\theta_n)+R_n^2(1+r_n^4),
\end{equation}
where the equality holds iff $x_0=pr_n= x+\frac{1-p}{r_n}$, i.e., $\omega_0^0|_\x=r_n$ and $\omega_0^1|_\x=-\frac{1}{r_n}$.
We can derive the tight lower bound for $S_0$ by evaluating \eqref{eq:S01} in a similar way.
It is given by 
\begin{equation}\label{eq:S_0_lower(cos>0)}
	S_0\geq R_n^2r_n(r_n^2-1)(-\cos(k\theta_n)+1)x-2R_n^2r_n^2\cos(k\theta_n)+2R_n^2r_n^2,
\end{equation}
where the equality holds iff $x_0=-\frac{p}{r_n}= x-(1-p)r_n$, i.e., $\omega_0^0|_\x=-\frac{1}{r_n}$ and $\omega_0^1|_\x=r_n$.

Let us consider optimizing
\begin{equation}\label{eq:S_1}
	S_1:=\sum_{a=0,1}p(a|1)\ang{\mathsf{Q}_1^a,\omega_1^a}.
\end{equation}
As discussed in \eqref{eq:Inv}, we can focus on an assemblage $(p'(a|1); {\omega'_1}^a)_a$ of the form
\begin{equation}\label{key}
	p'(0|1)=p'(1|1)=\frac{1}{2}
\end{equation}
and
\begin{equation}\label{eq:x}
	{\omega'_1}^0
	=
	\begin{pmatrix}
		x\\
		y\\
		1
	\end{pmatrix},\quad
{\omega'_1}^1
=
\begin{pmatrix}
	x\\
	-y\\
	1
\end{pmatrix},
\end{equation}
and obtain
\begin{equation}\label{eq:S1_y}
	S_1=R_nr_n\sin(k\theta_n)y+\frac{1}{2}.
\end{equation}
Let us fix $x$ in \eqref{eq:x}, and introduce $f(x)$ as the positive coordinate of the intersection of the line $x=x$ and the boundary of $\Omega_n$ in the hyperplane $z=1$.
The function $f$ is explicitly described as
\begin{equation}\label{eq:def_f}
	\begin{aligned}
		f(x)=-\frac{1}{\tan((2M+1)\theta_n)}x&+\frac{r_n\cos\theta_n}{\sin((2M+1)\theta_n)}
		\\
		&\quad\quad(x\in[r_n\cos((2M+2)\theta_n), r_n\cos(2M\theta_n)])
	\end{aligned}
\end{equation}
with $M=0,1,\ldots, \frac{n-3}{2}$ in accord with $-\frac{1}{r_n}\leq x\leq r_n$.
We have $-f(x)\le y\le f(x)$, and thus obtain tight relations
\begin{equation}\label{eq:S1_opt}
	-R_nr_n\sin(k\theta_n)f(x)+\frac{1}{2}\le S_1\le R_nr_n\sin(k\theta_n)f(x)+\frac{1}{2},
\end{equation}
where the equalities hold iff $y=\pm f(x)$.

\begin{figure}[h]
	\begin{minipage}[h]{1\linewidth}
		\centering
		\includegraphics[scale=0.42]{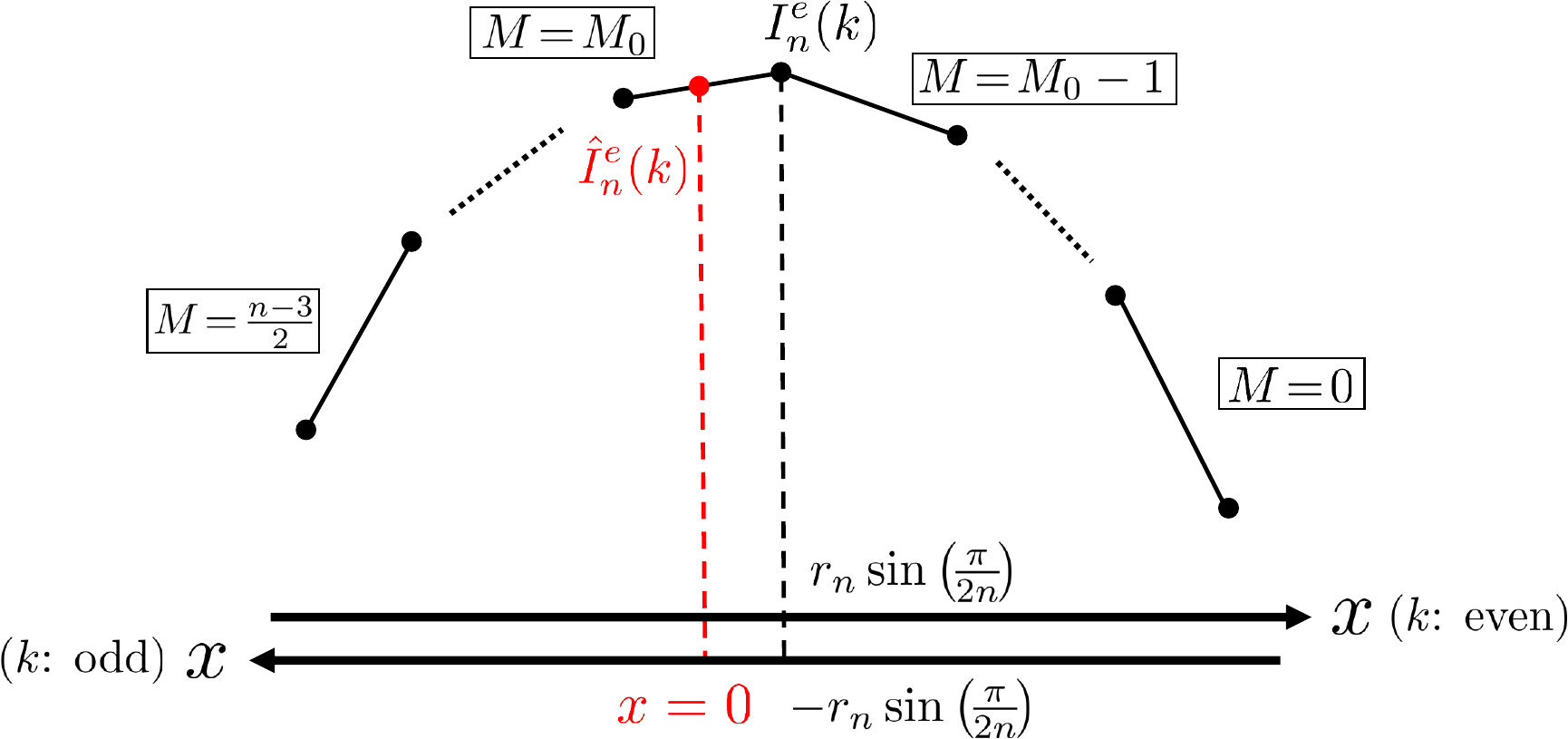}
		\subcaption{The case $n\equiv1,5$.}\label{subfig:1,5}
	\end{minipage}\vspace{5mm}\\
	\begin{minipage}[h]{1\linewidth}
		\centering
		\includegraphics[scale=0.42]{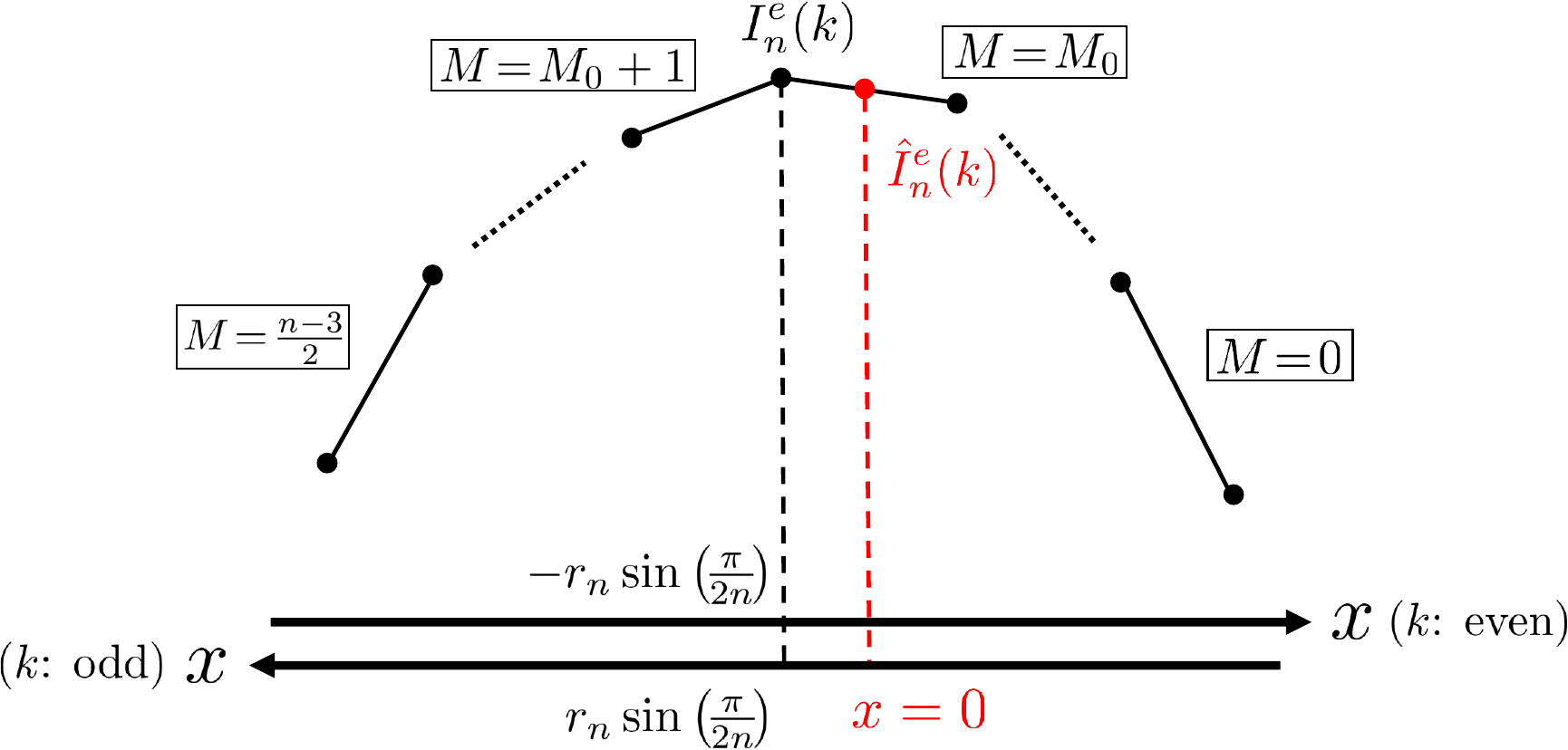}
		\subcaption{The case $n\equiv3,7$.}\label{subfig:3,7}
	\end{minipage}
	\caption{Description of the function $\beta_n^e(k;M)x+\hat{\beta}_n^e(k;M)$.  
		Each segment corresponds to $M=0,\ldots,\frac{n-3}{2}$.
	The function $\beta_n^o(k;M)x+\hat{\beta}_n^o(k;M)$ is obtained by the replacements $x\to -x$ and $\cos(k\theta_n)\to -\cos(k\theta_n)$.}\label{fig:graph}
\end{figure}

Overall, writing $2P_{\mathrm{win}}[\{(p(a|s); \omega^a_s)_a\}_{s}; \E_n(k), \E_n(0)]$ simply as $S[\{(p(a|s); \omega^a_s)_a\}_{s};k]$, we obtain from \eqref{eq:S_0_upper(cos>0)}, \eqref{eq:S_0_lower(cos>0)}, and \eqref{eq:S1_opt}
\begin{equation}\label{eq:S_bound}
		-\beta_n^e(k;M)x-\hat{\beta}_n^e(k;M)+\frac{3}{2}\leq S[\{(p(a|s); \omega^a_s)_a\}_{s};k]\le \beta_n^e(k;M)x+\hat{\beta}_n^e(k;M)+\frac{1}{2}
\end{equation}
with
\begin{equation}\label{eq:beta_e}
	\begin{aligned}
		&\beta_n^e(k;M)=R_nr_n\!\left[R_n(r_n^2-1)(\cos(k\theta_n)-1)-\frac{\sin(k\theta_n)}{\tan((2M+1)\theta_n)}\right],
		\\
		&\hat{\beta}_n^e(k;M)=
		R_n\!\left[2R_nr_n^2\cos(k\theta_n)+\frac{\sin(k\theta_n)}{\sin((2M+1)\theta_n)}\right]\!+R_n^2(1+r_n^4)
	\end{aligned}
\end{equation}
for even $k\in \inter{0,\frac{n-1}{2}}$ and assemblages $\{(p(a|s); \omega^a_s)_a\}_{s}$ such that $\sum_{a}p(a|0)\omega_0^a|_\x=\sum_{a}p(a|1)\omega_1^a|_\x=x$ and $x\in[r_n\cos((2M+2)\theta_n), r_n\cos(2M\theta_n)]$ ($M=0,\ldots, \frac{n-3}{2}$).
It indicates
\begin{equation}\label{eq:S_bound2}
	-\beta_n^e(k;M)x-\hat{\beta}_n^e(k;M)+\frac{1}{2}\leq \frac{C[\{(p(a|s); \omega^a_s)_a\}_{s};k]}{4}\le \beta_n^e(k;M)x+\hat{\beta}_n^e(k;M)-\frac{1}{2},
\end{equation}
or
\begin{equation}\label{eq:S_bound3}
	 |C[\{(p(a|s); \omega^a_s)_a\}_{s};k]|\le 4(\beta_n^e(k;M)x+\hat{\beta}_n^e(k;M))-2
\end{equation}
in terms of \eqref{win-CHSH}.
We evaluate the term $\beta_n^e(k;M)x+\hat{\beta}_n^e(k;M)$ in \eqref{eq:S_bound3} for even $k\in \inter{0,\frac{n-1}{2}}$.
With
	\begin{equation}\label{eq:M_0}
	M_0=\left\{
	\begin{aligned}
		&\frac{n-1}{4}\quad(n\equiv1,5)\\
		&\frac{n-3}{4}\quad(n\equiv3,7),
	\end{aligned}
	\right.
\end{equation}
it holds for any even $k\in \inter{0,\frac{n-1}{2}}$ that
\begin{itemize}
\item[(i)] ($n\equiv 1,5$)\quad $\beta_n^e(k;M)<0$ for $M< M_0$ and $\beta_n^e(k;M)>0$ for $M\ge M_0$;
\item[(ii)] ($n\equiv 3,7$)\quad $\beta_n^e(k;M)<0$ for $M\le M_0$ and $\beta_n^e(k;M)>0$ for $M> M_0$.
\end{itemize}
Here we only prove (i), but the proof of (ii) proceeds in the same way.
Let $n\equiv1,5$.
The coefficient $-\frac{1}{\tan((2M+1)\theta_n)}$ in $\beta_n^e(k,M)$ (see \eqref{eq:beta_e}) is an increasing function with respect to $M=0,\ldots,\frac{n-3}{2}$ and clearly satisfies $\frac{1}{\tan((2M+1)\theta_n)}<0$ (thus $g(k,M)<0$) for $M\leq M_0-1$. 
For $M=M_0$, we have 
\begin{align*}
	\frac{\beta_n^e(k,M_0)}{R_nr_n}
	&=\tan^2\frac{\theta_n}{2}(\cos(k\theta_n)-1)+\tan\frac{\theta_n}{2}\sin(k\theta_n)\\
	&=\frac{2\tan\frac{\theta_n}{2}}{\cos\frac{\theta_n}{2}}\cos\!\left(\frac{k\pi}{2n}+\frac{\theta_n}{2}\right)\sin\left(\frac{k\pi}{2n}\right)>0
\end{align*}
for any even $k\in \inter{0,\frac{n-1}{2}}$, which proves (i).
The observations (i) and (ii) enable us to plot the function $\beta_n^e(k;M)x+\hat{\beta}_n^e(k;M)$ of $x$ as Fig.~\ref{fig:graph}.
It shows
\begin{equation}\label{key}
	\beta_n^e(k;M)x+\hat{\beta}_n^e(k;M)\le I_n^{e}(k)
\end{equation}
with
\begin{align}\label{key}
	 I_n^{e}(k)=
		&\left\{
	\begin{aligned}
		&\beta_n^e(k;M_0)\left(r_n\sin\left(\frac{\pi}{2n}\right)\right)+\hat{\beta}_n^e(k;M_0) \quad&&(n\equiv1,5)\\
		&\beta_n^e(k;M_0)\left(-r_n\sin\left(\frac{\pi}{2n}\right)\right)+\hat{\beta}_n^e(k;M_0) \quad&&(n\equiv3,7)
	\end{aligned}
	\right.\\
	=&\left\{
	\begin{aligned}
	&\frac{R_n}{\cos^2\frac{\theta_n}{2}}\!\left(r_n^2\sin^3\frac{\theta_n}{2}+1\right)\cos(k\theta_n)+\frac{R_n}{\cos\frac{\theta_n}{2}}\!\left(r_n^2\sin^2\frac{\theta_n}{2}+1\right)\sin(k\theta_n)\\
		&\qquad\qquad\qquad\qquad\qquad\qquad\qquad
		-\frac{R_nr_n^2}{\cos^2\frac{\theta_n}{2}}\sin^3\frac{\theta_n}{2}+R_n^2(1+r_n^4)\qquad(n\equiv1,5)\\
		&\frac{R_n}{\cos^2\frac{\theta_n}{2}}\!\left(-r_n^2\sin^3\frac{\theta_n}{2}+1\right)\cos(k\theta_n)+\frac{R_n}{\cos\frac{\theta_n}{2}}\!\left(r_n^2\sin^2\frac{\theta_n}{2}+1\right)\sin(k\theta_n)\\
		&\qquad\qquad\qquad\qquad\qquad\qquad\qquad
		+\frac{R_nr_n^2}{\cos^2\frac{\theta_n}{2}}\sin^3\frac{\theta_n}{2}+R_n^2(1+r_n^4)\qquad(n\equiv3,7).
	\end{aligned}\label{eq:I_e}
	\right.
\end{align}

The above calculations for even $k$ can be applied to the case when $k$ is odd.
In fact, for odd $k$ the conditions \eqref{eq:bound1} and \eqref{eq:bound2} respectively become
\begin{equation}\label{}
	-pr_n\le x_0\le \frac{p}{r_n}\quad\mbox{and}\quad x-\frac{1-p}{r_n}\le x_0\le x+(1-p)r_n,
\end{equation}
i.e., 
\begin{equation}\label{}
	-\frac{p}{r_n}\le (-x_0)\le pr_n \quad\mbox{and}\quad  (-x)-(1-p)r_n\le(-x_0) \le (-x)+\frac{1-p}{r_n}.
\end{equation}
Rewriting the value $S_0$ in \eqref{eq:S_0} as
\begin{equation}\label{key}
		S_0=2R_nr_n(-\cos(k\theta_n))(-x_0)-R_nr_n(-\cos(k\theta_n))(-x)+(2p-1)R_n+1-p,
\end{equation}
we can apply the same evaluations as before by the replacement
\begin{equation}\label{key}
	\cos(k\theta_n)\to -\cos(k\theta_n),\quad x\to -x.
\end{equation}
In this way, we conclude
\begin{equation}\label{eq:S0_upper}
	S_0\leq 
		R_n^2r_n(r_n^2-1)(-\cos(k\theta_n)-1)x+2R_n^2r_n^2\cos(k\theta_n)+2R_n^2r_n^2
\end{equation}
and
\begin{equation}\label{eq:S0_lower}
	S_0\geq 
		R_n^2r_n(r_n^2-1)(\cos(k\theta_n)+1)x-2R_n^2r_n^2\cos(k\theta_n)+R_n^2(1+r_n^4)
\end{equation}
for odd $k$, where the former equality holds iff $\omega_0^0|_\x=\frac{1}{r_n}$ and $\omega_0^1|_\x=-r_n$ and the latter iff $\omega_0^0|_\x=-r_n$ and $\omega_0^1|_\x=\frac{1}{r_n}$.
The evaluation of $S_1$ for odd $k$ proceeds in the same way as \eqref{eq:S1_opt} to imply
\begin{equation}\label{eq:S1_opt2}
	-R_nr_n\sin(k\theta_n)f'(x)+\frac{1}{2}\le S_1\le R_nr_n\sin(k\theta_n)f'(x)+\frac{1}{2},
\end{equation}
where 
\begin{equation}\label{eq:def_f'}
	\begin{aligned}
		f'(x)=f(-x)=\frac{1}{\tan((2M+1)\theta_n)}x&+\frac{r_n\cos\theta_n}{\sin((2M+1)\theta_n)}
		\\
		&\quad\quad(x\in[-r_n\cos(2M\theta_n), -r_n\cos((2M+2)\theta_n)])
	\end{aligned}
\end{equation}
with $M=0,1,\ldots, \frac{n-3}{2}$ is used instead of \eqref{eq:def_f} (see Fig.~\ref{fig:graph}).
The relations \eqref{eq:S0_upper}, \eqref{eq:S0_lower}, and \eqref{eq:S1_opt2} imply
\begin{equation}\label{S_bound22}
	\begin{aligned}
		|C[\{(p(a|s); \omega^a_s)_a\}_{s};k]|
		&\le 4(\beta_n^o(k;M)x+\hat{\beta}_n^o(k;M))-2\\
		&(=4((-\beta_n^o(k;M))(-x)+\hat{\beta}_n^o(k;M))-2)
	\end{aligned}
\end{equation}
for odd $k$ and assemblages $\{(p(a|s); \omega^a_s)_a\}_{s}$ such that $\sum_{a}p(a|0)\omega_0^a|_\x=\sum_{a}p(a|1)\omega_1^a|_\x=x$ and $x\in[-r_n\cos(2M\theta_n), -r_n\cos((2M+2)\theta_n)]$ ($M=0,\ldots, \frac{n-3}{2}$)
with
\begin{equation}\label{eq:beta_o}
	\begin{aligned}
		&\beta_n^o(k;M)=
	R_nr_n\!\left[R_n(r_n^2-1)(-\cos(k\theta_n)-1)+\frac{\sin(k\theta_n)}{\tan((2M+1)\theta_n)}\right],\\
		&\hat{\beta}_n^o(k;M)=
			R_n\!\left[2R_nr_n^2\cos(k\theta_n)+\frac{\sin(k\theta_n)}{\sin((2M+1)\theta_n)}\right]\!+2R_n^2r_n^2.
	\end{aligned}
\end{equation}
We note that thanks to the similarity of $\beta_n^o(k;M)$ to $(-\beta_n^e(k;M))$, we can evaluate the term $\beta_n^o(k;M)x+\hat{\beta}_n^o(k;M)$ in a similar way to \eqref{eq:I_e}.
It results in
\begin{equation}\label{key}
	\beta_n^o(k;M)x+\hat{\beta}_n^o(k;M)\le I_n^{o}(k)
\end{equation}
for odd $k\in\inter{0,\frac{n-1}{2}}$ with
\begin{align}\label{key}
	I_n^{o}(k)=
	&\left\{
	\begin{aligned}
		&\beta_n^o(k;M_0)\left(-r_n\sin\left(\frac{\pi}{2n}\right)\right)+\hat{\beta}_n^o(k;M_0) \quad&&(n\equiv1,5)\\
		&\beta_n^o(k;M_0)\left(r_n\sin\left(\frac{\pi}{2n}\right)\right)+\hat{\beta}_n^o(k;M_0) \quad&&(n\equiv3,7)
	\end{aligned}
	\right.\\
	=&\left\{
	\begin{aligned}
		&\frac{R_n}{\cos^2\frac{\theta_n}{2}}\!\left(r_n^2\sin^3\frac{\theta_n}{2}+1\right)\cos(k\theta_n)+\frac{R_n}{\cos\frac{\theta_n}{2}}\!\left(r_n^2\sin^2\frac{\theta_n}{2}+1\right)\sin(k\theta_n)\\
		&\qquad\qquad\qquad\qquad\qquad\qquad\qquad
		+\frac{R_nr_n^2}{\cos^2\frac{\theta_n}{2}}\sin^3\frac{\theta_n}{2}+2R_n^2r_n^2\qquad(n\equiv1,5)\\
		&\frac{R_n}{\cos^2\frac{\theta_n}{2}}\!\left(-r_n^2\sin^3\frac{\theta_n}{2}+1\right)\cos(k\theta_n)+\frac{R_n}{\cos\frac{\theta_n}{2}}\!\left(r_n^2\sin^2\frac{\theta_n}{2}+1\right)\sin(k\theta_n)\\
		&\qquad\qquad\qquad\qquad\qquad\qquad\qquad
		-\frac{R_nr_n^2}{\cos^2\frac{\theta_n}{2}}\sin^3\frac{\theta_n}{2}+2R_n^2r_n^2\qquad(n\equiv3,7),
	\end{aligned}\label{eq:I_o}
	\right.
\end{align}
where $M_0$ is the same as \eqref{eq:M_0}.
Now we have obtained the optimal CHSH value 
\begin{equation}
	G_n(k)=\underset{\{(p(a|s); \omega^a_s)_a\}_{s}\in\mathsf{Ens}^{\mathrm{All}}}{\max}~|C[\{(p(a|s); \omega^a_s)_a\}_{s};\E_n(k),\E_n(0)]|
\end{equation}
among all possible assemblages $\{(p(a|s); \omega^a_s)_a\}_{s}$ summarized as Table~\ref{table:G(k)}.
\begin{table}[h]
	\begin{minipage}[t]{0.48\textwidth}
		\begin{center}
			\begin{tabular}{|c||c|c|}
				\hline
				$G_n(k)$ & \mbox{$k$: even}  & \mbox{$k$: odd}    \rule[-3.5mm]{0mm}{10.5mm}                       
				\\ \hhline{|=#=|=|}
				$k\in\inter{0, \frac{n-1}{2}}$ & $4I_n^{e}(k)-2$  & $4I_n^{o}(k)-2$    \rule[-3.5mm]{0mm}{10.5mm}          
				\\ \hline
			\end{tabular}
		\end{center}
		\caption{The value $G_n(k)$.}
		\label{table:G(k)}
	\end{minipage}
	\hfill
	\begin{minipage}[t]{0.48\textwidth}
		\begin{center}
			\begin{tabular}{|c||c|c|}
				\hline
				$H_n(k)$ & \mbox{$k$: even}  & \mbox{$k$: odd}    \rule[-3.5mm]{0mm}{10.5mm}                       
				\\ \hhline{|=#=|=|}
				$k\in\inter{0, \frac{n-1}{2}}$ & $4\hat{I}_n^{e}(k)-2$  & $4\hat{I}_n^{o}(k)-2$    \rule[-3.5mm]{0mm}{10.5mm}          
				\\ \hline
			\end{tabular}
		\end{center}
		\caption{The value $H_n(k)$.}
		\label{table:H(k)}
	\end{minipage}
\end{table}

We calculate the optimal CHSH value 
\begin{equation}
	H_n(k)=\underset{\{(p(a|s); \omega^a_s)_a\}_{s}\in\mathsf{Ens}^{\mathrm{ME}}}{\max}~|C[\{(p(a|s); \omega^a_s)_a\}_{s};\E_n(k),\E_n(0)]|
\end{equation}
among assemblages realized by maximally entangled states.
For a maximally entangled state $\hat{\eta}\in GL(\Omega_n)$, its inducing assemblages $\{(p(a|s); \omega^a_s)_a\}_{s}$ satisfy
\[
\sum_{a=0,1}p(a|s)\omega_s^a=\hat{\eta}(u)=
\begin{pmatrix}
	0\\
	0\\
	1
\end{pmatrix},
\]
Thus, letting $x=0$ (i.e., $M=M_0$) in \eqref{eq:S_bound} (\eqref{eq:S_bound2}) and \eqref{S_bound22}, we have
\begin{equation}\label{eq:S_bound_ME}
	-\hat{\beta}_n^e(k;M_0)+\frac{1}{2}\leq \frac{C[\{(p(a|s); \omega^a_s)_a\}_{s};k]}{4}\le \hat{\beta}_n^e(k;M_0)-\frac{1}{2}
\end{equation}
for even $k$ and 
\begin{equation}\label{eq:S_bound_ME2}
	-\hat{\beta}_n^o(k;M_0)+\frac{1}{2}\leq \frac{C[\{(p(a|s); \omega^a_s)_a\}_{s};k]}{4}\le \hat{\beta}_n^o(k;M_0)-\frac{1}{2}
\end{equation}
for odd $k$.
We can prove that the upper equality in \eqref{eq:S_bound_ME} and the lower equality in \eqref{eq:S_bound_ME2} are saturated by certain maximally entangled states.
To see this, we choose $\hat{\eta}\in GL(\Omega_n)$ and Alice's observables $(\E_n(i), \E_n(j))$ by
\begin{equation}\label{eq:ens opt1}
	\begin{aligned}
		&p(0|0)\omega_0^0=\hat{\eta}(e_n(i))=R_n
		\begin{pmatrix}
			r_n\\
			0\\
			1
		\end{pmatrix},\ \ 
		p(1|0)\omega_0^1=\hat{\eta}(\overline{e_n(i)})=(1-R_n)
		\begin{pmatrix}
			-\frac{1}{r_n}\\
			0\\
			1
		\end{pmatrix},\\
		&p(0|1)\omega_1^0=\hat{\eta}(e_n(j))=R_n
		\begin{pmatrix}
			r_n\cos(2M'_0\theta_n)\\
			r_n\sin(2M'_0\theta_n)\\
			1
		\end{pmatrix},\ \ 
		p(1|1)\omega_1^1=\hat{\eta}(\overline{e_n(j)})=(1-R_n)
		\begin{pmatrix}
			-\frac{1}{r_n}\cos(2M'_0\theta_n)\\
			-\frac{1}{r_n}\sin(2M'_0\theta_n)\\
			1
		\end{pmatrix}
	\end{aligned}
\end{equation}
when $k$ is even and
\begin{equation}\label{eq:ens opt2}
	\begin{aligned}
		&p(0|0)\omega_0^0=\hat{\eta}(e_n(i))=R_n
		\begin{pmatrix}
			-r_n\\
			0\\
			1
		\end{pmatrix},\ \ 
		p(1|0)\omega_0^1=\hat{\eta}(\overline{e_n(i)})=(1-R_n)
		\begin{pmatrix}
			\frac{1}{r_n}\\
			0\\
			1
		\end{pmatrix},\\
		&p(0|1)\omega_1^0=\hat{\eta}(e_n(j))=R_n
		\begin{pmatrix}
			-r_n\cos(2M'_0\theta_n)\\
			-r_n\sin(2M'_0\theta_n)\\
			1
		\end{pmatrix},\ \ 
		p(1|1)\omega_1^1=\hat{\eta}(\overline{e_n(j)})=(1-R_n)
		\begin{pmatrix}
			\frac{1}{r_n}\cos(2M'_0\theta_n)\\
			\frac{1}{r_n}\sin(2M'_0\theta_n)\\
			1
		\end{pmatrix}
	\end{aligned}
\end{equation}
when $k$ is odd, where
\begin{equation}\label{key}
	M'_0=\left\{
	\begin{aligned}
		&\frac{n-1}{4}\quad(n\equiv1,5)\\
		&\frac{n+1}{4}\quad(n\equiv3,7).
	\end{aligned}
	\right.
\end{equation}
To see this, we choose $\hat{\eta}\in GL(\Omega_n)$ and Alice's observables $(\E_n(i), \E_n(j))$ satisfying 
It is not difficult to confirm that these assemblages respectively realize the upper equality in \eqref{eq:S_bound_ME} and the lower equality in \eqref{eq:S_bound_ME2}.
In fact, they satisfy the conditions such that the equalities in \eqref{eq:S_0_upper(cos>0)} and \eqref{eq:S1_opt}, and \eqref{eq:S0_lower} and \eqref{eq:S1_opt2} hold to derive
\begin{equation}\label{eq:opt even}
	\frac{C[\{(p(a|s); \omega^a_s)_a\}_{s};k]}{4}= \hat{\beta}_n^e(k;M_0)-\frac{1}{2}
\end{equation}
and 
\begin{equation}\label{eq:opt odd}
	\frac{C[\{(p(a|s); \omega^a_s)_a\}_{s};k]}{4}= -\hat{\beta}_n^o(k;M_0)+\frac{1}{2}
\end{equation}
respectively.
There are indeed triples $(\hat{\eta}; \E_n(i), \E_n(j))$  that induce the assemblages \eqref{eq:ens opt1} or \eqref{eq:ens opt2} (and realize \eqref{eq:opt even} or \eqref{eq:opt odd}): for example,
\[
(\hat{\eta}; \E_n(i), \E_n(j))=\left\{
\begin{aligned}
	&(\id; \E_n(k), \E_n(k+M'_0))&&(\mbox{$k$: even})\\
	&(\id; \E_n(n-k), \E_n(n-(k+M'_0))&&(\mbox{$k$: odd})
\end{aligned}
\right.
\]
with the identity map $\id\in GL(\Omega_n)$ on $\R^3$.
We have now obtained the optimum $H_n(k)$ summarized as Table \ref{table:H(k)}, where we introduced
\begin{align}
	&\hat{I}_n^{e}(k)=	\hat{\beta}_n^e(k;M_0)=	\frac{R_n}{\cos^2\frac{\theta_n}{2}}\cos(k\theta_n)+\frac{R_n}{\cos\frac{\theta_n}{2}}\sin(k\theta_n)+R_n^2(1+r_n^4),\label{eq:I_hat_e+}\\
	&\hat{I}_n^{o}(k)=	\hat{\beta}_n^o(k;M_0)=	\frac{R_n}{\cos^2\frac{\theta_n}{2}}\cos(k\theta_n)+\frac{R_n}{\cos\frac{\theta_n}{2}}\sin(k\theta_n)+2R_n^2r_n^2\label{eq:J_hat_o+}
\end{align}
in terms of \eqref{eq:beta_e} and \eqref{eq:beta_o}.
\begin{rmk}
	\label{rmk:equality}
While the upper equality in \eqref{eq:S_bound} is realized by a certain maximally entangled state, it can be verified that no $\hat{\eta}\in GL(\Omega_n)$ can satisfy the other equality in \eqref{eq:S_bound}.
In fact, the condition $\frac{\hat{\eta}(e_n(i))}{\ang{u,\hat{\eta}(e_n(i))}}|_\mathrm{x}=\omega_0^0|_\mathrm{x}=-\frac{1}{r_n}$ for $\hat{\eta}\in GL(\Omega_n)$ contradicts the fact that it maps extremal rays of the dual cone $V_+^*$ to extremal rays of the positive cone $V_+$ (similarly, there is no maximally entangled state saturating the upper equality in \eqref{eq:S_bound2}).
\end{rmk}

We consider maximizing $H_n(k)$ over $k\in\inter{0,\ldots, \frac{n-1}{2}}$, where $H_n(k)$ is either $4\hat{I}_n^{e}(k)-2$ or $4\hat{I}_n^{o}(k)-2$.
We focus on maximizing the term 
\[
\frac{R_n}{\cos^2\frac{\theta_n}{2}}\cos(k\theta_n)+\frac{R_n}{\cos\frac{\theta_n}{2}}\sin(k\theta_n)=\frac{R_n}{\cos^2\frac{\theta_n}{2}}\left[\cos(k\theta_n)+\cos\frac{\theta_n}{2}\sin(k\theta_n)\right]
\]
of $\hat{I}_n^{e}$ and $\hat{I}_n^{o}$ in \eqref{eq:I_hat_e+} and \eqref{eq:J_hat_o+} respectively.
It holds for any $k\in\inter{0,\frac{n-1}{2}}$ that
\begin{equation}\label{eq:ineq_H}
\sqrt{2}\cos\frac{\theta_n}{2}\sin\left(k\theta_n+\frac{\pi}{4}\right)
<\cos(k\theta_n)+\cos\frac{\theta_n}{2}\sin(k\theta_n)
<\sqrt{2}\sin\left(k\theta_n+\frac{\pi}{4}\right).
\end{equation}
Let $n_\star, {\tilde{n}}^+_\star\in\inter{0,\frac{n-1}{2}}$ be the (unique) integers such that $n_\star\theta_n+\frac{\pi}{4}$ and ${\tilde{n}}^+_\star\theta_n+\frac{\pi}{4}$ is respectively the closest and the second closest to $\frac{\pi}{2}$.
That is, 
\begin{equation}\label{eq:n_star_app}
	n_\star=\left\{
	\begin{aligned}
		&\frac{n-1}{4}\quad(n\equiv 1,5)\\
		&\frac{n+1}{4}\quad(n\equiv 3,7),
	\end{aligned}
	\right.
	\qquad
	{\tilde{n}}_\star=\left\{
	\begin{aligned}
		&\frac{n+3}{4}\quad(n\equiv 1,5)\\
		&\frac{n-3}{4}\quad(n\equiv 3,7).
	\end{aligned}
	\right.
\end{equation}
Because 
\begin{equation}
\sin\left({\tilde{n}}^+_\star\theta_n+\frac{\pi}{4}\right)=\cos\frac{3\theta_n}{4}
\end{equation}
and
\begin{equation}
	\cos\frac{\theta_n}{2}\sin\left(n_\star\theta_n+\frac{\pi}{4}\right)=\cos\frac{\theta_n}{2}\cos\frac{\theta_n}{4}=\frac{1}{2}\left(\cos\frac{\theta_n}{4}+\cos\frac{3\theta_n}{4}\right)>\cos\frac{3\theta_n}{4}
\end{equation}
hold, \eqref{eq:ineq_H} proves
\begin{equation}\label{eq:max1}
	\max\left\{H_n(k)\ \middle|\ k\in\inter{0,\frac{n-1}{2}},\ \mbox{$k$: even}\right\}=
	\left\{
		\begin{aligned}
	&4\hat{I}_n^{e}\left(\frac{n-1}{4}\right)-2\ &&(n\equiv1)\\
	&4\hat{I}_n^{e}\left(\frac{n+1}{4}\right)-2\ &&(n\equiv7)
		\end{aligned}
	\right.
\end{equation}
and
\begin{equation}\label{eq:max2}
	\max\left\{H_n(k)\ \middle|\ k\in\inter{0,\frac{n-1}{2}},\ \mbox{$k$: odd}\right\}=
	\left\{
	\begin{aligned}
&4\hat{I}_n^{o}\left(\frac{n+1}{4}\right)-2\ &&(n\equiv3)\\
&4\hat{I}_n^{o}\left(\frac{n-1}{4}\right)-2\ &&(n\equiv5),
	\end{aligned}
	\right.
\end{equation}
where the parity of $n_\star$ is concerned and the maximum is attained iff $k=n_\star$.
It still remains to reveal whether \eqref{eq:max1} is greater than
\begin{equation}\label{eq:max11}
	\max\left\{H_n(k)\ \middle|\ k\in\inter{0,\frac{n-1}{2}},\ \mbox{$k$: odd}\right\}
\end{equation}
for $n\equiv1,7$
and \eqref{eq:max2} is greater than
\begin{equation}\label{eq:max21}
	\max\left\{H_n(k)\ \middle|\ k\in\inter{0,\frac{n-1}{2}},\ \mbox{$k$: even}\right\}
\end{equation}
for $n\equiv3,5$.
Because the other cases can be calculated in similar ways, here we only show that \eqref{eq:max2}$>$\eqref{eq:max21} holds for $n\equiv3$, i.e., 
\begin{equation}\label{eq:max22}
	\hat{I}_n^{o}\left(\frac{n+1}{4}\right)>	\max\left\{\hat{I}_n^{e}(k)\ \middle|\ k\in\inter{0,\frac{n-1}{2}},\ \mbox{$k$: even}\right\}.
\end{equation} 
We rewrite $\hat{I}_n^{e}(k)$ in \eqref{eq:I_hat_e+} as
\begin{equation}\label{eq:I_another}
	\hat{I}_n^{e}(k)=\frac{R_n}{\cos^2\frac{\theta_n}{2}}\left(\cos(k\theta_n)+\cos\frac{\theta_n}{2}\sin(k\theta_n)+2r_n^4\sin^4\frac{\theta_n}{2}\right)+2R_n^2r_n^2,
\end{equation}
where we used
\[
R_n^2(1+r_n^4)-2R_n^2r_n^2=R_n^2(r_n^2-1)^2=\frac{R_n}{\cos^2\frac{\theta_n}{2}}\cdot2r_n^4\sin^4\frac{\theta_n}{2}.
\]
To verify \eqref{eq:max22}, it is enough to show
\begin{equation}\label{eq:max222}
	\sqrt{2}\cos\frac{\theta_n}{2}\cos\frac{\theta_n}{4}>\max\left\{\sqrt{2}\sin\left(k\theta_n+\frac{\pi}{4}\right)+2r_n^4\sin^4\frac{\theta_n}{2}\ \middle|\ k\in\inter{0,\frac{n-1}{2}},\ \mbox{$k$: even}\right\}
\end{equation}
by means of \eqref{eq:ineq_H}.
The maximum in the r.h.s. is realized by $k=\frac{n-3}{4}$ (see \eqref{eq:n_star_app}), and thus it can be rewritten as
\begin{equation}\label{key}
	\sqrt{2}\cos\frac{\theta_n}{2}\cos\frac{\theta_n}{4}>\sqrt{2}\cos\frac{3\theta_n}{4}+2r_n^4\sin^4\frac{\theta_n}{2}.
\end{equation}
Since we have
\begin{align*}
	\sqrt{2}\cos\frac{\theta_n}{2}\cos\frac{\theta_n}{4}-\left(\sqrt{2}\cos\frac{3\theta_n}{4}+2r_n^2\sin^4\frac{\theta_n}{2}\right)
	&=r_n^2\left(\sqrt{2}\cos\theta_n\sin\frac{\theta_n}{2}\sin\frac{\theta_n}{4}-2\sin^4\frac{\theta_n}{2}\right)\\
	&=r_n^2\sin\frac{\theta_n}{2}\sin\frac{\theta_n}{4}\left(\sqrt{2}\cos\theta_n-4\sin^2\frac{\theta_n}{2}\cos\frac{\theta_n}{4}\right)
\end{align*}
and 
\begin{align*}
	\sqrt{2}\cos\theta_n-4\sin^2\frac{\theta_n}{2}\cos\frac{\theta_n}{4}&>	\sqrt{2}\cos\theta_n-4\sin^2\frac{\theta_n}{2}\\
	&>\sqrt{2}\cos\frac{\pi}{5}-4\sin^2\frac{\pi}{10}=0.76...>0
\end{align*}
for $n\ge5$, we can conclude that \eqref{eq:max222}, i.e., \eqref{eq:max22} holds.
In this way, we obtain
\begin{equation}\label{eq:max+1}
	\max\left\{H_n(k)\ \middle|\ k\in\inter{0,\frac{n-1}{2}}\right\}=
	\left\{
	\begin{aligned}
		&4\hat{I}_n^{e}\left(\frac{n-1}{4}\right)-2\ &&(n\equiv1)\\
		&4\hat{I}_n^{o}\left(\frac{n+1}{4}\right)-2\ &&(n\equiv3)\\
		&4\hat{I}_n^{o}\left(\frac{n-1}{4}\right)-2\ &&(n\equiv5)\\
		&4\hat{I}_n^{e}\left(\frac{n+1}{4}\right)-2\ &&(n\equiv7),
	\end{aligned}
	\right.
\end{equation}
where the maximum is attained iff $k=n_\star$ given by
\begin{equation}\label{eq:n_star_App}
	n_\star=\left\{
	\begin{aligned}
		&\frac{n-1}{4}\quad(n\equiv 1,5)\\
		&\frac{n+1}{4}\quad(n\equiv 3,7),
	\end{aligned}
	\right.
\end{equation}
and it proves \eqref{eq:lem0}, \eqref{eq:lem1}, and \eqref{eq:n_star}.

\subsection{Part 2: Proof of (\ref{eq:lem2})}
Here we prove the claim for the case $n\equiv3$ (similar proofs can be given for the other cases).
According to \eqref{eq:max+1}, what we should show is
\begin{equation}
	\hat{I}_n^{o}\left(\frac{n+1}{4}\right)>\max\left\{G_n(k)\ \middle|\ k\in\inter{0,\frac{n-1}{2}},\ k\neq\frac{n+1}{4}\right\},\label{eq:claim+}
\end{equation}
equivalently
\begin{align}
	&\hat{I}_n^{o}\left(\frac{n+1}{4}\right)>\max\left\{I_n^{e}(k)\ \middle|\ k\in\inter{0,\frac{n-1}{2}},\ k\neq\frac{n+1}{4},\ \mbox{$k$: even}\right\},\label{eq:claim1}\\
	&\hat{I}_n^{o}\left(\frac{n+1}{4}\right)>\max\left\{I_n^{o}(k)\ \middle|\ k\in\inter{0,\frac{n-1}{2}},\ k\neq\frac{n+1}{4},\ \mbox{$k$: odd}\right\}\label{eq:claim2}
\end{align}
(see Table~\ref{table:G(k)}).
To show these, we rewrite $I_n^{e}(k)$ and $I_n^{o}(k)$ (see \eqref{eq:I_e} and \eqref{eq:I_o}) respectively as
\begin{equation}\label{key}
		I_n^{e}(k)=\frac{R_n}{\cos^2\frac{\theta_n}{2}}J_{n(3)}^{e}(k)+2R_n^2r_n^2
\end{equation}
and
\begin{equation}\label{key}
	I_n^{o}(k)=\frac{R_n}{\cos^2\frac{\theta_n}{2}}J_{n(3)}^{o}(k)+2R_n^2r_n^2
\end{equation}
with
\begin{equation}
	\begin{aligned}
		J_{n(3)}^{e}(k)	=	&\left(-r_n^2\sin^3\frac{\theta_n}{2}+1\right)\cos(k\theta_n)+\cos\frac{\theta_n}{2}\left(r_n^2\sin^2\frac{\theta_n}{2}+1\right)\sin(k\theta_n)
		\\
		&\qquad\qquad\qquad\qquad\qquad\qquad\qquad\qquad\qquad+r_n^2\sin^3\frac{\theta_n}{2}+2r_n^2\sin^4\frac{\theta_n}{2}
	\end{aligned}
\end{equation}
and
\begin{equation}
	J_{n(3)}^{o}(k)	=	\left(-r_n^2\sin^3\frac{\theta_n}{2}+1\right)\cos(k\theta_n)+\left(r_n^2\sin^2\frac{\theta_n}{2}+1\right)\sin(k\theta_n)
	-r_n^2\sin^3\frac{\theta_n}{2}.
\end{equation}
Similarly, $\hat{I}_n^{o}(\frac{n+1}{4})$ can be rewritten as  (see \eqref{eq:I_another})
\begin{equation}\label{key}
	\hat{I}_n^{o}\left(\frac{n+1}{4}\right)=\frac{R_n}{\cos^2\frac{\theta_n}{2}}\hat{J}_{n(3)}^{o}\left(\frac{n+1}{4}\right)+2R_n^2r_n^2
\end{equation}
with
\begin{equation}\label{key}
\hat{J}_{n(3)}^{o}\left(\frac{n+1}{4}\right)
=\cos\left(\frac{\pi}{4}+\frac{\theta_n}{4}\right)+\cos\frac{\theta_n}{2}\left(\frac{\pi}{4}+\frac{\theta_n}{4}\right).
\end{equation}
The problems \eqref{eq:claim1} and \eqref{eq:claim2} now become
\begin{align}
	&\hat{J}_{n(3)}^{o}\left(\frac{n+1}{4}\right)>\max\left\{J_{n(3)}^{e}(k)\ \middle|\ k\in\inter{0,\frac{n-1}{2}},\ k\neq\frac{n+1}{4},\ \mbox{$k$: even}\right\},\label{eq:claim1'}\\
	&\hat{J}_{n(3)}^{o}\left(\frac{n+1}{4}\right)>\max\left\{J_{n(3)}^{o}(k)\ \middle|\ k\in\inter{0,\frac{n-1}{2}},\ k\neq\frac{n+1}{4},\ \mbox{$k$: odd}\right\}\label{eq:claim2'}
\end{align}
respectively.
The latter problem is easier to prove than the former one, so we here demonstrate \eqref{eq:claim1'} explicitly described as
\begin{equation}\label{eq:lem_main1}
	\begin{aligned}
		&\cos\left(\frac{\pi}{4}+\frac{\theta_n}{4}\right)+\cos\frac{\theta_n}{2}\left(\frac{\pi}{4}+\frac{\theta_n}{4}\right)\\
		&\qquad\qquad
		>\max\left\{\left(-r_n^2\sin^3\frac{\theta_n}{2}+1\right)\cos(k\theta_n)+\cos\frac{\theta_n}{2}\left(r_n^2\sin^2\frac{\theta_n}{2}+1\right)\sin(k\theta_n)\right\}\\
		&\qquad\qquad\qquad\qquad\qquad\qquad\qquad\qquad\qquad\qquad\qquad\qquad
		+r_n^2\sin^3\frac{\theta_n}{2}+2r_n^2\sin^4\frac{\theta_n}{2},
	\end{aligned}
\end{equation}
where the maximum is taken over $k\in\inter{0,\frac{n-1}{2}},\ k\neq\frac{n+1}{4},\ \mbox{$k$: even}$.
We note that the r.h.s. is smaller than 
\begin{equation}
	\begin{aligned}
		&\max\left\{\left(-r_n^2\sin^2\frac{\theta_n}{2}+1\right)\cos(k\theta_n)+\left(r_n^2\sin^2\frac{\theta_n}{2}+\cos\frac{\theta_n}{2}\right)\sin(k\theta_n)\right\}\\
		&\qquad\qquad\qquad\qquad\qquad\qquad\qquad\qquad\qquad\qquad\qquad\qquad
		+r_n^2\sin^2\frac{\theta_n}{2}+2r_n^2\sin^4\frac{\theta_n}{2},
	\end{aligned}
\end{equation}
and thus it is enough to show 
\begin{equation}\label{eq:lem_main2}
	\begin{aligned}
		&\cos\left(\frac{\pi}{4}+\frac{\theta_n}{4}\right)+\cos\frac{\theta_n}{2}\sin\left(\frac{\pi}{4}+\frac{\theta_n}{4}\right)\\
		&\qquad\qquad
		>\max\left\{\left(-r_n^2\sin^2\frac{\theta_n}{2}+1\right)\cos(k\theta_n)+\left(r_n^2\sin^2\frac{\theta_n}{2}+\cos\frac{\theta_n}{2}\right)\sin(k\theta_n)\right\}\\
		&\qquad\qquad\qquad\qquad\qquad\qquad\qquad\qquad\qquad\qquad\qquad\qquad
		+r_n^2\sin^2\frac{\theta_n}{2}+2r_n^2\sin^4\frac{\theta_n}{2}
	\end{aligned}
\end{equation}
instead of \eqref{eq:lem_main1}.
We consider maximizing the term 
\begin{equation}\label{eq:lem_main''}
	\left(-r_n^2\sin^2\frac{\theta_n}{2}+1\right)\cos(k\theta_n)+\left(r_n^2\sin^2\frac{\theta_n}{2}+\cos\frac{\theta_n}{2}\right)\sin(k\theta_n)
\end{equation}
over $k\in\inter{0,\frac{n-1}{2}},\ k\neq\frac{n+1}{4},\ \mbox{$k$: even}$.
We can reveal that the maximum is attained iff $k=\frac{n-3}{4}$ (remember that $\frac{n-3}{4}$ is an even integer for $n\equiv3$) in terms of the relation 
\begin{equation}\label{eq:lem_main'}
	\tan\left(\frac{\pi}{4}-\frac{\theta_n}{4}\right)<\frac{-r_n^2\sin^2\frac{\theta_n}{2}+1}{r_n^2\sin^2\frac{\theta_n}{2}+\cos\frac{\theta_n}{2}}\left(<\tan\frac{\pi}{4}\right)
\end{equation}
confirmed by
\begin{equation}
	\frac{1-r_n^2\sin^2\frac{\theta_n}{2}}{\cos\frac{\theta_n}{2}+r_n^2\sin^2\frac{\theta_n}{2}}-\tan\left(\frac{\pi}{4}-\frac{\theta_n}{4}\right)
	=\frac{\cos\theta_n-\sin^2\frac{\theta_n}{2}}{\cos\theta_n\cos\frac{\theta_n}{2}+\sin^2\frac{\theta_n}{2}}-\frac{\cos\frac{\theta_n}{4}-\sin\frac{\theta_n}{4}}{\cos\frac{\theta_n}{4}+\sin\frac{\theta_n}{4}}
\end{equation}
and
\begin{align*}
	&\left(\cos\theta_n-\sin^2\frac{\theta_n}{2}\right)\left(\cos\frac{\theta_n}{4}+\sin\frac{\theta_n}{4}\right)-\left(\cos\theta_n\cos\frac{\theta_n}{2}+\sin^2\frac{\theta_n}{2}\right)\left(\cos\frac{\theta_n}{4}-\sin\frac{\theta_n}{4}\right)\\
	&\qquad\qquad\qquad\qquad\qquad\qquad\qquad
	=\sin\frac{\theta_n}{2}\cos\theta_n\sin\frac{\theta_n}{4}+\sin\frac{\theta_n}{2}\cos\frac{\theta_n}{4}\left(\cos\theta_n-2\sin\frac{\theta_n}{2}\right)\\
	&\qquad\qquad\qquad\qquad\qquad\qquad\qquad
	>\sin\frac{\theta_n}{2}\cos\theta_n\sin\frac{\theta_n}{4}+\sin\frac{\theta_n}{2}\cos\frac{\theta_n}{4}\left(\cos\frac
	{\pi}{5}-2\sin\frac{\pi}{10}\right)>0
\end{align*}
for $n\ge5$.
In fact, by virtue of \eqref{eq:lem_main'}, the term \eqref{eq:lem_main''} is found to be proportional to $\sin(k\theta_n+\alpha_n)$ with $\frac{\pi}{4}-\frac{\theta_n}{4}<\alpha_n<\frac{\pi}{4}$, which is maximized by the even integer $k=\frac{n-3}{4}$.
The claim \eqref{eq:lem_main2} becomes
\begin{equation}\label{eq:lem_main3}
	\begin{aligned}
		&\cos\left(\frac{\pi}{4}+\frac{\theta_n}{4}\right)+\cos\frac{\theta_n}{2}\sin\left(\frac{\pi}{4}+\frac{\theta_n}{4}\right)\\
		&\qquad\qquad
		>\left(-r_n^2\sin^2\frac{\theta_n}{2}+1\right)\cos\left(\frac{\pi}{4}-\frac{3\theta_n}{4}\right)+\left(r_n^2\sin^2\frac{\theta_n}{2}+\cos\frac{\theta_n}{2}\right)\sin\left(\frac{\pi}{4}-\frac{3\theta_n}{4}\right)\\
		&\qquad\qquad\qquad\qquad\qquad\qquad\qquad\qquad\qquad\qquad\qquad\qquad
		+r_n^2\sin^2\frac{\theta_n}{2}+2r_n^2\sin^4\frac{\theta_n}{2}
	\end{aligned}
\end{equation}
or
\begin{equation}
	\begin{aligned}
		r_n^2\sin^2\frac{\theta_n}{2}\left[
		2\cos\theta_n\cos\left(\frac{\theta_n}{4}+\frac{\pi}{4}\right)+\sqrt{2}\sin\frac{3\theta_n}{4}+\cos\theta_n-2
		\right]>0.
	\end{aligned}
\end{equation}
Taking derivatives with respect to $\theta_n$, we can show 
\[
	2\cos\theta_n\cos\left(\frac{\theta_n}{4}+\frac{\pi}{4}\right)+\sqrt{2}\sin\frac{3\theta_n}{4}+\cos\theta_n-2>0
\]
for $0\le\theta_n\le\frac{\pi}{5}$ and thus \eqref{eq:lem_main2} and \eqref{eq:claim1'} hold.
As mentioned above, we can develop similar arguments for the other cases and prove 
\begin{equation}\label{key}
	\underset{k\neq n_\star}{\max}~G_n(k)< H_n(n_\star),
\end{equation}
which completes the proof of Lemma~\ref{lem:max_k0}.
\qed

\def\thesection{Appendix\ \Alph{section}}
\section{Solving the linear programming (\ref{primal_main})}
\label{app_LP}
\renewcommand{\theequation}{C.\arabic{equation}}
\renewcommand{\thesection}{\Alph{section}}
\setcounter{equation}{0}
\setcounter{subsection}{0}
\renewtheorem{lem}[subsection]{Lemma}
In this appendix, we show that \eqref{ME_elements2_main} is an optimal solution for the linear programming problem \eqref{primal_main}.
Here we use $(x_1,x_2,x_3,x_4,x_5,x_6)$ instead of $(a',b',c',d'_1,e'_1, e'_2)$ to simplify the notation.
The problem is
\begin{align}
	\label{primal}
	&\mbox{maximize}&&\vec{C}'^{T}\cdot(x_1,x_2,x_3,x_4,x_5,x_6)^{T}\\
	\label{cond_positive}
	&\mbox{subject to}
	&&x_1\ge0,\ \ x_2\ge0,\ \ x_3\ge0,\ \ x_4\ge0, \ x_5\ge0,\ \ x_6\ge0,\\
	&
	&&\Gamma\cdot(x_1,x_2,x_3,x_4,x_5,x_6)^{T}\le\vec{r}.\label{cond_tot}
\end{align}
We note that the equality in \eqref{cond_tot} holds if
\begin{equation}\label{ME_elements2}
	\begin{aligned}
		(x_1,x_2,x_3,x_4,x_5,x_6)=(a_\star,b_\star,c_\star,-d_\star,0,0)=(\cos(2m\theta_n),\sin(2m\theta_n),0,\cos(2m\theta_n),0,0).
	\end{aligned}
\end{equation}
In fact, \eqref{ME_elements2} corresponds to the case $\hat{\eta}_\star^\epsilon=\hat{\eta}_\star^{\mathrm{ME}}$, where all equalities in \eqref{cond1} and \eqref{cond2,3} hold because $\hat{\eta}_\star^\epsilon\circ\hat{\eta}_\star^{\mathrm{ME}}$ is the identity operator on $\R^3$.
The dual problem \cite{LP_Vanderbei} is important to verify our claim:
\begin{align}
	\label{dual}
	&\mbox{minimize}&&\vec{r}\!~^{T}\cdot(y_1,y_2,y_3,y_4)^{T}\\
	\label{dual_positive}
	&\mbox{subject to}
	&&y_1\ge0,\ \ y_2\ge0,\ \ y_3\ge0,\ \ y_4\ge0,\\
	&
	&&\Gamma^T\cdot
	(y_1,y_2,y_3,y_4)^{T}
	\ge \vec{C}'.
	\label{dual_tot}
\end{align}
It is known that solutions $(x_1^\circ,x_2^\circ,x_3^\circ,x_4^\circ,x_5^\circ,x_6^\circ)$ and $(y_1^\circ,y_2^\circ,y_3^\circ,y_4^\circ)$ are optimal respectively for the primal and dual problem if and only if they satisfy the complementary slackness condition (Theorem~5.3 in \cite{LP_Vanderbei})
\begin{align}
	&x_j^\circ=0\quad\mbox{or}\quad \sum_i\Gamma_{ij}y_i^\circ=(\vec{C}')_j\quad( j=1,\ldots,6),\quad\mbox{and}\label{slaclnes_primal}\\
	&y_i^\circ=0\quad\mbox{or}\quad \sum_j\Gamma_{ij}x_j^\circ=(\vec{r})_i\quad( i=1,\ldots,4)\label{slaclnes_dual},
\end{align}
where the indices represent the corresponding elements of the vectors and matrix.
As we have seen, $(x_1,x_2,x_3,x_4,x_5,x_6)=(a_\star,b_\star,0,-d_\star,0,0)$ satisfies the second condition of \eqref{slaclnes_dual}.
Thus if we can find $(y_1, y_2, y_3,y_4)$ such that
	\begin{align}
		&y_1\ge0,\ \ y_2\ge0,\ \ y_3\ge0,\ \ y_4\ge0,\label{problem_app}\\
		&\sum_i\Gamma_{ij}y_i=(\vec{C}')_j \quad(j=1,2,4),\label{problem_app2}
	\end{align}
then \eqref{slaclnes_primal} is satisfied and $(x_1,x_2,x_3,x_4,x_5,x_6)=(a_\star,b_\star,0,-d_\star,0,0)$ are verified to be optimal for the primal problem.
Requiring an additional relation $y_1=y_2$, we can explicitly solve the simultaneous equations \eqref{problem_app2} as
\begin{align}
	&y_3=\frac{2\sin(2m\theta_n)}{r_n^2[\sin(2m\theta_n)+\sin(6m\theta_n)]}\left[\frac{\sin(2m\theta_n)}{\sin((4m-1)\theta_n)}-\frac{\sin(6m\theta_n)\cos(4m\theta_n)}{\sin\theta_n}\right],\\
	&y_4=\frac{2\sin(2m\theta_n)}{r_n^2[\sin(2m\theta_n)+\sin(6m\theta_n)]}\left[\frac{\sin(2m\theta_n)}{\sin((4m-1)\theta_n)}+\frac{\sin(2m\theta_n)\cos(4m\theta_n)}{\sin\theta_n}\right],\\
	&y_1=y_2=2\cos(2m\theta_n)-\frac{r_n^2\sin((2m-1)\theta_n)y_3+r_n^2\sin((2m+1)\theta_n)y_4}{2\sin(2m\theta_n)}.
\end{align}
It is not difficult to see that they are all positive, and thus $(x_1,x_2,x_3,x_4,x_5,x_6)=(a_\star,b_\star,0,-d_\star,0,0)$ is an optimal solution for \eqref{primal}, \eqref{cond_positive}, \eqref{cond_tot}.

\def\thesection{Appendix\ \Alph{section}}
\section{Optimization for the cases $n\equiv3,5,7$}
\label{app_LP_others}
\renewcommand{\theequation}{D.\arabic{equation}}
\renewcommand{\thesection}{\Alph{section}}
\setcounter{equation}{0}
\setcounter{subsection}{0}
\renewtheorem{lem}[subsection]{Lemma}
Here we show that the maximally entangled state
\begin{align}
	\hat{\eta}^{\mathrm{ME}}_\star
	&=\begin{pmatrix}
		\cos(2(K_n-n_\star)\theta_n) & \sin(2(K_n-n_\star)\theta_n) & 0\\
		\sin(2(K_n-n_\star)\theta_n)& -\cos(2(K_n-n_\star)\theta_n) & 0\\
		0 & 0 & 1
	\end{pmatrix}\\
&=\left\{
\begin{aligned}
	&\begin{pmatrix}
		\cos\left(\frac{-3n+1}{4}\theta_n\right) & \sin\left(\frac{-3n+1}{4}\theta_n\right) & 0\\
		\sin\left(\frac{-3n+1}{4}\theta_n\right)& -\cos\left(\frac{-3n+1}{4}\theta_n\right) & 0\\
		0 & 0 & 1
		\end{pmatrix}&&\quad(n\equiv3)\\
	&\begin{pmatrix}
		\cos\left(\frac{-3n-1}{4}\theta_n\right) & \sin\left(\frac{-3n-1}{4}\theta_n\right) & 0\\
		\sin\left(\frac{-3n-1}{4}\theta_n\right)& -\cos\left(\frac{-3n-1}{4}\theta_n\right) & 0\\
		0 & 0 & 1
		\end{pmatrix}&&\quad(n\equiv5)\\
	&\begin{pmatrix}
		\cos\left(\frac{n+1}{4}\theta_n\right) & \sin\left(\frac{n+1}{4}\theta_n\right) & 0\\
		\sin\left(\frac{n+1}{4}\theta_n\right)& -\cos\left(\frac{n+1}{4}\theta_n\right) & 0\\
		0 & 0 & 1
		\end{pmatrix}&&\quad(n\equiv7)
\end{aligned}\label{ME_explicit}
\right.
\end{align}
gives the optimal CHSH value for each case $n\equiv3,5,7$.
Note that we follow the coordinates $\{\mathbf{f}_\mathrm{x}, \mathbf{f}_\mathrm{y},\mathbf{f}_\mathrm{z}\}$ throughout this part.
\subsection{The case $n\equiv7$}
For $n\equiv7$, we can apply the same method as the case $n\equiv1$ in Subsec.~\ref{subsec:proof for odd}.
We introduce
\begin{equation}\label{eq:eta_epsilon_app}
	\hat{\eta}^\epsilon_\star=\begin{pmatrix}
		a' & b' & c'\\
		b' & d' & e'\\
		c' & e' & 1
	\end{pmatrix}
\end{equation}
with
\begin{equation}\label{cond0_app}
	a'\ge0,\quad  b'\ge0,\quad  d'\le0
\end{equation}
and set $m=\frac{n-7}{8}$ $(m=0,1,2,\ldots)$.
For this state $\hat{\eta}^\epsilon_\star$ to give a greater CHSH value than $\hat{\eta}^{\mathrm{ME}}_\star$, in this case it should hold that 
\begin{equation}\label{key}
	c'\le 0
\end{equation}
(see Fig.~\ref{fig:graph} and remember that $n_\star=\frac{n+1}{4}$ is even).
By means of normal vectors
\begin{equation}\label{normal 0_app}
	l_1\!\left(\frac{n_\star}{2}\right)\!=\begin{pmatrix}
		1\\
		\frac{1}{\sin2\theta_n}-\frac{1}{\tan2\theta_n}\\
		-r_n
	\end{pmatrix},\quad
	l_2\!\left(\frac{n_\star}{2}\right)\!=\begin{pmatrix}
		1\\
		-\frac{1}{\sin2\theta_n}+\frac{1}{\tan2\theta_n}\\
		-r_n
	\end{pmatrix}
\end{equation}
similar to \eqref{normal 0}, we require
\begin{equation}\label{key}
	\begin{aligned}
		&\ang{l_1\!\left(\frac{n_\star}{2}\right)\!,~ \hat{\eta}_\star^\epsilon\circ\hat{\eta}_\star^{\mathrm{ME}}\left(e_n\!\left(\frac{n_\star}{2}\right)\!\right)}\le0,\quad
		\ang{l_2\!\left(\frac{n_\star}{2}\right)\!,~ \hat{\eta}_\star^\epsilon\circ\hat{\eta}_\star^{\mathrm{ME}}\left(e_n\!\left(\frac{n_\star}{2}\right)\!\right)}\le0,\\
		&\ang{l_1\!\left(\frac{n_\star}{2}+\alpha_1\right)\!,~ \hat{\eta}_\star^\epsilon\circ\hat{\eta}_\star^{\mathrm{ME}}\left(e_n\!\left(\frac{n_\star}{2}+\alpha_1\right)\!\right)}\le0,\quad
		\ang{l_2\!\left(\frac{n_\star}{2}+\alpha_2\right)\!,~ \hat{\eta}_\star^\epsilon\circ\hat{\eta}_\star^{\mathrm{ME}}\left(e_n\!\left(\frac{n_\star}{2}+\alpha_2\right)\!\right)}\le0
	\end{aligned}
\end{equation}
with 
\begin{equation}\label{alpha_i_7}
	\alpha_1=3m+2,\quad \alpha_2=-m-3
\end{equation}
instead of \eqref{cond1} and \eqref{cond2,3}.
They are explicitly expressed as
\begin{align}\label{d3,4_0_app}
	&\vec{\gamma}^{(7)T}_{1}\cdot(a',b',c',d',e')^T\le r_n,\quad \vec{\gamma}_{2}^{(7)T}\cdot(a',b',c',d',e')^T\le r_n,\\
	&\vec{\gamma}^{(7)T}_{3}\cdot(a',b',c',d',e')^T\le r_n,\quad \vec{\gamma}_{4}^{(7)T}\cdot(a',b',c',d',e')^T\le r_n
\end{align}
with
\begin{equation}
	\begin{aligned}
		\vec{\gamma}_{1}^{(7)}
		=
		\begin{pmatrix}
			r_n\cos((2m+2)\theta_n)\\
			r_n^3\sin((2m+3)\theta_n)\\
			1-r_n^2\cos((2m+2)\theta_n)\\
			r_n^3\sin((2m+2)\theta_n)\sin\theta_n\\
			-r_n^2[\sin((2m+2)\theta_n)-\sin\theta_n]
		\end{pmatrix}, \quad 
		\vec{\gamma}^{(7)}_{2}
		=\begin{pmatrix}
			r_n\cos((2m+2)\theta_n)\\
			r_n^3\sin((2m+1)\theta_n)\\
			1-r_n^2\cos((2m+2)\theta_n)\\
			-r_n^3\sin((2m+2)\theta_n)\sin\theta_n\\
			-r_n^2[\sin((2m+2)\theta_n)+\sin\theta_n]
		\end{pmatrix},
	\end{aligned}
\end{equation}
\begin{align}\label{}
	&\vec{\gamma}_{3}^{(7)}
	=\begin{pmatrix}
		-r_n^3\cos((4m+2)\theta_n)\cos((2m+2)\theta_n)\\
		r_n^3\sin((2m+3)\theta_n)\\
		-r_n^2[\cos((4m+2)\theta_n)+\cos((2m+2)\theta_n)]\\
		-r_n^3\sin((4m+2)\theta_n)\sin((2m+2)\theta_n)\\
		r_n^2[\sin((4m+2)\theta_n)+\sin((2m+2)\theta_n)]
	\end{pmatrix},\\
&	\vec{\gamma}_{4}^{(7)}
=\begin{pmatrix}
	-r_n^3\cos((4m-1)\theta_n)\cos((2m+7)\theta_n)\\
	r_n^3\sin((2m+1)\theta_n)\\
	r_n^2[\cos((4m-1)\theta_n)+\cos((2m+7)\theta_n)]\\
	-r_n^3\sin((4m-1)\theta_n)\sin((2m+7)\theta_n)\\
	-r_n^2[\sin((4m-1)\theta_n)+\sin((2m+7)\theta_n)]
\end{pmatrix}.
\end{align}
Now a linear programming problem 
\begin{equation}\label{key}
	\left[\quad
	\begin{aligned}
		&\mbox{maximize}&&\vec{C}^{T}\cdot(a',b',c',d',e')^{T}\\
		&\mbox{subject to}
		&&a'\ge0,\ \ b'\ge0,\ \ c'\le0,\ \ d'\le0,\\
		&&
		&\begin{pmatrix}
			& \vec{\gamma}_1^{(7)T} &\\
			& \vec{\gamma}_2^{(7)T} &\\
			& \vec{\gamma}_3^{(7)T} &\\
			& \vec{\gamma}_4^{(7)T} &
		\end{pmatrix}\cdot
		(a',b',c',d',e')^{T}
		\le
		\begin{pmatrix}
			r_n\\
			r_n\\
			r_n\\
			r_n
		\end{pmatrix}
	\end{aligned}\quad
\right]
\end{equation}
is defined.
Applying a similar consideration to the case $n\equiv1$ presented in \ref{app_LP}, we can confirm that 
\begin{equation}
	\begin{aligned}
		(a',b',c',d',e')=(a_\star,b_\star,c_\star,d_\star,e_\star)
		=(\cos((2m+2)\theta_n), \sin((2m+2)\theta_n), 0, -\cos((2m+2)\theta_n),0)
	\end{aligned}
\end{equation}
induced from the maximally entangled state $\hat{\eta}^{\mathrm{ME}}_\star$ is its optimal solution.

\subsection{The cases $n\equiv3,5$}
The cases $n\equiv3,5$ can be treated in the same way as $n\equiv1,7$.
The maximally entangled state $\hat{\eta}^{\mathrm{ME}}_\star$ for $n\equiv3,5$ is
\begin{equation}
	\hat{\eta}^{\mathrm{ME}}_\star=
	\left\{
	\begin{aligned}
		&\begin{pmatrix}
			\cos\left(\frac{-3n+1}{4}\theta_n\right) & \sin\left(\frac{-3n+1}{4}\theta_n\right) & 0\\
			\sin\left(\frac{-3n+1}{4}\theta_n\right)& -\cos\left(\frac{-3n+1}{4}\theta_n\right) & 0\\
			0 & 0 & 1
		\end{pmatrix}\quad&&(n\equiv3)\\
	&\begin{pmatrix}
			\cos\left(\frac{-3n-1}{4}\theta_n\right) & \sin\left(\frac{-3n-1}{4}\theta_n\right) & 0\\
		\sin\left(\frac{-3n-1}{4}\theta_n\right)& -\cos\left(\frac{-3n-1}{4}\theta_n\right) & 0\\
		0 & 0 & 1
	\end{pmatrix}&&(n\equiv5).
	\end{aligned}
\right.
	\label{opt ME2_n=3,5}
\end{equation}
We again consider
\begin{equation}\label{key}
	\eta_\star^\epsilon=(1-\epsilon)\eta^{\mathrm{ME}}_{\star}+\epsilon\eta_{\star}=
	\begin{pmatrix}
		a' & b' & c'\\
		b' & d' & e'\\
		c' & e' & 1
	\end{pmatrix}
\end{equation}
with sufficiently small $\epsilon\in(0,1)$ so that 
\begin{equation}\label{cond00}
	a'\le0,\quad  b'\le0,\quad  d'\ge0.
\end{equation}
For $n\equiv3,5$, we introduce normal vectors
\begin{equation}\label{normal 0_app3}
	l_1\!\left(\frac{n-n_\star}{2}\right)\!=\begin{pmatrix}
		-1\\
		\frac{1}{\sin2\theta_n}-\frac{1}{\tan2\theta_n}\\
		-r_n
	\end{pmatrix},\quad
	l_2\!\left(\frac{n-n_\star}{2}\right)\!=\begin{pmatrix}
		-1\\
		-\frac{1}{\sin2\theta_n}+\frac{1}{\tan2\theta_n}\\
		-r_n
	\end{pmatrix}
\end{equation}
with respect to the hyperplanes $L_1(\frac{n-n_\star}{2})$ and $L_2(\frac{n-n_\star}{2})$ spanned by $\{\omega_n(\frac{n-n_\star}{2}), \omega_n(\frac{n-n_\star}{2}+1), O\}$ and $\{\omega_n(\frac{n-n_\star}{2}), \omega_n(\frac{n-n_\star}{2}-1), O\}$ respectively.
These normal vectors are set as references instead of \eqref{normal 0}.
Similarly to the previous cases $n\equiv1,7$, we define 
\begin{equation}\label{key}
	m=\left\{
	\begin{aligned}
		\frac{n-3}{8}\qquad(n\equiv3)\\
		\frac{n-5}{8}\qquad(n\equiv5)
	\end{aligned}
	\right.
\end{equation}
and require
\begin{equation}\label{cond_app1}
	\begin{aligned}
		&\ang{l_1\!\left(\frac{n-n_\star}{2}+\alpha_1\right)\!,~ \hat{\eta}_\star^\epsilon\circ\hat{\eta}_\star^{\mathrm{ME}}\left(e_n\!\left(\frac{n-n_\star}{2}+\alpha_1\right)\!\right)}\le0,\\
		&\ang{l_2\!\left(\frac{n-n_\star}{2}+\alpha_1\right)\!,~ \hat{\eta}_\star^\epsilon\circ\hat{\eta}_\star^{\mathrm{ME}}\left(e_n\!\left(\frac{n-n_\star}{2}+\alpha_1\right)\!\right)}\le0,\\
		&\ang{l_i\!\left(\frac{n-n_\star}{2}+\alpha_2\right)\!,~ \hat{\eta}_\star^\epsilon\circ\hat{\eta}_\star^{\mathrm{ME}}\left(e_n\!\left(\frac{n-n_\star}{2}+\alpha_2\right)\!\right)}\le0,\\
		&\ang{l_j\!\left(\frac{n-n_\star}{2}+\alpha_3\right)\!,~ \hat{\eta}_\star^\epsilon\circ\hat{\eta}_\star^{\mathrm{ME}}\left(e_n\!\left(\frac{n-n_\star}{2}+\alpha_3\right)\!\right)}\le0
	\end{aligned}
\end{equation}
with 
\begin{equation}\label{alpha_i_3}
	(i,j; \alpha_1,\alpha_2,\alpha_3)=\left\{
	\begin{aligned}
		&(1,1; 2m+1,0,-3m-1)\quad&&(n\equiv3)\\
		&(1,2;2m+1,0,-3m-2)\quad&&(n\equiv5,\ n>5).
	\end{aligned}
	\right.
\end{equation}
The conditions \eqref{cond_app1} can be explicitly rewritten as
\begin{align}\label{d3,4_app3}
	&\vec{\gamma}^{(3)T}_{1}\cdot(a',b',c',d',e')^T\le r_n,\quad \vec{\gamma}_{2}^{(3)T}\cdot(a',b',c',d',e')^T\le r_n,\\
	&\vec{\gamma}^{(3)T}_{3}\cdot(a',b',c',d',e')^T\le r_n,\quad \vec{\gamma}_{4}^{(3)T}\cdot(a',b',c',d',e')^T\le r_n
\end{align}
with
\begin{align}
		&\vec{\gamma}_{1}^{(3)}
		=
		\begin{pmatrix}
			-r_n^3\cos((4m+1)\theta_n)\cos((2m+1)\theta_n)\\
			-r_n^3\sin((2m+1)\theta_n)\\
			-r_n^2[\cos((4m+1)\theta_n)+\cos((2m+1)\theta_n)]\\
			r_n^3\sin((4m+1)\theta_n)\sin((2m+1)\theta_n)\\
			-r_n^2[\sin((4m+1)\theta_n)-\sin((2m+1)\theta_n)]
		\end{pmatrix},\\
		&\vec{\gamma}^{(3)}_{2}
		=\begin{pmatrix}
			r_n^3\cos(4m\theta_n)\cos((2m+1)\theta_n)\\
		-r_n^3\sin((2m+2)\theta_n)\\
		-r_n^2[\cos(4m\theta_n)+\cos((2m+1)\theta_n)]\\
		r_n^3\sin(4m\theta_n)\sin((2m+1)\theta_n)\\
		-r_n^2[\sin(4m\theta_n)-\sin((2m+1)\theta_n)]
		\end{pmatrix},\\
	\end{align}
\begin{equation}\label{key}
	\vec{\gamma}_{3}^{(3)}
	=\begin{pmatrix}
		-r_n\cos((2m+1)\theta_n)\\
		-r_n^3\sin(2m\theta_n)\\
		-1-r_n^2\cos((2m+1)\theta_n)\\
		r_n^3\sin((2m+1)\theta_n)\sin\theta_n\\
		-r_n^2[\sin((2m+1)\theta_n)-\sin\theta_n]
	\end{pmatrix},\quad
	\vec{\gamma}_{4}^{(3)}
	=\begin{pmatrix}
		-r_n^3\cos(2m\theta_n)\\
		-r_n^3\sin(2m\theta_n)\\
		r_n^2[1+\cos(2m\theta_n)]\\
		0\\
		r_n^2\sin(2m\theta_n)
	\end{pmatrix}
\end{equation}
for $n\equiv3$, and 
\begin{align}\label{d3,4_app5}
	&\vec{\gamma}^{(5)T}_{1}\cdot(a',b',c',d',e')^T\le r_n,\quad \vec{\gamma}_{2}^{(5)T}\cdot(a',b',c',d',e')^T\le r_n,\\
	&\vec{\gamma}^{(5)T}_{3}\cdot(a',b',c',d',e')^T\le r_n,\quad \vec{\gamma}_{4}^{(5)T}\cdot(a',b',c',d',e')^T\le r_n
\end{align}
with
\begin{align}
		&\vec{\gamma}_{1}^{(5)}
		=
		\begin{pmatrix}
			-r_n^3\cos((4m+1)\theta_n)\cos((2m+1)\theta_n)\\
			-r_n^3\sin(2m\theta_n)\\
			-r_n^2[\cos((4m+1)\theta_n)+\cos((2m+1)\theta_n)]\\
			r_n^3\sin((4m+1)\theta_n)\sin((2m+1)\theta_n)\\
			-r_n^2[\sin((4m+1)\theta_n)-\sin((2m+1)\theta_n)]
		\end{pmatrix},\\
		&\vec{\gamma}^{(5)}_{2}
		=\begin{pmatrix}
			r_n^3\cos((4m+2)\theta_n)\cos((2m+1)\theta_n)\\
			-r_n^3\sin((2m+2)\theta_n)\\
			-r_n^2[\cos((4m+2)\theta_n)+\cos((2m+1)\theta_n)]\\
			r_n^3\sin((4m+2)\theta_n)\sin((2m+1)\theta_n)\\
			-r_n^2[\sin((4m+2)\theta_n)-\sin((2m+1)\theta_n)]
		\end{pmatrix},\\
	\end{align}
\begin{equation}\label{key}
	\vec{\gamma}_{3}^{(5)}
	=\begin{pmatrix}
		-r_n\cos((2m+1)\theta_n)\\
		-r_n^3\sin(2m\theta_n)\\
		-1-r_n^2\cos((2m+1)\theta_n)\\
		r_n^3\sin((2m+1)\theta_n)\sin\theta_n\\
		-r_n^2[\sin((2m+1)\theta_n)-\sin\theta_n]
	\end{pmatrix},\quad
	\vec{\gamma}_{4}^{(5)}
	=\begin{pmatrix}
		-r_n^3\cos((2m+2)\theta_n)\\
		-r_n^3\sin((2m+2)\theta_n)\\
		r_n^2[1+\cos((2m+2)\theta_n)]\\
		0\\
		r_n^2\sin((2m+2)\theta_n)
	\end{pmatrix}
\end{equation}
for $n\equiv5$ $(n>5)$.
When $n=5$ ($m=0$), we require
\begin{equation}\label{cond_app5}
	\begin{aligned}
		&\ang{l_1\!\left(\frac{n-n_\star}{2}\right)\!,~ \hat{\eta}_\star^\epsilon\circ\hat{\eta}_\star^{\mathrm{ME}}\left(e_n\!\left(\frac{n-n_\star}{2}\right)\!\right)}\le0,\\
		&\ang{l_2\!\left(\frac{n-n_\star}{2}\right)\!,~ \hat{\eta}_\star^\epsilon\circ\hat{\eta}_\star^{\mathrm{ME}}\left(e_n\!\left(\frac{n-n_\star}{2}\right)\!\right)}\le0,\\
		&\ang{l_2\!\left(\frac{n-n_\star}{2}+\alpha_1\right)\!,~ \hat{\eta}_\star^\epsilon\circ\hat{\eta}_\star^{\mathrm{ME}}\left(e_n\!\left(\frac{n-n_\star}{2}+\alpha_1\right)\!\right)}\le0,\\
		&\ang{l_2\!\left(\frac{n-n_\star}{2}+\alpha_2\right)\!,~ \hat{\eta}_\star^\epsilon\circ\hat{\eta}_\star^{\mathrm{ME}}\left(e_n\!\left(\frac{n-n_\star}{2}+\alpha_2\right)\!\right)}\le0
	\end{aligned}
\end{equation}
with 
\begin{equation}\label{alpha_i_5}
	(\alpha_1,\alpha_2)=(2m+2,-3m-1)=(2,-1).
\end{equation}
They are explicitly expressed as
\begin{align}\label{d3,4_0_app3}
	&\vec{\gamma}^{(n=5)T}_{1}\cdot(a',b',c',d',e')^T\le r_n,\quad \vec{\gamma}_{2}^{(n=5)T}\cdot(a',b',c',d',e')^T\le r_n,\\
	&\vec{\gamma}^{(n=5)T}_{3}\cdot(a',b',c',d',e')^T\le r_n,\quad \vec{\gamma}_{4}^{(n=5)T}\cdot(a',b',c',d',e')^T\le r_n
\end{align}
with
\begin{align}
	&\vec{\gamma}_{1}^{(n=5)}
	=
	\begin{pmatrix}
		-r_5\cos\theta_5\\
	0\\
	-1-r_5^2\cos\theta_5\\
	0\\
	0
	\end{pmatrix},\quad
	\vec{\gamma}^{(n=5)}_{2}
	=\begin{pmatrix}
		-r_5\cos\theta_5\\
		-r_5^3\sin2\theta_5\\
		-1-r_5^2\cos\theta_5\\
	0\\
		-2r_5^2\sin\theta_5
	\end{pmatrix},\\
	&\vec{\gamma}_{3}^{(n=5)}
	=\begin{pmatrix}
		-r_5^3\cos3\theta_5\\
		-r_5^3\sin3\theta_5\\
		r_5^2[1-\cos3\theta_5]\\
		0\\
		r_5^2\sin3\theta_5
	\end{pmatrix},\quad
	\vec{\gamma}_{4}^{(n=5)}
	=\begin{pmatrix}
	r_5^3\cos2\theta_5\cos\theta_5\\
	-r_5^3\sin3\theta_5\\
	r_5^2[\cos2\theta_5-\cos\theta_5]\\
	-r_5^3\sin2\theta_5\sin\theta_5\\
		-r_5^2[\sin2\theta_5-\sin\theta_5]
	\end{pmatrix}.
\end{align}
Now linear programming problems 
\begin{align}\label{key}
	(n\equiv3)\qquad&\left[\quad
	\begin{aligned}
		&\mbox{minimize}&&\vec{C}^{T}\cdot(a',b',c',d',e')^{T}\\
		&\mbox{subject to}
		&&a'\le0,\ \ b'\le0,\ \ c'\ge0,\ \ d'\ge0,\\
		&&
		&\begin{pmatrix}
			& \vec{\gamma}_1^{(3)T} &\\
			& \vec{\gamma}_2^{(3)T} &\\
			& \vec{\gamma}_3^{(3)T} &\\
			& \vec{\gamma}_4^{(3)T} &
		\end{pmatrix}\cdot
		(a',b',c',d',e')^{T}
		\le
		\begin{pmatrix}
			r_n\\
			r_n\\
			r_n\\
			r_n
		\end{pmatrix}
	\end{aligned}
	\quad\right],\\
	(n\equiv5,\ n>5)\qquad
	&\left[\quad
	\begin{aligned}
		&\mbox{minimize}&&\vec{C}^{T}\cdot(a',b',c',d',e')^{T}\\
		&\mbox{subject to}
		&&a'\le0,\ \ b'\le0,\ \ c'\le0,\ \ d'\ge0,\\
		&&
		&\begin{pmatrix}
			& \vec{\gamma}_1^{(5)T} &\\
			& \vec{\gamma}_2^{(5)T} &\\
			& \vec{\gamma}_3^{(5)T} &\\
			& \vec{\gamma}_4^{(5)T} &
		\end{pmatrix}\cdot
		(a',b',c',d',e')^{T}
		\le
		\begin{pmatrix}
			r_n\\
			r_n\\
			r_n\\
			r_n
		\end{pmatrix}
	\end{aligned}
	\quad\right],\\
	(n=5)\qquad
	&\left[\quad
	\begin{aligned}
		&\mbox{minimize}&&\vec{C}^{T}\cdot(a',b',c',d',e')^{T}\\
		&\mbox{subject to}
		&&a'\le0,\ \ b'\le0,\ \ c'\le0,\ \ d'\ge0,\\
		&&
		&\begin{pmatrix}
			& \vec{\gamma}_1^{(n=5)T} &\\
			& \vec{\gamma}_2^{(n=5)T} &\\
			& \vec{\gamma}_3^{(n=5)T} &\\
			& \vec{\gamma}_4^{(n=5)T} &
		\end{pmatrix}\cdot
		(a',b',c',d',e')^{T}
		\le
		\begin{pmatrix}
			r_n\\
			r_n\\
			r_n\\
			r_n
		\end{pmatrix}
	\end{aligned}
	\quad\right].
\end{align}
are defined.
Applying a similar consideration to the case $n\equiv1$ (see \ref{app_LP}), we can confirm that 
\begin{equation}
	\begin{aligned}
		(a',b',c',d',e')=(a_\star,b_\star,c_\star,d_\star,e_\star)
		=(\cos((2m+2)\theta_n), \sin((2m+2)\theta_n), 0, -\cos((2m+2)\theta_n),0)
	\end{aligned}
\end{equation}
induced from the maximally entangled state $\hat{\eta}^{\mathrm{ME}}_\star$ is its optimal solution.

\def\thesection{Appendix\ \Alph{section}}
\section{Proof of Proposition~\ref{prop:min_max}}
\label{app_min}
\renewcommand{\theequation}{B.\arabic{equation}}
\renewcommand{\thesection}{\Alph{section}}
\setcounter{equation}{0}
\setcounter{subsection}{0}
\renewtheorem{lem}[subsection]{Lemma}
The proof of Proposition~\ref{prop:min_max} proceeds in the same way as that of Theorem~\ref{thm:main} presented in Subsec.~\ref{subsec:proof for odd} and \ref{app_LP_others}.
We continue using notations and coordinates introduced there.
\subsection{The cases $n\equiv1,7$}
We first consider the case $n\equiv1,7$.
Suppose that 
\begin{equation}\label{app_suppose}
		H_n(n_\star)\le-C[\eta_{\diamond};\E_n(i_{\diamond}),\E_n(j_{\diamond});\E_n(k_{\diamond}),\E_n(l_{\diamond})](=|C[\eta_{\diamond};\E_n(i_{\diamond}),\E_n(j_{\diamond});\E_n(k_{\diamond}),\E_n(l_{\diamond})]|)
\end{equation}
holds with some $\eta_{\diamond}\in\Omega_n\otimes_{max}\Omega_n$ and $i_{\diamond},j_{\diamond},k_{\diamond},l_{\diamond}\in\{0,\ldots,n-1\}$.
Because we have \eqref{eq:lem2}, the observables are of the form $(\E_n(n_\star),\E_n(0); \E_n(n_\star),\E_n(0))$ and the state can be set to be self-adjoint:
\begin{equation}\label{key}
	\eta_{\diamond}=
	\begin{pmatrix}
		a & b & c\\
		b & d & e\\
		c & e & 1
	\end{pmatrix}.
\end{equation}
Introducing a maximally entangled state
\begin{align}\label{key}
	\hat{\eta}^{\mathrm{ME}}_{\diamond}
	&=\begin{pmatrix}
		\cos(2K_n\theta_n) & -\sin(2K_n\theta_n) & 0\\
		-\sin(2K_n\theta_n)& -\cos(2K_n\theta_n) & 0\\
		0 & 0 & 1
	\end{pmatrix}\\
&=\left\{
\begin{aligned}
	&\begin{pmatrix}
		\cos\left(\frac{3n-3}{4}\theta_n\right) & -\sin\left(\frac{3n-3}{4}\theta_n\right) & 0\\
		-\sin\left(\frac{3n-3}{4}\theta_n\right)& -\cos\left(\frac{3n-3}{4}\theta_n\right) & 0\\
		0 & 0 & 1
	\end{pmatrix}&&\quad(n\equiv1)\\
	&\begin{pmatrix}
	\cos\left(\frac{3n+3}{4}\theta_n\right) & -\sin\left(\frac{3n+3}{4}\theta_n\right) & 0\\
	-\sin\left(\frac{3n+3}{4}\theta_n\right)& -\cos\left(\frac{3n+3}{4}\theta_n\right) & 0\\
		0 & 0 & 1
	\end{pmatrix}&&\quad(n\equiv7),
\end{aligned}
\right.
\end{align}
we can derive 
\begin{equation}\label{prop_claim}
	C[\eta_{\diamond};\E_n(i_{\diamond}),\E_n(j_{\diamond});\E_n(k_{\diamond}),\E_n(l_{\diamond})]
	=C[\eta^{\mathrm{ME}}_{\diamond};\E_n(n_\star),\E_n(0); \E_n(n_\star),\E_n(0)]
\end{equation}
with a similar method to Theorem~\ref{thm:main}.
To see this, we consider
\begin{equation}\label{key}
	\eta_\diamond^\epsilon=(1-\epsilon)\eta^{\mathrm{ME}}_{\diamond}+\epsilon\eta_{\diamond}=
	\begin{pmatrix}
		a' & b' & c'\\
		b' & d' & e'\\
		c' & e' & 1
	\end{pmatrix}.
\end{equation}
In this expression, $c'\ge0$ for $n\equiv1$ or $c'\le0$ for $n\equiv7$ holds due to the assumption \eqref{app_suppose} (see \eqref{eq:cond0}), and $\epsilon$ is taken sufficiently small so that $a'\le0,\ b'\le0,\ d'\ge0$ hold for both cases.
For $\eta_\diamond^\epsilon$ to be a valid state, we require
\begin{equation}\label{key}
	\begin{aligned}
		&\ang{l_1\!\left(\frac{n_\star}{2}+\alpha_1\right)\!,~ \hat{\eta}_\diamond^\epsilon\circ\hat{\eta}_{\diamond}^{\mathrm{ME}}\left(e_n\!\left(\frac{n_\star}{2}+\alpha_1\right)\!\right)}\le0,\ \ 
		\ang{l_2\!\left(\frac{n_\star}{2}+\alpha_1\right)\!,~ \hat{\eta}_\diamond^\epsilon\circ\hat{\eta}_{\diamond}^{\mathrm{ME}}\left(e_n\!\left(\frac{n_\star}{2}+\alpha_1\right)\!\right)}\le0,\\
		&\ang{l_i\!\left(\frac{n_\star}{2}+\alpha_2\right)\!,~ \hat{\eta}_\diamond^\epsilon\circ\hat{\eta}_{\diamond}^{\mathrm{ME}}\left(e_n\!\left(\frac{n_\star}{2}+\alpha_2\right)\!\right)}\le0,\ \ 
		\ang{l_j\!\left(\frac{n_\star}{2}+\alpha_3\right)\!,~ \hat{\eta}_\diamond^\epsilon\circ\hat{\eta}_{\diamond}^{\mathrm{ME}}\left(e_n\!\left(\frac{n_\star}{2}+\alpha_3\right)\!\right)}\le0
	\end{aligned}
\end{equation}
with 
\begin{equation}\label{app_alpha_i_3}
	(i,j; \alpha_1,\alpha_2,\alpha_3)=\left\{
	\begin{aligned}
		&\left(2,2; \frac{n+1}{2}, \frac{n+3}{4},\frac{n-1}{8}\right)\quad&&(n\equiv1)\\
		&\left(1,1; \frac{-3n-3}{8},\frac{n+1}{4},\frac{n-7}{8}\right) \quad&&(n\equiv7).
	\end{aligned}
	\right.
\end{equation}
We express these conditions in a simpler form as
\begin{align}\label{d3,4_0_app(1)}
	&\vec{\delta}^{(1)T}_{1}\cdot(a',b',c',d',e')^T\le r_n,\quad \vec{\delta}_{2}^{(1)T}\cdot(a',b',c',d',e')^T\le r_n,\\
	&\vec{\delta}^{(1)T}_{3}\cdot(a',b',c',d',e')^T\le r_n,\quad \vec{\delta}_{4}^{(1)T}\cdot(a',b',c',d',e')^T\le r_n
\end{align}
and
\begin{align}\label{d3,4_0_app(7)}
	&\vec{\delta}^{(7)T}_{1}\cdot(a',b',c',d',e')^T\le r_n,\quad \vec{\delta}_{2}^{(7)T}\cdot(a',b',c',d',e')^T\le r_n,\\
	&\vec{\delta}^{(7)T}_{3}\cdot(a',b',c',d',e')^T\le r_n,\quad \vec{\delta}_{4}^{(7)T}\cdot(a',b',c',d',e')^T\le r_n
\end{align}
respectively for the cases $n\equiv1$ and $n\equiv7$.
With a similar method in \ref{app_LP}, it can be shown that the maximally entangled state $\eta^{\mathrm{ME}}_{\diamond}$ gives optimal solutions of the linear programming problems
\begin{equation}\label{key}
	(n\equiv1)\qquad\left[\quad
	\begin{aligned}
		&\mbox{minimize}&&\vec{C}^{T}\cdot(a',b',c',d',e')^{T}\\
		&\mbox{subject to}
		&&a'\le0,\ \ b'\le0,\ \ c'\ge0,\ \ d'\ge0,\\
		&&
		&\begin{pmatrix}
			& \vec{\delta}_1^{(1)T} &\\
			& \vec{\delta}_2^{(1)T} &\\
			& \vec{\delta}_3^{(1)T} &\\
			& \vec{\delta}_4^{(1)T} &
		\end{pmatrix}\cdot
		(a',b',c',d',e')^{T}
		\le
		\begin{pmatrix}
			r_n\\
			r_n\\
			r_n\\
			r_n
		\end{pmatrix}
	\end{aligned}
\quad\right],
\end{equation}
\begin{equation}\label{key}
		(n\equiv7)\qquad
		\left[\quad
	\begin{aligned}
		&\mbox{minimize}&&\vec{C}^{T}\cdot(a',b',c',d',e')^{T}\\
		&\mbox{subject to}
		&&a'\le0,\ \ b'\le0,\ \ c'\le0,\ \ d'\ge0,\\
		&&
		&\begin{pmatrix}
			& \vec{\delta}_1^{(7)T} &\\
			& \vec{\delta}_2^{(7)T} &\\
			& \vec{\delta}_3^{(7)T} &\\
			& \vec{\delta}_4^{(7)T} &
		\end{pmatrix}\cdot
		(a',b',c',d',e')^{T}
		\le
		\begin{pmatrix}
			r_n\\
			r_n\\
			r_n\\
			r_n
		\end{pmatrix}
	\end{aligned}
\quad\right].
\end{equation}
Now \eqref{prop_claim} is verified, but it contradicts \eqref{app_suppose} because
\begin{equation}\label{key}
	H_n(n_\star)>-C[\eta^{\mathrm{ME}}_{\diamond};\E_n(n_\star),\E_n(0); \E_n(n_\star),\E_n(0)]
\end{equation}
holds for each $n\equiv1,7$.

\subsection{The cases $n\equiv3,5$}
We can make similar arguments for the cases $n\equiv3,5$.
In these cases, we introduce maximally entangled states
\begin{align}\label{ME37}
	\hat{\eta}^{\mathrm{ME}}_{\diamond}
	&=\begin{pmatrix}
		\cos(2K_n\theta_n) & -\sin(2K_n\theta_n) & 0\\
		-\sin(2K_n\theta_n)& -\cos(2K_n\theta_n) & 0\\
		0 & 0 & 1
	\end{pmatrix}\\
	&=\left\{
	\begin{aligned}
		&\begin{pmatrix}
			\cos\left(\frac{n-3}{4}\theta_n\right) & \sin\left(\frac{n-3}{4}\theta_n\right) & 0\\
			\sin\left(\frac{n-3}{4}\theta_n\right)& -\cos\left(\frac{n-3}{4}\theta_n\right) & 0\\
			0 & 0 & 1
		\end{pmatrix}&&\quad(n\equiv3)\\
		&\begin{pmatrix}
			\cos\left(\frac{n+3}{4}\theta_n\right) & \sin\left(\frac{n+3}{4}\theta_n\right) & 0\\
			\sin\left(\frac{n+3}{4}\theta_n\right)& -\cos\left(\frac{n+3}{4}\theta_n\right) & 0\\
			0 & 0 & 1
		\end{pmatrix}&&\quad(n\equiv5).
	\end{aligned}
	\right.
\end{align}
They satisfy 
\begin{equation}\label{key}
	H_n(n_\star)>C[\eta^{\mathrm{ME}}_{\diamond};\E_n(n_\star),\E_n(0); \E_n(n_\star),\E_n(0)].
\end{equation}
We again consider
\begin{equation}\label{key}
	\eta_\diamond^\epsilon=(1-\epsilon)\eta^{\mathrm{ME}}_{\diamond}+\epsilon\eta_{\diamond}=
	\begin{pmatrix}
		a' & b' & c'\\
		b' & d' & e'\\
		c' & e' & 1
	\end{pmatrix}.
\end{equation}
with sufficiently small $\epsilon$.
For this state $\eta_\diamond^\epsilon$, we impose
\begin{equation}\label{key}
	\begin{aligned}
		&\ang{l_1\!\left(\frac{n-n_\star}{2}+\alpha_1\right)\!,~ \hat{\eta}_\diamond^\epsilon\circ\hat{\eta}_{\diamond}^{\mathrm{ME}}\left(e_n\!\left(\frac{n-n_\star}{2}+\alpha_1\right)\!\right)}\le0,\\
		&\ang{l_2\!\left(\frac{n-n_\star}{2}+\alpha_1\right)\!,~ \hat{\eta}_\diamond^\epsilon\circ\hat{\eta}_{\diamond}^{\mathrm{ME}}\left(e_n\!\left(\frac{n-n_\star}{2}+\alpha_1\right)\!\right)}\le0,\\
		&\ang{l_i\!\left(\frac{n-n_\star}{2}+\alpha_2\right)\!,~ \hat{\eta}_\diamond^\epsilon\circ\hat{\eta}_{\diamond}^{\mathrm{ME}}\left(e_n\!\left(\frac{n-n_\star}{2}+\alpha_2\right)\!\right)}\le0,\\
		&\ang{l_j\!\left(\frac{n-n_\star}{2}+\alpha_3\right)\!,~ \hat{\eta}_\diamond^\epsilon\circ\hat{\eta}_{\diamond}^{\mathrm{ME}}\left(e_n\!\left(\frac{n-n_\star}{2}+\alpha_3\right)\!\right)}\le0
	\end{aligned}
\end{equation}
with 
\begin{equation}\label{app_alpha_i}
	(i,j; \alpha_1,\alpha_2,\alpha_3)=\left\{
	\begin{aligned}
		&\left(2,2; -\frac{n+5}{8},\frac{n-3}{8},\frac{3n-1}{8}\right)\quad&&(n\equiv3)\\
		&\left(1,1; ,-\frac{n+11}{8},\frac{n+3}{8},\frac{n+3}{4}\right) \quad&&(n\equiv5,\ n>5)\\
		&\left(1,1; -1,1,2\right) \quad&&(n=5).
	\end{aligned}
	\right.
\end{equation}
They are rewritten simply as
\begin{align}\label{d3,4_0_app(3)}
	&\vec{\delta}^{(3)T}_{1}\cdot(a',b',c',d',e')^T\le r_n,\quad \vec{\delta}_{2}^{(3)T}\cdot(a',b',c',d',e')^T\le r_n,\\
	&\vec{\delta}^{(3)T}_{3}\cdot(a',b',c',d',e')^T\le r_n,\quad \vec{\delta}_{4}^{(3)T}\cdot(a',b',c',d',e')^T\le r_n,
\end{align}
\begin{align}\label{d3,4_0_app(5)}
	&\vec{\delta}^{(5)T}_{1}\cdot(a',b',c',d',e')^T\le r_n,\quad \vec{\delta}_{2}^{(5)T}\cdot(a',b',c',d',e')^T\le r_n,\\
	&\vec{\delta}^{(5)T}_{3}\cdot(a',b',c',d',e')^T\le r_n,\quad \vec{\delta}_{4}^{(5)T}\cdot(a',b',c',d',e')^T\le r_n,
\end{align}
and 
\begin{align}\label{d3,4_0_app(5)}
	&\vec{\delta}^{(n=5)T}_{1}\cdot(a',b',c',d',e')^T\le r_n,\quad \vec{\delta}_{2}^{(n=5)T}\cdot(a',b',c',d',e')^T\le r_n,\\
	&\vec{\delta}^{(n=5)T}_{3}\cdot(a',b',c',d',e')^T\le r_n,\quad \vec{\delta}_{4}^{(n=5)T}\cdot(a',b',c',d',e')^T\le r_n
\end{align}
respectively.
They induce the following linear programming problems
\begin{align}\label{key}
	(n\equiv3)\qquad&\left[\quad
	\begin{aligned}
		&\mbox{minimize}&&\vec{C}^{T}\cdot(a',b',c',d',e')^{T}\\
		&\mbox{subject to}
		&&a'\le0,\ \ b'\le0,\ \ c'\ge0,\ \ d'\ge0,\\
		&&
		&\begin{pmatrix}
			& \vec{\delta}_1^{(3)T} &\\
			& \vec{\delta}_2^{(3)T} &\\
			& \vec{\delta}_3^{(3)T} &\\
			& \vec{\delta}_4^{(3)T} &
		\end{pmatrix}\cdot
		(a',b',c',d',e')^{T}
		\le
		\begin{pmatrix}
			r_n\\
			r_n\\
			r_n\\
			r_n
		\end{pmatrix}
	\end{aligned}
	\quad\right],\\
	(n\equiv5,\ n>5)\qquad
	&\left[\quad
	\begin{aligned}
		&\mbox{minimize}&&\vec{C}^{T}\cdot(a',b',c',d',e')^{T}\\
		&\mbox{subject to}
		&&a'\le0,\ \ b'\le0,\ \ c'\le0,\ \ d'\ge0,\\
		&&
		&\begin{pmatrix}
			& \vec{\delta}_1^{(5)T} &\\
			& \vec{\delta}_2^{(5)T} &\\
			& \vec{\delta}_3^{(5)T} &\\
			& \vec{\delta}_4^{(5)T} &
		\end{pmatrix}\cdot
		(a',b',c',d',e')^{T}
		\le
		\begin{pmatrix}
			r_n\\
			r_n\\
			r_n\\
			r_n
		\end{pmatrix}
	\end{aligned}
	\quad\right],\\
	(n=5)\qquad
	&\left[\quad
	\begin{aligned}
		&\mbox{minimize}&&\vec{C}^{T}\cdot(a',b',c',d',e')^{T}\\
		&\mbox{subject to}
		&&a'\ge0,\ \ b'\ge0,\ \ c'\le0,\ \ d'\ge0,\\
		&&
		&\begin{pmatrix}
			& \vec{\delta}_1^{(n=5)T} &\\
			& \vec{\delta}_2^{(n=5)T} &\\
			& \vec{\delta}_3^{(n=5)T} &\\
			& \vec{\delta}_4^{(n=5)T} &
		\end{pmatrix}\cdot
		(a',b',c',d',e')^{T}
		\le
		\begin{pmatrix}
			r_n\\
			r_n\\
			r_n\\
			r_n
		\end{pmatrix}
	\end{aligned}
	\quad\right].
\end{align}
We can confirm that the maximally entangled states \eqref{ME37} are solutions of these problems, and it proves Proposition~\ref{prop:min_max} for $n\equiv3,5$.
\qed

\bibliographystyle{hieeetr_url} 
\bibliography{ref_polygon}

\end{document}